# Deep versus Broad Technology Search and the Timing of Innovation Impact

Likun Cao[†*], James Evans[*‡1]

[†]Purdue University
[*]University of Chicago
[‡]Santa Fe Institute

**Abstract:** This study offers a new perspective on the depth-versus-breadth debate in innovation strategy, by modeling inventive search within dynamic collective knowledge systems, and underscoring the importance of timing for technological impact. Using frontier machine learning to project patent citation networks in hyperbolic space, we analyze 4.9 million U.S. patents to examine how search strategies give rise to distinct temporal patterns in impact accumulation. We find that inventions based on deep search, which relies on a specialized understanding of complex recombination structures, drive higher short-term impact through early adoption within specialized communities, but face diminishing returns as innovations become "locked-in" with limited diffusion potential. Conversely, when inventions are grounded in broad search that spans disparate domains, they encounter initial resistance but achieve wider diffusion and greater long-term impact by reaching cognitively diverse audiences. Individual inventions require both depth and breadth for stable impact. Organizations can strategically balance approaches across multiple inventions: using depth to build reliable technological infrastructure while pursuing breadth to expand applications. We advance innovation theory by demonstrating how deep and broad search strategies distinctly shape the timing and trajectory of technological impact, and how individual inventors and organizations can leverage these mechanisms to balance exploitation and exploration.

[1] Correspondence to James Evans, Knowledge Lab and Sociology, University of Chicago at jevans@uchicago.edu.

**Highlights**

- Using hyperbolic embedding, we model technological evolution as a hierarchical tree.
- This approach captures depth and breadth as independent features of the search path.
- Depth fosters short-term recognition within a local community,
- Breadth predicts long-term impact by reaching a wider audience.
- Organizations can extend robust components from deep search to broad applications.

## 1. Introduction

The recombination framework, which stems from the theory of complex, adaptive systems (Simon, 1991 [1962]), has become a dominant perspective for understanding and predicting technological evolution (Brian Arthur, 2009; Fleming, 2001; Weitzman, 1998; Xiao et al., 2022). Successful recombination requires assembling components into an appropriate architecture that yields novel capability, such as the right level of modularity (Ethiraj et al., 2008) and diversity (Dhanaraj and Parkhe, 2006). In the recombination literature, one of the most established yet debated questions concerns the relative influence of depth and breadth in technological search. Here, depth refers to the intensive use of knowledge within a single or a few closely related domains for the development of new technologies. Breadth involves integrating knowledge from distant and diverse sources into novel inventions. Prior research has reported conflicting evidence regarding their effects, with the "tension" perspective on recombination arguing that distant or diverse knowledge breaks cognitive inertia and stimulates breakthrough ideas (Hall et al., 2001; Trajtenberg et al., 1997; Verhoeven et al., 2016a). Scholarship on the "foundational" perspective, by contrast, celebrates deep immersion in a particular domain, preparing inventors with expertise requisite to identify anomalies and resolve them (Csikszentmihalyi, 2009; Fleming, 2001; Kaplan and Vakili, 2015a; Sternberg and Lubart, 1995). To date, findings from empirical studies remain inconsistent, and underlying mechanisms remain unclear. In this article, we revisit this core topic and offer a solution that reconciles ongoing debate between two seemingly conflicting stances by incorporating the time dimension.

The dual concepts of depth and breadth received earliest attention in the foundational work of March and Simon (1993 [1958]), and continue to inform the practical decision-making of knowledge workers. Despite their theoretical and practical significance, prior research has reported conflicting evidence regarding their effects. Kaplan and Vakili (2015b) suggest that the benefits of these strategies may differ across the types of returns, either cognitive or economic. While this distinction partly addresses the conflicting findings, it does not explain why opposite effects can still occur for the same outcome variable, particularly citations (Fleming, 2001; Uzzi et al., 2013). Due to this empirical fuzziness, subsequent works view depth and breadth as more or less strategic contingent on context, such as the pace of change in knowledge (Teodoridis et al., 2019) and institutional environment (Zheng et al., 2022).

To advance this line of research, we turn to computational methods, formalizing the hierarchical relationships among knowledge components, and modeling depth and breadth as independent search dimensions within a technological landscape that evolves over time. We do this using the Poincaré embedding method, a machine learning algorithm (Nickel and Kiela, 2017) that facilitates empirical-based construction of a hierarchical map from the network of dependencies. This tree-like technological map, constructed from patent citation networks, captures practical proximity between technological categories in collective invention, and provides a stronger foundation for measuring depth and breadth. We precisely operationalize *depth* as distance from the location of the knowledge components to the root of the tree, and *breadth* as the distance between tree branches combined. Through the precise, intuitive, and independent measures of *depth* and *breadth*, we investigate the impact and mechanisms underlying these two strategies, from the perspective of collective intelligence and innovation.

Our results show that the effects of depth and breadth operate across different time scales. While depth fosters short-term recognition within a local community, breadth predicts long-term impact by reaching a wider audience. Based on these empirical patterns and our exploration of underlying mechanisms, we argue that the inconsistency in empirical findings likely stems from two sources: the reliance on small, limited samples from specific industries (Terjesen and Patel, 2017; Zhou and Li, 2012), and the common assumption that technological search is an individual activity, overlooking the context of collective invention (Kaplan and Vakili, 2015a). Together, these challenges have led to theoretical ambiguity in the literature today: a proliferation of conflicting findings, limited clarity about the mechanisms, and insufficient attention to collective dynamics. These limitations are difficult to overcome using a traditional, small-data approach, but pose an opportunity for computationally supported social science exploration.

Our work serves as the first to systematically examine deep and broad search and their temporal outcomes using precise, high-resolution technological mappings (Aharonson and Schilling, 2016; Escolar et al., 2023). Existing measures, such as Herfindahl diversity and technological distance (Hall et al., 2001; Trajtenberg et al., 1997), primarily capture conceptual distance rather than the functional distance of technologies in practice (Benson and Magee, 2015; Magee et al., 2016). The static nature of conceptual taxonomies makes it difficult to identify the temporal patterns, as we observe here. Moreover, previous studies treat depth and breadth as opposite ends of a single continuum, rather than as two independent dimensions (Kaplan and Vakili, 2015a; Trajtenberg et al., 1997). In contrast, our novel approach uses the structure of the entire knowledge system as a reference frame, accurately capturing technological functions in practice, clearly distinguishing between depth and breadth, and considering collective invention contexts. This method enables us to clearly observe the social dynamics of knowledge diffusion and accumulation in an intuitive and trackable manner, thereby revealing the critical role of time in reconciling the depth versus breadth debate. Beyond interpretation, it also opens opportunities for prediction, making it especially valuable for policymakers and industry decision-makers seeking timely and quantifiable insights into the frontier or technological progress.

The remainder of our paper is organized as follows. We begin by reviewing the literature on recombination theory, focusing on debates surrounding deep versus broad search in technological recombination. Next, we introduce the representation learning framework and show how it can be applied to model the knowledge landscape, and construct measures of depth and breadth. We then assess how depth and breadth influence the impact of patented technologies over time. Finally, we explore mechanisms that may drive the observed empirical patterns.

## 2. Depth versus Breadth

### 2.1 Depth and Breadth Search Strategies in Technological Recombination

The well-known dichotomy of exploration versus exploitation (March, 1991) separates those who burrow deep into a singular domain from those who travel and link many, memorably characterized by Isaiah Berlin as "hedgehogs" and "foxes" (Berlin, 2013 [1953]; also see Dyson, 2015). Recombinant depth has become associated with local search and the exploitation of prior technological success. Deep strategies foster robust understanding, anomaly detection, and the integration of components underlying a technology (Mannucci and Yong, 2018; Teodoridis et al., 2019). By contrast, recombinant breadth has become associated with broad and even global exploratory search, achieved through the radical combination of technological components in novel and unexpected ways (Kaplan and Vakili, 2015b; Trajtenberg et al., 1997).

In the context of technological and commercial development, this distinction was rearticulated as the "foundational" and the "tension" perspectives, which respectively favor depth versus breadth of technological search (Kaplan and Vakili, 2015a). The "foundational" view celebrates deep immersion in a particular domain, as it prepares inventors with an expert mindset to identify anomalies worthy of resolution (Csikszentmihalyi, 2009; Kaplan and Vakili, 2015a;

Sternberg and Lubart, 1995). The "tension" view, by contrast, claims that recombination of distant or diverse knowledge breaks cognitive inertia and stimulates breakthrough ideas (Fleming, 2001; Hall et al., 2001; Trajtenberg et al., 1997; Verhoeven et al., 2016b).

Early studies aimed to determine whether deep versus broad search strategies were more strongly associated with success, sparking sustained debate (Fleming, 2001; Kaplan and Vakili, 2015b; Uzzi et al., 2013). To reconcile these conflicting findings, later research focused on identifying key contingencies. Contexts that mediate the success of search depth versus breadth include the pace of change within a knowledge domain (Teodoridis et al., 2019), network structure shaping information flows (Ter Wal et al., 2016) and the broader institutional environment (Zheng et al., 2022). Depth and breadth might also beneficially be balanced within a firm—a principle called "ambidexterity" (Brahm et al., 2021; Raisch and Birkinshaw, 2008; Stettner and Lavie, 2014). (Terjesen and Patel, 2017; Zhou and Li, 2012)

Conflicting evidence may stem from both the modest scale of data, and wide variation in how depth and breadth are measured. Qualitative information (Boh et al., 2014) and survey methods (Terjesen and Patel, 2017; Zhou and Li, 2012) provide more direct and accurate assessments of search paths, but they are costly and cover only a limited number of respondents. Quantitative measures based on taxonomies, such as the United States Patent Classification (USPC) or International Patent Classification (IPC) for patents (Moorthy and Polley, 2010; Nakamura et al., 2015), can be applied to large-scale datasets, but remain imperfect and oversimplified proxies for their underlying concepts. For example, depth is often calculated as the inverse of breadth (e.g., Kaplan and Vakili, 2015a; Lodh and Battaggion, 2015) or as a use of the same category (Chattopadhyay and Bercovitz, 2020; Kim et al., n.d.; Paruchuri and Awate, 2017), but neither shallowness nor repetition necessarily imply a deep understanding of knowledge in theory. Moreover, because taxonomies reflect the logic of technological components embedded in theoretical knowledge rather than their functional dependencies in practice (Benson and Magee, 2015; Magee et al., 2016; Triulzi et al., 2020), measures based on these taxonomies may not be firmly grounded in the empirical history of invention.

Therefore, due to limitations in both data and measurement, the effects and underlying mechanisms of depth and breadth remain insufficiently understood. Such empirical simplifications also constrain the ability of existing studies to advance theory. To improve on these measures and provide a stronger empirical foundation for depth versus breadth discussion, we introduce collective invention into our modeling, a form of knowledge production that has become increasingly prominent in both academia and industry.

**2.2 Technological Invention as a Collective Effort**

Between the end of the 19th and the early 20th century, the nature of invention underwent systematic transformation, from an individual pursuit to a corporate activity. During this period, inventions previously pursued and legally assigned to individual inventors became predominantly targeted, managed and owned by firms and related research organizations (Lamoreaux and Sokoloff, 1999). These organizations were driven by incentives to make strategic technological investments that would bolster the success of their research and patenting activities. They produced not only novel investigations and landmark patents, but also clusters of defensive patents to legally protect the former (Shapiro, 2000). These developments transformed invention into a collective endeavor, where organizations and communities, rather than individual inventors, now primarily drive both knowledge production and consumption.

Although not fully modeled or measured, the collective nature of invention has been recognized in the literature. For example, research in the economics of innovation highlights social returns to the public as an important component of invention returns when estimating the knowledge spillovers within and across organizational boundaries (Bernstein and Nadiri, 1989; Bloom et al., 2013; Guillard et al., n.d.).

Our collective perspective has two key implications for understanding deep versus broad search. First, from the perspective of knowledge production, depth and breadth should be seen as properties underlying the search paths involved in collective practice. Modern inventions are predominantly discovered through continuous exchanges of information and ideas with colleagues and clients, and not inventors reviewing past technology records or experimenting in isolation. As a result, breadth of search is determined not by conceptual differences between categories, but by how closely or distantly they are applied in industrial practice and technological consensus (Benson and Magee, 2015; Magee et al., 2016). To capture this practical proximity in collective understanding, existing taxonomies are insufficient and alone would distort distance estimation (Ghosh et al., 2016; Teodoridis et al., 2022). A technological mapping that accounts for how technologies are used is required (Aharonson and Schilling, 2016; Escolar et al., 2023; Linzhuo et al., 2020; Teodoridis et al., 2019). In this study, we address this problem by using a representation learning algorithm to incorporate information within citation networks.

Second, and more importantly, if search has an impact on knowledge consumption, it unfolds through collective evaluations and reactions. Both the trajectory and timing of follow-up work are essential for understanding the character of an invention's impact, and both are shaped by social dynamics within technology communities. Imagine two patents with equal citations but which follow different patterns—one remains unnoticed for years before gaining recognition—a "sleeping beauty" (Ke et al., 2015)—while the other receives immediate but short-lived attention (Kang et al., 2024; Silva et al., 2020). These two forms of knowledge obviously have different nature, play different roles in construction of knowledge systems, and have distinct values for both inventors and organizations. To better understand this complexity, we observe knowledge evolution as a dynamic process, and systematically track how an invention is received by its audience, including the speed and spread of recognition, intensity and shifts in audience attention, and patterns of diffusion, etc. To date, these dimensions remain largely underexplored due to methodological limitations in prior work.

In summary, collectives play a central role in shaping how knowledge is produced and consumed through both competitive and coordinated search. By modeling the landscape of collective knowledge and the structure of search paths, we can construct more accurate measures of depth and breadth in practical contexts and, importantly, track how their impacts accumulate dynamically. This approach allows us to examine time as a key dimension separating the distinction between depth and breadth effects. We next explain why building such a model with machine learning algorithms is desirable, guided by theoretical insights from technological evolution and complexity theory.

### 2.3 Theoretical Bases for a Hyperbolic Representation of Collective Knowledge Systems

We base our method on theories of technological evolution, which explain how past inventions shape the emergence of future knowledge. They provide a theoretical basis for empirically analyzing knowledge as a collective system.

Theoretical discussion of technological evolution (Arthur and Polak, 2006; Brian Arthur, 2009) and empirical work on technological trajectories (Fontana et al., 2008; Huenteler et al., 2016) both frame technologies as evolving progressively and recursively. Some technological components serve as the basis for higher-level systems that emerge from them. Based on this insight, technological knowledge, which is created and stored within human collective understanding, should have a hierarchical, tree-like structure. This notion echoes philosophical discussions about knowledge ranging from Aristotle and Augustine of Hippo to the "tree of knowledge" structuring Diderot and D'Alembert's encyclopedia (Aristotle, 1999 [350BCE]; Augustine, 2003 [426]; Weingart, n.d.).

This hierarchy goes hand in hand with "modularity," introduced by Simon (1991 [1962]) in the study of complex systems and later expanded on by Arthur (2009) in the context of technology. In hierarchical, modular knowledge systems, lower level components integrate into robust, tightly connected subsystems, which are recursively combined to create more complex technological designs. One example might be a wind turbine, which is made up of four subsystems: the turbine rotor, power train, mounting and grid connection. The rotor, in turn, is made up of more granular components such as blades, hub and rotor control systems, etc. (Huenteler et al., 2016). Inventing in the field of wind turbines involves altering connections between existing subcomponents, redesigning a subcomponent, or modifying lower-level elements within it. This process can be traced downward to the lowest level where actual technological changes occur.

In this context, deep local search involves delving into the underlying technological infrastructure—refining lower-level knowledge components within submodules. The lower the level inventors explore, the more deeply they engage in the search. This aligns with the "foundational view", which posits that deep local search enables individuals and firms to better understand the underlying connections between components, and to cultivate the familiarity and expertise required to identify anomalies (Mannucci and Yong, 2018; Teodoridis et al., 2019).

Broad search, on the other hand, incorporates knowledge from other domains. In a multi-layered, recursive knowledge system, broad search involves not only different fields, but fields falling on diverging branches of the knowledge tree. For example, integrating quantitative social science with statistics is not as broad as combining it with biology, even if both may seem equally distant in the conceptual taxonomy. This is because statistics has historically contributed to the foundation, or "knowledge roots", of applied social science, whereas biology belongs to a distinctive branch of knowledge.

Depth does not imply narrowness, just as breadth does not imply shallowness. For example, an electrical engineer and a material scientist may form a team bringing together deep expertise in processors and 2D materials to invent a next generation semiconductor. In this case, the team's collective deep understanding of two very different fields may combine into an invention both deep and broad. Meanwhile, some scientists and inventors are generalists who master multiple areas of knowledge and can work across them creatively (Milojević, 2015; Risha et al., 2023; Stark, 2011). Therefore, invented technologies may be broad, deep, both or neither in knowledge recombination. From this theoretical standpoint, we base our measures of depth and breadth on an empirical mapping of technology, and operationalize depth and breadth as independent dimensions of technological recombination. We detail this method in section 2.4.

### 2.4 Constructing a Hyperbolic Representation of the Technological Tree

Innovation theory calls for a model that goes beyond *a priori* classification systems and enables direct measurement of collective invention trajectories. Ideally, such a model captures the hierarchy and practical proximity of technological knowledge, while also distinguishing depth and breadth as independent dimensions. To this end, we turn to hyperbolic geometric representations that natively encode hierarchy and so allow for modeling recursive recombinations in technology.

Hyperbolic space exhibits negative curvature, with the distance between data points increasing exponentially from the center to the margin. Consider a hierarchical tree mapped to a flat, two dimensional Euclidean space. As hierarchical layers ($x$) increase, the number of leaves at each layer grows exponentially ($4^x$) and the distance between leaves at each layer must shrink in the Euclidean space, as the area of a circle grows quadratically relative to its radii. The trade-off between depth (number of layers $x$) and breadth (distance between same-layer leaves) in Euclidean space also finds frequent expression in the operationalization of depth and breadth from existing literature described above (Mueller et al., 2021; Nakamura et al., 2015; Papazoglou and Spanos, 2018; Teodoridis et al., 2019). More Euclidean dimensions cannot solve this problem, failing to keep pace with exponential growth.

Fortunately, Poincaré embedding is designed to capture hierarchical structures and can avoid this distortion: it naturally represents horizontal distances as growing exponentially with depth. Embedding algorithms that project hierarchies within hyperbolic space do so without distortion, and have successfully captured many significant cultural and social hierarchies (Chamberlain et al., 2017; Chami et al., 2019; Fellbaum, 2005; Linzhuo et al., 2020).

Figure 1 offers a conceptual explanation of hyperbolic geometry using popular hyperbolic embedding representation (a 2-D Poincaré disk). When lines and points on a hyperboloid surface are projected onto such a disk representation, lines transform into circular arcs, interior angles within a triangle sum to less than 180°, and distances increase exponentially. This model separates depth and breadth as two separate dimensions. In section 3, we explain the technical details of this algorithm and how it may help generate new theoretical insights.

FIGURE 1 Hyperbolic, Poincaré Disk Embedding

**2.5 Temporal Hypotheses of Depth and Breadth Effects and Potential Mechanisms**

Our method allows for a comprehensive analysis of the temporal dynamics that unfold with patent citation growth as outcomes of depth and breadth in search. We hypothesize that deep search produces short-term impact, whereas broad search contributes to long-term impact, driven by two underlying mechanisms suggested in prior research: lock-in and bridging.

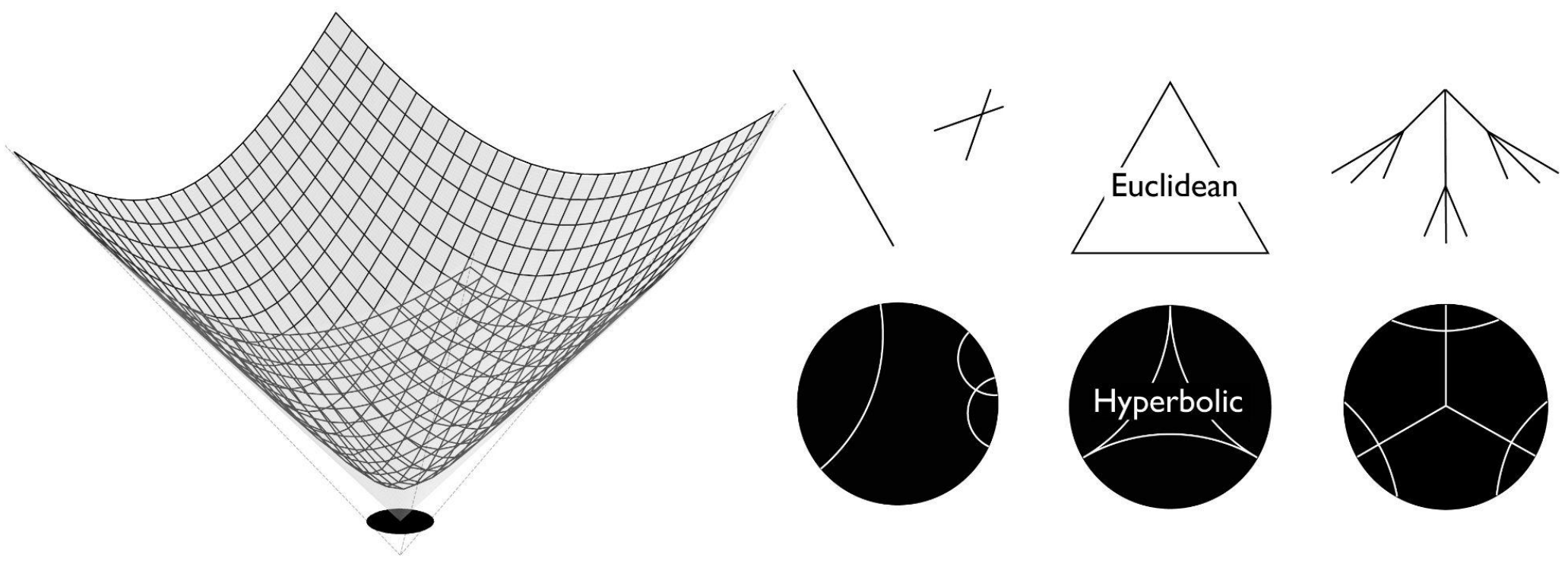

Deep search can generate short-term impact through lock-in. This mechanism has been discussed in various forms in the literature, including, but not limited to: *cognitive lock-in*, where inventors who search deeply develop specialized mental models and routines that make it easier to identify anomalies within their field but harder to recognize opportunities outside it (Dougherty, 1992; Leonard-Barton, 1993; Levinthal and March, 1993; Nelson and Winter, 1990; Tripsas and Gavetti, 2000)*; resource lock-in*, where organizations investing in deep capabilities build specialized assets, skills, and relationships that create switching costs (Christensen and Bower, 1996; Ghemawat, 1991; Henderson and Clark, 1990; Teece, 1986; Williamson, 1985)*; and network lock-in*, where deep search creates dense local networks that reinforce existing trajectories but limit exposure to distant knowledge (Granovetter, 1973; Rosenkopf and Nerkar, 2001; Stuart and Podolny, 2007; Uzzi, 1997).

Conversely, broad search enhances impact through exploratory bridging. It allows inventors to build structural connections across knowledge hierarchies *(structural bridging)* (Fleming and Waguespack, 2007; Hargadon and Sutton, 1997; Tushman and Scanlan, 1981), develop translational capabilities that facilitate knowledge integration *(cognitive bridging) (Carlile, 2004; Hargadon, 2002; Kellogg et al., 2006; Taylor and Greve, 2006)*, and create new diffusion pathways that may require time to mature and become recognized *(temporal bridging)* (Navis and Glynn, 2010; Rogers et al., 2014; Santos and Eisenhardt, 2009; Van de Ven, 1986).

We discuss these mechanisms and empirically explore them, using computationally derived proxies. Although the roles of depth and breadth have been conceived as theoretically contradictory, our computational models demonstrate how these mechanisms could coexist across different time scales and jointly contribute to observed patterns of technological progress.

## 3. Assessing Search Strategies in Patents with Hyperbolic Geometry

In this section, we propose new measures of depth and breadth that account for latent knowledge hierarchies. We first demonstrate this approach by using U.S. patent data to construct Poincaré embedding spaces, then explain how this embedding captures depth and breadth while addressing the theoretical requirements outlined above.

### 3.1 Representation Learning and the Embedding Method

Machine learning has become increasingly popular in social science research in order to (1) aid human coders in processing raw data (Harrison et al., n.d.), (2) extract subtle features from complex unstructured data that are otherwise difficult to quantify (Choi et al., 2021; Choudhury et al., 2019; Sgourev et al., 2023; Vicinanza et al., 2023), and (3) reveal robust patterns from structured data to support theory-building, as a supplement to traditional statistical estimation (Choudhury et al., 2021; Kumar et al., 2022; Shrestha et al., 2021). Among machine learning methods, embedding algorithms have become a powerful tool, which automatically reduces high-dimensional inputs into lower-dimensional expressions, and discover meaningful and predictive representations of raw data that enable feature extraction for subsequent analysis (Aceves and Evans, 2024).

Word embedding is the most prominent embedding method in social science literature, given the prevalent usage of text data and natural language processing tools (Aceves and Evans, 2024). It uses precise word proximity and sequence information to capture semantic relationships, representing words as high-dimensional vectors (Joulin et al., 2017; Kozlowski et al., 2019; Mikolov et al., 2013). Poincaré embedding, on the other hand, belongs to another branch of representation learning—network embedding, which simplifies a raw network for analysis and visualization. Like word embedding, network embedding aims to learn simple and interpretable continuous representations from complex discrete data. While word embedding uses words or words-in-context as basic units for representation, network embedding is designed to represent nodes, edges, or entire networks as representational units.

Many popular network embedding methods, such as node2vec, still represent networks in flat, Euclidean space, but require many dimensions (Grover and Leskovec, 2016; Hamilton et al., 2017; Perozzi et al., 2014). Insofar as knowledge diffusion trajectories manifest hierarchical structure, with core technology categories serving as foundations and building blocks for others, we adopt the Poincaré embedding with hyperbolic geometry, which is designed to naturally encode hierarchy (Chamberlain et al., 2017; Nickel and Kiela, 2017).

### 3.2 Technological Evolution, Citation Networks, and Poincaré Embedding

A common approach to quantify technological evolution trajectories is to construct and analyze citation networks tracing knowledge diffusion between patented inventions (Chang et al., 2009; Feldman and Yoon, 2012). Citation networks reveal the inheritance of knowledge inheritance, as citing patents draw inspiration from those cited.

The knowledge networks have a clear hierarchical structure, stemming from the history of technological evolution. Within such networks, when a class is cited by and combined with diverse other classes, the technology it represents has been deployed in a wide variety of products and processes, suggesting a higher level of "generality" (Feldman and Yoon, 2012; Petralia, 2020; Sterzi, 2013; Trajtenberg et al., 1997). At the extreme, such technologies become "general purpose" technologies like electricity, the internet, and most recently artificial intelligence techniques that have application across the productive economy (Crafts, 2021; Petralia, 2020). By contrast, when a technology class is cited less, but cites other classes more, this suggests a higher level of specificity, such as an implementation or application of an upstream technological element (Fontana et al., 2008).

This hierarchical structure of the knowledge tree can be effectively captured by Poincaré embedding, which is specifically designed to model discrete hierarchies with minimal distortion. Compared to the raw network of citations, where even distant nodes remain connected by edges, Poincaré embedding offers a clearer and more parsimonious representation. Relative to Euclidean embeddings, it requires far fewer dimensions to capture significantly more complex information. In the top three panels of Figure 2, we contrast a network representation with hyperbolic and Euclidean embeddings. With the same dimensionality, hyperbolic embeddings more effectively capture hierarchical relationships within data.

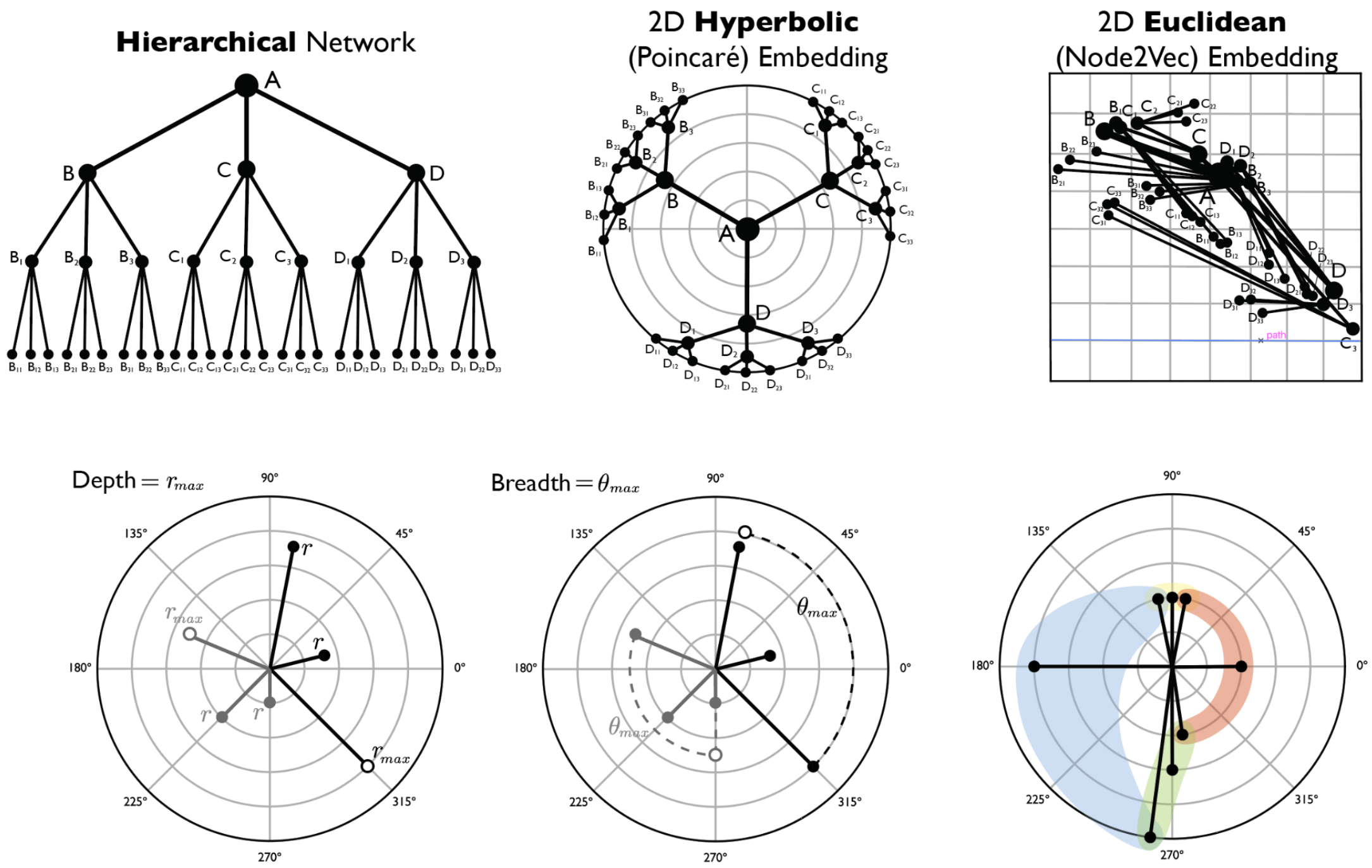

FIGURE 2 Hierarchical Depth and Breadth on a Poincaré Disk

In a 2-D Poincaré embedding disk, nodes possess two dimensions: the "generality" dimension pointing from most to least centered components, and the "similarity" dimension along the periphery where related components are locally clustered, thereby exhibiting a high level of homophily. This is illustrated in the lower 3 panels of Figure 2. The left panel illustrates differences in the radius (*r*) of the Poincaré disk at which embedded technology classes lie. Higher *r* suggests a "deeper" or more specific technology class and its associated patents. The middle panel illustrates differences in the angle ($\theta$) across the Poincaré disk that assess distance across the hierarchy of technology anchoring classes with a patent. Different combinations of depth and breadth generate four types of inventions: deep broad (blue), deep narrow (green), shallow broad (red), and shallow narrow, as illustrated in the right panel. Utilizing this embedding, we re-examine the classical hypothesis of deep versus wide recombination.

### 3.3 Constructing Measures for Depth and Breadth for Patents

We investigate the implications of deep and wide recombination based on 4,992,915 granted patents with identifiable prior art categories in the Poincaré disk. For our analysis, we only utilize patents granted between 1976 and 2017 to facilitate the calculation of outcome variables such as five-year forward citations. We first construct weighted, directed networks between technological categories based on their citation within patents for each 3-year period (with 1976-1978 as the first period, and 2015-2017 as the last), and embed them into Poincaré disks. Then, we deploy measures for recombination depth ($r$) and breadth ($\theta$), and estimate the effects of deep and broad recombination on patent short and long term impact.

The citation network and Poincaré embedding space are built using Python 3, and regression analyses are conducted with Stata 18.

#### 3.3.1 Citation Networks and Poincaré Disks for U.S. Patents, 1976-2017

To construct first-step citation networks, we use the CPC schema, treating each level-4 classification (e.g., A01B33) as an analytical unit. This yields roughly 20,000 technological categories that serve as nodes in the citation network. CPC taxonomies serve as the foundation for stable embedding spaces, with validation tests presented in the Appendix.

We first construct weighted, directed knowledge diffusion networks between CPC technological categories for 14 periods: 1976-1978, 1979-1981, 1982-1984, and so on, with the last period 2015-2017. We eliminate self-loops, as they are irrelevant to positioning classes in relation to one another. In this network, the most frequently and diversely cited categories occupy the center of the hierarchy, indicating a high potential for downstream application use.

We then embed this network into a 2-D disk using the Poincaré Embedding algorithm with training parameter settings that follow (Linzhuo et al., 2020)[2].

This embedding space parsimoniously represents the hierarchical structure of technical knowledge. Each node is assigned polar coordinates ($r$, $\theta$) on the disk. A node near the center (small $r$) receives citations from diverse fields and serves many applications. A node near the edge (large $r$) is used in a focused set of applications. Small differences in $\theta$ with follow-up patents indicate narrow applications, while large $\theta$ differences indicate wide diversity. Because our embeddings cover 14 periods and are time-sensitive, node positions may shift across embeddings. For instance, an integrative technology that evolves into a platform for future inventions may move from the periphery toward the center of the disk, reflecting its changing role over time.

Figure 3 (left panel) presents a real-world example, showing the Poincaré embedding results for patented technologies from 2015 to 2017. Coloring of the nodes is based on the "section symbol"—the highest-level category according to the CPC system. Results indicate that physics (G) and electricity (H) are the most prevalent foundations for other technologies, and they also form an important cluster in downstream applications. Beyond these, chemistry/metallurgy (C), mechanical engineering (F) and performing operations/transportation (B) constitute three other major clusters for applications. Performing operations/transporting (B) are dispersed throughout the space, acting as an adhesive for other technological components. The right panel of figure 3 illustrates three examples for technological tree branches, with subclass families C02F (black), D01D (blue), and G02B (red). Nodes within each family scatter across the disk in cones, representing the evolutionary history through which general-purpose components have combined into increasingly complex applications. For example, within G02B (optical elements), simple and compound lenses (G02B3) make the basis for microscopes (G02B21) and telescopes (G02B23). Within D01D (methods or apparatus in the manufacture of filaments, threads, fibers), formation of filaments (D01D5) is the basis for treatment of filaments-forming (D01D1) and physical treatment of filaments in manufacturing (D01D10).

[2] Batchsize=30, learning rate=0.2, the model is evaluated every 10 epochs. Given the asymmetrical nature of the citation network, we set the parameter symmetrize to be False. We also set the training epochs to be 1000 to guarantee sufficient training, of which the first 20 were "burn-in" epochs.

Statistical validations of this method are provided in Appendix A. These tests show the embedding spaces remain stable across all 14 modeling periods, as visualized using the Procrustes procedure (Hamilton et al., 2016) (Appendix A1), and have efficiently and accurately captured distances between technological categories (Appendix A2). Distances between knowledge branches, as identified by our machine learning algorithm, reflect practical proximities between technological applications (Benson and Magee, 2015; Triulzi et al., 2020). Although our model does not explicitly represent semantic distance, as text-embedding approaches often do, it does account for the cognitive span between domains and enables the construction of measures that incorporate this dimension (Appendix A3).

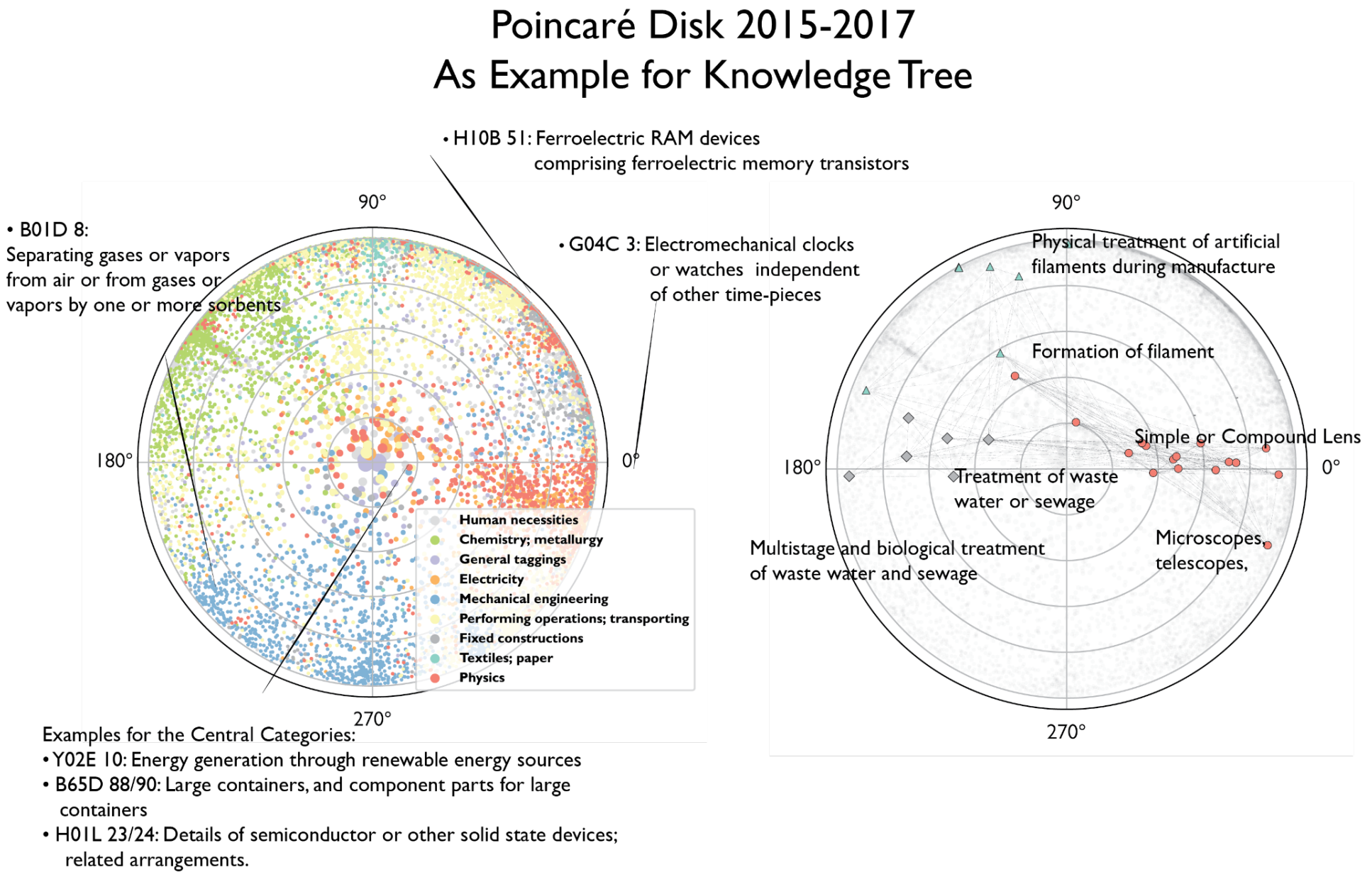

FIGURE 3 Poincare Disks for Embedding Technological Categories

### 3.3.2 Constructing Measures for Deep and Broad Recombination

Following prior literature, we treat the technical categories of one-step prior art (cited patents) as representing the space of possible combinations for each focal patent (Kaplan and Vakili, 2015a; Uzzi et al., 2013; Verhoeven et al., 2016c). We first identify all combined categories and determine their positions on the Poincaré disk, and then calculate depth and breadth measures based on their positions and geometric relationships within the embedding space.

In a 2-D Poincaré disk, each node is represented by a polar coordinate ($r$, $\theta$). As $r$ measures vertical depth in the knowledge tree and $\theta$ represents diversity, we construct depth and breadth measures to capture the maximum depth and breadth of the invention based on hyperbolic embeddings as follows:

$Depth = 90th\ Percentile(r_1, r_2, r_3, ...r_n)$, where $r_n = |r_n - r_{origin}|$

$Breadth = 90th\ Percentile(\theta_1, \theta_2, \theta_3, ...\theta_n)$, where $\theta_n = |\theta_i - \theta_j|, \forall i = i_1...i_k, \forall j = j_1...j_k, i \neq j$

Depth measures the specificity of knowledge elements within the knowledge tree, while breadth evaluates the span of knowledge elements across domains. We use the 90th percentile as a threshold to address potential over-dispersion, but our results are robust to this choice. Robustness checks using the 95th percentile and the maximum value as alternative thresholds yield similar results (Appendix B4). We provide two patent examples to illustrate our measures, and compare them with classic measures in Appendix D2 and D3.

In contrast to previous studies that calculate recombination depth and width based on existing classification systems, our measures consider search styles as a joint distribution in a polar 2-D space. The Pearson correlation between depth and breadth in our data is approximately 0.05, indicating near conceptual independence. This allows for high depth and high breadth (or low depth and low breadth) to occur simultaneously. Their correlation with the number of prior art categories are 0.09 and 0.20, respectively, indicating only a weak association.

Building on these robust, machine learning–based metrics, we estimate regressions to evaluate effects of deep versus broad search on technological impact.

## 4. Linking Depth and Breadth with Short versus Long Term Impact

### 4.1 Variables

***Dependent Variables.*** We use forward patent citations, one of the most widely accepted metrics, to measure patent impact. This refers to the number of citations a focal patent receives in subsequent years, which correlates with patent licensing revenue (Sampat and Ziedonis, 2005). More citations suggest the technology underlying the patent is attributed significant influence and is frequently reused, revised, or expanded upon in subsequent inventive endeavors to which its knowledge diffuses (Fontana et al., 2008). Approximately half of a patent's citations are added by patent inventors, the others coming from patent examiners through a process that critically assesses the novelty of the invention (Alcácer and Gittelman, 2006).

Due to the right-censored nature of citation growth, technological impact is measured by the number of citations a patent receives within a specific period after its grant, typically 3–7 years in empirical studies (Fleming, 2001; Verhoeven et al., 2016c). Here we use 5 years as the time frame for short-term citations. Meanwhile, accumulating a high number of short-term citations may not ensure long-term success. Collective invention mechanisms—like preferential attachment and the incorporation of ideas into later work—exert influence over the long term (Wang et al., 2013). To account for differences between short and long-term impact, we constructed four citation variables: 0-5 year, 5-10 year, 10-15 year, and 15-20 year forward citations. These variables trace impacts ranging from the most immediate to the most enduring.

***Independent Variables.*** We employ depth and breadth measures based on hyperbolic embedding space as independent variables, as operationalized in Section 3.3.2.

***Control Variables.*** We include the most important control variables identified in prior research to account for potential confounding factors. Controls include (1) familiarity of components and combinations from prior inventions (Fleming, 2001); (2) number of technological categories credited within both focal patents and prior art (Verhoeven et al., 2016c); (3) inventor context, including indicator variables for organizational assignee, team collaboration, and inventor experience; and (4) bibliometric indicators including non-patent references, number of claims, and family size (number of patents within a collection covering the same or similar technical content) of the focal patent. As the number of prior art citations is almost perfectly correlated with the total number of technological categories represented in prior art ($r=0.93$), we do not include the former in the regression. We also control for year fixed effects and the average domain location $\theta$ of all patent categories to account for heterogeneity across tech domains.

**4.2 Data and Model Specification**

In our first analysis, we estimate the impact of depth and breadth on citations with all US patents granted between 1976-2017 whose prior categories are identifiable. In total, this yields a sample of 4,992,915 patents. Citations are characterized by over-dispersed non-negative integer values. In accordance with Fleming (2001) and Verhoeven et al. (2016), we utilize a generalized negative binomial regression to estimate both the average and dispersion of citations, and use Poisson Maximum Likelihood Estimator ("QMLE") as a robustness test in the Appendix B3. The first set of models are conducted at the patent level:

$$log(Citations_i) = \beta_1 \times Depth_i + \beta_2 \times Breadth_i + \beta_3 \times Controls_i + \epsilon_i,$$

where $Depth_i$ is the depth measure for the focal patent i, and $Breadth_i$ captures patent breadth. This model estimates the combinatorial effects across the broadest samples available. Here, $Citations_i$ includes $Citations_{i,0-5}$, $Citations_{i,5-10}$, $Citations_{i,10-15}$ and $Citations_{i,15-20}$, allowing us to estimate the direct effects of depth and breadth on short- and long-term impacts, respectively. When we compare effects across time, we include only patents granted before 2002 to ensure model comparability.

To address the potential endogeneity problem, we also conducted two additional analyses. First, we adopted a twin-patent design (Bikard, 2020; Kovács et al., 2021). For this analysis, we only retained patents submitted to both the USPTO and EPO, and granted at both patent offices. These patents have identical quality and patent features, but are granted and evaluated within different knowledge systems with different histories, conditioning distinct expectations. Based on our theoretical prediction, we expect depth and breadth effects to persist, even after controlling for the intrinsic quality of an invention, as citations are shaped by the perceptions and decisions of other inventors through a process of collective intelligence. Second, we used an instrumental variable—changes in the USPTO taxonomy system—to address potential confounding effects. Statistical tests confirm that it is a valid and strong instrument. Details and results of both the twin-design tests and the IV analyses are reported in Appendix B5.

In the second set of analysis, we aim to explore mechanisms by which depth and breadth differentially associate with short and long-term impact. For this part of the analysis, we perform a mediation analysis based on the full USPTO sample (N=4,992,915). We construct four mediator variables from our embedding spaces to capture potential effect pathways: (1) *cognitive lock-in*, calculated as the cognitively accessible citations an invention receives within its own domain (i.e., citations divided by the standard deviation of $\Delta\theta$ between focal and citing patents); (2) *cognitive bridging*, computed as the average diffusion range of follow-up work across the space of inventions (i.e., the average of $\Delta\theta$ between focal and citing patents); (3) *network lock-in*, measured by the average path length between the inventors of the focal patent and inventors of follow-up works within the same component of collaboration networks, indicating the extent to which knowledge is recognized and accepted within local communities through social relationships; and (4) *structural bridging*, measured by the average diffusion range of follow-up works across levels of the knowledge tree (i.e., the average of $|\Delta r|$ between focal and citing patents), which reflects the formation of structural connections across knowledge hierarchies. While these variables are imperfect proxies and may be partly influenced by external factors, they allow us to explore the intellectual and social processes behind observed patterns, generating meaningful and promising hypotheses for future investigation. We estimate mediating regressions using citations received in different periods after the focal patent is granted, including 0–5, 5–10, 10–15, and 15–20 year forward citations, respectively.

### 4.3 Results for the Negative Binomial Regressions

Table 1 reports basic descriptive statistics for all variables in different samples. Table 2 presents the correlation matrix. And Table 3 reports results from our negative binomial models.

Our negative binomial models indicate that, overall, depth positively predicts short-term technological impact, while breadth exhibits a contrasting effect. A one-unit increase in depth (from theoretical minimum to maximum) results in roughly a 0.196 increase in the five-year forward citation count. At first glance, this might seem unimpressive. However, considering the highly skewed nature of the forward-citation distribution, with 1 as the median, a 0.196 increase represents an approximately 20% rise for more than half of the patents in the sample. In contrast, a one-unit increase in recombination breadth corresponds to a decrease of 0.043 in citations. Given that the theoretical maximum change in breadth is about 3.14 radians (180°) in the Poincaré disk, the maximum increase in recombination breadth would lead to an approximate 0.135 decrease in short-term citations.

Models 2-5 present regressions based on samples from before 2002, allowing 20 years (2002-2022) to observe short- and long-term citations. For this sample, we constructed four dependent variables for citations from different periods and estimated the effects of depth and breadth over time: 0-5, 5-10, 10-15, and 15-20 year citations. While the 0-5 year citations model (model 2) aligns with the general model (model 1) results, effects for both depth and breadth are reversed for long-term citations (models 3-5). Over time, the depth effect transitions from positive to negative (from $\beta = 0.095$ to -0.434, both $p<10^{-5}$), whereas the breadth effect turns from negative to positive (from $\beta = -0.036$ to 0.037, both $p<10^{-5}$). The shift in coefficients for depth and breadth, as citation lags grows, are both monotonic, with a modest acceleration from model 2 to model 5. We visualize the temporal change of depth and breadth effects in Figure 4.

We have also estimated models 1-5 on only within-organization citations, and other sub-samples (patents before 2012, and patents before 2007) for robustness, manifesting a similar pattern of results. These results are reported in the Appendix B1 and B2. In Appendix E, we also report the dispersion model (Table E9), showing the effects of depth and breadth on citation dispersion. In general, both depth and breadth reduce the standard deviation for citation ($\beta_{depth}$ = -0.117, $\beta_{breadth}$= -0.031, as seen in model 1). Nevertheless, the two variables manifest different cross-time variation: depth boosts variance first and then reduces monotonically as it turns negative; breadth remains stable and negative throughout.

Our models show that a patent's depth and breadth influence how inventors perceive its significance, and decide whether to adopt it in their own work. Consistent with category-spanning theory (Zuckerman, 1999), our models show a short-term punishment for inventions that span many categories distant from one another. Nevertheless, our work also uncovers temporal dynamics less explored in prior studies: category-spanning inventions may be initially underrecognized due to their ambiguity, but tend to gain attention over time, whereas deep inventions stand out early, but exhaust their potential quickly. To explore the processes underlying these patterns, we turn to an analysis of mechanisms.

**TABLE 1** Descriptive Statistics

| | Full Samples (N=4,992,915) | | | | Twin Patent: US Sample (N=84,916) | | | | Twin Patent: European Sample (N=84,916) | | | |
|---|---|---|---|---|---|---|---|---|---|---|---|---|
| | **MEAN** | **SD** | **MIN** | **MAX** | **MEAN** | **SD** | **MIN** | **MAX** | **MEAN** | **SD** | **MIN** | **MAX** |
| 0-5 Year Citations | 3.443 | 10.156 | 0 | 1142 | 1.087 | 1.789 | 0 | 45 | 0.521 | 1.592 | 0 | 45 |
| 5-10 Year Citations | 5.320 | 19.205 | 0 | 2555 | 1.144 | 2.059 | 0 | 125 | 0.888 | 3.025 | 0 | 365 |
| 10-15 Year Citations | 4.115 | 18.029 | 0 | 2542 | 0.639 | 1.476 | 0 | 82 | 0.709 | 4.011 | 0 | 379 |
| 15-20 Year Citations | 2.926 | 16.013 | 0 | 2468 | 0.242 | 0.935 | 0 | 47 | 0.357 | 2.880 | 0 | 319 |
| Depth | 0.817 | 0.160 | .034 | .999 | 0.680 | 0.172 | 0 | 0.999 | 0.671 | 0.245 | 0 | 1 |
| Breadth | 1.052 | 0.995 | 0.000 | 3.141 | 0.923 | 0.908 | 0 | 3.140 | 1.004 | 0.984 | 0 | 3.14 |
| Horizontal Location (mean theta) | 0.201 | 1.551 | -3.139 | 3.138 | -0.072 | 1.653 | -3.313 | 3.129 | 0.211 | 1.549 | -3.127 | 3.134 |
| Ln (Component Familiarity) | 11.426 | 2.546 | 0 | 17.494 | 7.850 | 1.483 | 0 | 11.614 | 10.389 | 1.376 | 4.359 | 14.299 |
| Ln (Combination Familiarity) | 4.279 | 3.358 | 0 | 13.496 | 2.723 | 2.223 | 0 | 8.782 | 5.852 | 2.661 | 0 | 12.174 |
| Number of Tech Classes | 5.372 | 4.352 | 0 | 17 | 5.345 | 4.040 | 0 | 16 | 4.586 | 3.096 | 1 | 12 |
| Number of Prior Art Classes | 56.605 | 75.070 | 2 | 295 | 47.200 | 50.063 | 0 | 196 | 206.335 | 159.531 | 9 | 626 |
| Assigned to Organizations | .897 | 0.304 | 0 | 1 | 0.973 | 0.162 | 0 | 1 | 0.973 | 0.162 | 0 | 1 |
| Invented by Team | .628 | 0.483 | 0 | 1 | 0.709 | 0.454 | 0 | 1 | 0.709 | 0.454 | 0 | 1 |
| Ln (Inventor Experience) | 2.662 | 1.854 | 0 | 12.219 | 1.975 | 1.179 | 0 | 7.510 | 1.850 | 1.295 | 0.693 | 6.685 |
| Non-Patent References | 3.081 | 5.999 | 0 | 23 | 1.997 | 3.211 | 0 | 12 | 0.481 | 0.911 | 0 | 3 |
| Number of Claims | 15.272 | 8.907 | 1 | 36 | 15.724 | 8.982 | 1 | 37 | 4.43 | 6.461 | 0 | 20 |
| Family Size | .439 | 0.611 | 0 | 2 | 1 | 0 | 1 | 1 | 1 | 0 | 1 | 1 |

Note: *N*=4,992,915 for all patents granted by USPTO between 1976-2017, and *N*=84,916 for patents granted at both USPTO and EPO between 2001-2012. Patents with < 2 prior art categories among the patents they cite are excluded as we cannot construct meaningful independent variables.

**TABLE 2** Correlation Matrix of Major Variables (Full Sample, N=4,992,915)

| | (1) | (2) | (3) | (4) | (5) | (6) | (7) | (8) | (9) | (10) | (11) | (12) | (13) | (14) |
|---|---|---|---|---|---|---|---|---|---|---|---|---|---|---|
| (1) 5-Year Citations | 1.000 | | | | | | | | | | | | | |
| (2) Depth | 0.017 | 1 | | | | | | | | | | | | |
| | [0.000] | | | | | | | | | | | | | |
| (3) Breadth | 0.034 | -0.160 | 1 | | | | | | | | | | | |
| | [0.000] | [0.000] | | | | | | | | | | | | |
| (4) Horizontal Location (mean theta) | 0.018 | -0.060 | 0.010 | 1 | | | | | | | | | | |
| | [0.000] | [0.000] | [0.000] | | | | | | | | | | | |
| (5) Ln (Component Familiarity) | 0.059 | -0.421 | 0.003 | 0.037 | 1 | | | | | | | | | |
| | [0.000] | [0.000] | [0.000] | [0.000] | | | | | | | | | | |
| (6) Ln (Combination Familiarity) | 0.013 | -0.223 | -0.122 | 0.016 | 0.314 | 1 | | | | | | | | |
| | [0.000] | [0.000] | [0.000] | [0.000] | [0.000] | | | | | | | | | |
| (7) Number of Tech Classes | 0.086 | -0.118 | 0.122 | 0.029 | 0.337 | -0.371 | 1 | | | | | | | |
| | [0.000] | [0.000] | [0.000] | [0.000] | [0.000] | [0.000] | | | | | | | | |
| (8) Number of Prior Art Classes | 0.169 | 0.043 | 0.274 | 0.016 | 0.254 | -0.063 | 0.382 | 1 | | | | | | |
| | [0.000] | [0.000] | [0.000] | [0.000] | [0.000] | [0.000] | [0.000] | | | | | | | |
| (9) Assigned to Organizations | 0.025 | -0.094 | -0.049 | 0.017 | 0.228 | 0.079 | 0.094 | 0.084 | 1 | | | | | |
| | [0.000] | [0.000] | [0.000] | [0.000] | [0.000] | [0.000] | [0.000] | [0.000] | | | | | | |
| (10) Invented by Team | 0.035 | -0.067 | 0.004 | 0.002 | 0.174 | 0.008 | 0.128 | 0.113 | 0.291 | 1 | | | | |
| | [0.000] | [0.000] | [0.000] | [0.000] | [0.000] | [0.000] | [0.000] | [0.000] | [0.000] | | | | | |
| (11) Ln (Inventor Experience) | 0.040 | -0.139 | 0.036 | 0.005 | 0.305 | 0.088 | 0.201 | 0.223 | 0.224 | 0.178 | 1 | | | |
| | [0.000] | [0.000] | [0.000] | [0.000] | [0.000] | [0.000] | [0.000] | [0.000] | [0.000] | [0.000] | | | | |
| (12) Non-Patent References | 0.113 | 0.050 | 0.014 | 0.007 | 0.235 | -0.030 | 0.235 | 0.456 | 0.112 | 0.149 | 0.165 | 1 | | |
| | [0.000] | [0.000] | [0.000] | [0.000] | [0.000] | [0.000] | [0.000] | [0.000] | [0.000] | [0.000] | [0.000] | | | |
| (13) Number of Claims | 0.098 | -0.001 | 0.040 | -0.011 | 0.154 | 0.039 | 0.097 | 0.214 | 0.093 | 0.098 | 0.089 | 0.183 | 1 | |
| | [0.000] | [0.000] | [0.000] | [0.000] | [0.000] | [0.000] | [0.000] | [0.000] | [0.000] | [0.000] | [0.000] | [0.000] | | |
| (14) Family Size | -0.071 | -0.086 | -0.042 | 0.014 | -0.000 | 0.008 | 0.002 | -0.184 | 0.158 | 0.056 | 0.081 | -0.145 | -0.166 | 1 |
| | [0.000] | [0.000] | [0.000] | [0.000] | [0.544] | [0.000] | [0.000] | [0.000] | [0.000] | [0.000] | [0.000] | [0.000] | [0.000] | |

Note: *p*-values are between square brackets. All tests are two-tailed.

**TABLE 3** Effects of Recombination Depth vs. Breadth on Patent Short- and Long-Term Impact

| | Model with All Patents: 1976-2017 | Model with Early Patents: 1976-2002 | | | |
|---|---|---|---|---|---|
| | **Model 1: Predicting 0-5 Year Citation** | **Model 2 Predicting 0-5 Year Citation** | **Model 3 Predicting 5-10 Year Citation** | **Model 4 Predicting 10-15 Year Citation** | **Model 5 Predicting 15-20 Year Citation** |
| ***Main Models*** | | | | | |
| Depth | 0.196*** | 0.095*** | -0.000 | -0.164*** | -0.434*** |
| | [0.000] | [0.000] | [0.950] | [0.000] | [0.000] |
| | (0.005) | (0.006) | (0.007) | (0.008) | (0.009) |
| Breadth | -0.043*** | -0.036*** | -0.019*** | -0.001 | 0.037*** |
| | [0.000] | [0.000] | [0.000] | [0.549] | [0.000] |
| | (0.001) | (0.001) | (0.001) | (0.001) | (0.001) |
| Horizontal Location (mean theta) | 0.035*** | 0.077*** | 0.073*** | 0.069*** | 0.087*** |
| | [0.000] | [0.000] | [0.000] | [0.000] | [0.000] |
| | (0.000) | (0.001) | (0.001) | (0.001) | (0.001) |
| Ln (Component Familiarity) | 0.033*** | 0.029*** | 0.033*** | 0.034*** | 0.023*** |
| | [0.000] | [0.000] | [0.000] | [0.000] | [0.000] |
| | (0.000) | (0.001) | (0.001) | (0.001) | (0.001) |
| Ln (Combination Familiarity) | 0.030*** | 0.035*** | 0.025*** | 0.019*** | 0.015*** |
| | [0.000] | [0.000] | [0.000] | [0.000] | [0.000] |
| | (0.000) | (0.000) | (0.000) | (0.000) | (0.001) |
| Number of Tech Classes | 0.028*** | 0.032*** | 0.040*** | 0.046*** | 0.050*** |
| | [0.000] | [0.000] | [0.000] | [0.000] | [0.000] |
| | (0.000) | (0.000) | (0.000) | (0.000) | (0.000) |
| Number of Prior Art Classes | 0.003*** | 0.003*** | 0.004*** | 0.004*** | 0.005*** |
| | [0.000] | [0.000] | [0.000] | [0.000] | [0.000] |
| | (0.000) | (0.000) | (0.000) | (0.000) | (0.000) |
| Assigned to Organizations | 0.075*** | 0.115*** | 0.072*** | -0.011*** | -0.085*** |
| | [0.000] | [0.000] | [0.000] | [0.000] | [0.000] |
| | (0.002) | (0.002) | (0.003) | (0.003) | (0.003) |
| Invented by Team | 0.076*** | 0.086*** | 0.094*** | 0.086*** | 0.084*** |
| | [0.000] | [0.000] | [0.000] | [0.000] | [0.000] |
| | (0.001) | (0.002) | (0.002) | (0.002) | (0.002) |
| Ln (Inventor Experience) | 0.010*** | -0.000 | -0.023*** | -0.039*** | -0.047*** |
| | [0.000] | [0.880] | [0.000] | [0.000] | [0.000] |
| | (0.000) | (0.001) | (0.001) | (0.001) | (0.001) |
| Non Patent References | 0.009*** | 0.007*** | 0.017*** | 0.019*** | 0.017*** |
| | [0.000] | [0.000] | [0.000] | [0.000] | [0.000] |
| | (0.000) | (0.000) | (0.000) | (0.000) | (0.000) |
| Number of Claims | 0.018*** | 0.018*** | 0.021*** | 0.023*** | 0.024*** |
| | [0.000] | [0.000] | [0.000] | [0.000] | [0.000] |
| | (0.000) | (0.000) | (0.000) | (0.000) | (0.000) |
| Family Size | -0.169*** | -0.101*** | -0.203*** | -0.268*** | -0.308*** |
| | [0.000] | [0.000] | [0.000] | [0.000] | [0.000] |
| | (0.001) | (0.001) | (0.001) | (0.002) | (0.002) |
| Year Fixed Effects | YES | YES | YES | YES | YES |
| Constant | 0.049 | 0.099 | 0.036 | 0.206*** | 0.439*** |
| | [0.367] | [0.068] | [0.525] | [0.001] | [0.000] |
| | (0.054) | (0.054) | (0.057) | (0.061) | (0.060) |
| ***Dispersion Model*** | | ***With Same Independent Variables, Reported in Appendix E*** | | | |
| Observations | 4992915 | 2010031 | 2010031 | 2010031 | 2010031 |

Note: *p*-values are between square brackets. Robust standard errors are in parentheses. All tests are two-tailed.

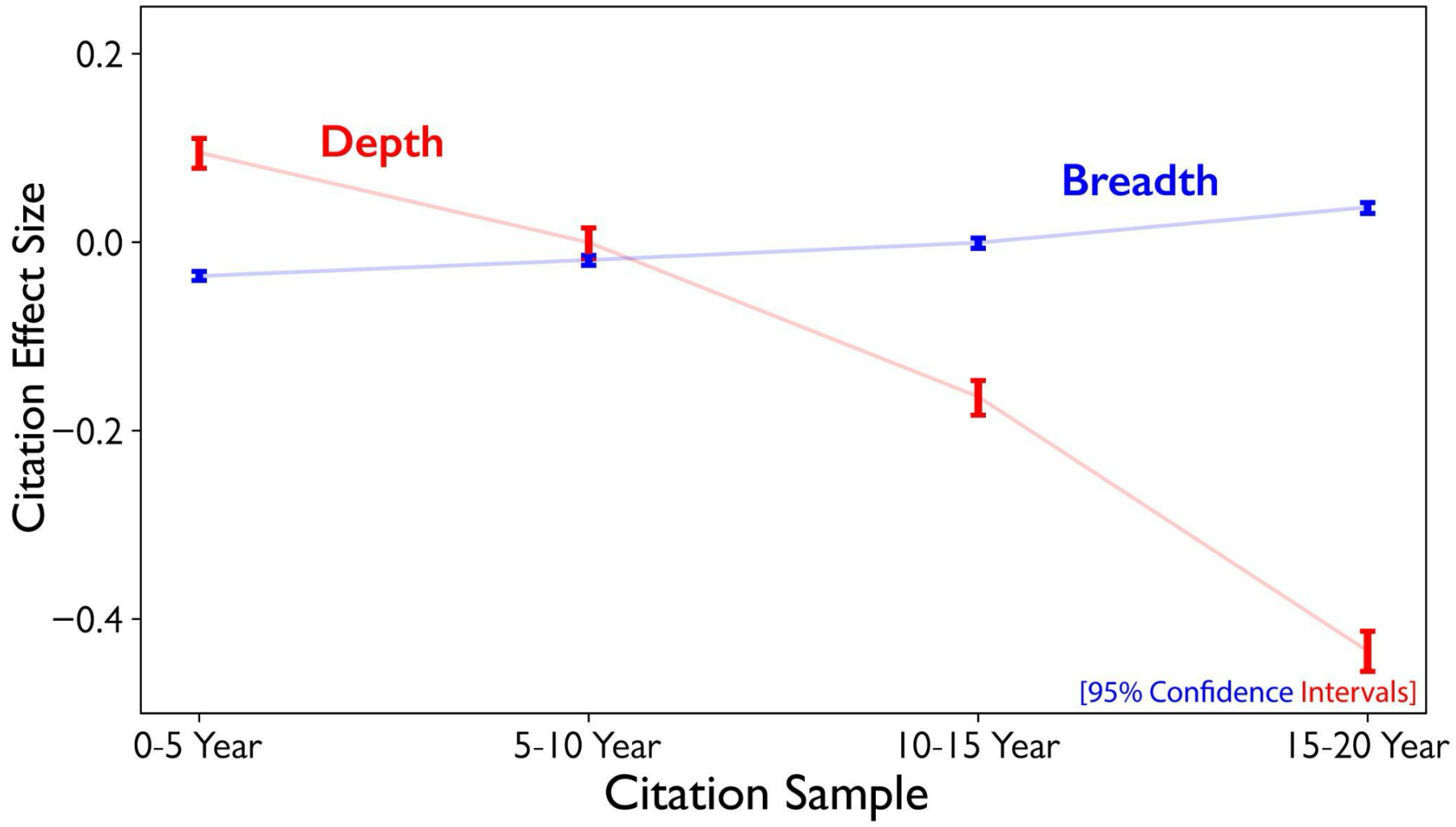

FIGURE 4 Depth and Breadth Effects on Short-Term and Long-Term Patent Impact

## 5. Exploring Mechanisms with Knowledge Flows across Embedding Space

Our analysis indicates that search depth correlates with higher forward citations in the short term, while breadth exhibits the opposite effect, with the direction of these effects reversing further into the future. Why does this occur? As the path and pattern of citation growth can be tracked in the Poincaré embedding space, our embedding model provides a window for exploring the underlying mechanisms.

We first construct proxy measures for the cognitive lock-in and bridging mechanisms, based on our embedding space. The left panel of Figure 5 shows a conceptual illustration of the collective invention process. After the emergence of a focal invention (dark blue dots), subsequent patents may arise either along the same intellectual vein (light blue dots) or from different knowledge domains (yellow dots). We use the absolute value of the difference between the focal patent's $\theta$ and a follow-up patent's $\theta$ (denoted $|\Delta\theta|$) to measure the horizontal diffusion distance between domains. As illustrated by the conceptual distribution of this distance (right panel of Figure 5), follow-up works generally tend to cluster around the focal invention and diffuse toward more distant areas with decreasing probability, suggesting a process of lock-in.

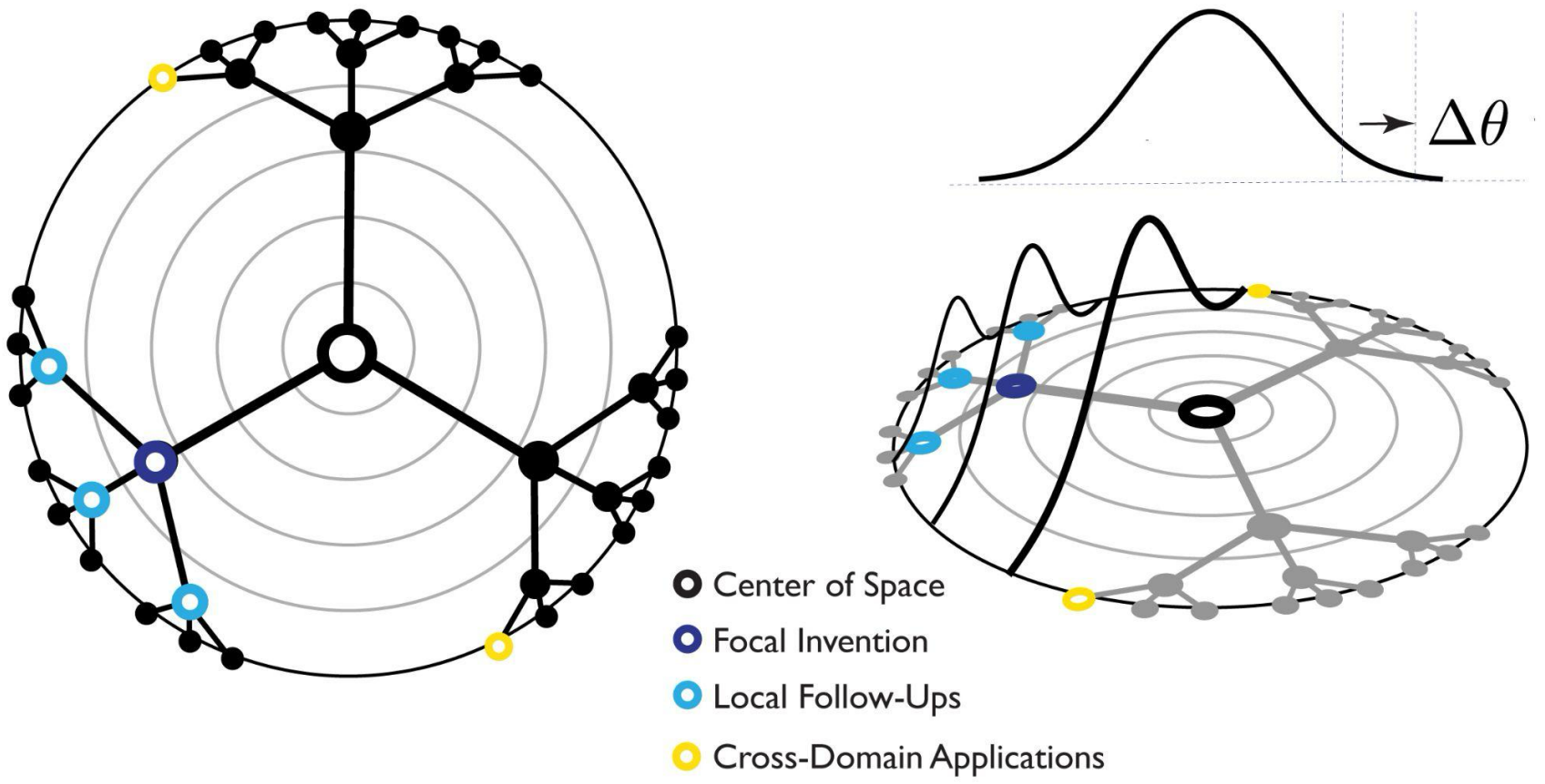

FIGURE 5 Conceptual Figure for Post-Publication Diffusion

This framework allows us to construct four proxies for the mediating pathways, as discussed in Section 4.2: (1) *cognitive lock-in*, (2) *cognitive bridging*, (3) *network lock-in*, and (4) *structural bridging*. Based on these mediators, we conduct a mediation analysis.

The results in Table 4 show that, in the short term, depth is associated with higher local impact through both cognitive and network lock-in ($\beta$=1.214*** for cognitive lock-in and $\beta$=0.034*** for network lock-in), while breadth is linked to a broader diffusion range in both vertical and horizontal directions ($\beta$=0.171*** for cognitive bridging and $\beta$=0.026*** for structural bridging). Depth also contributes to more structural bridging, though its effect size is smaller than that of breadth ($\beta$=0.019 < 0.026).

TABLE 4 Mediation Analysis for Impact Accumulation (Five Years) based on GSEM

| | Cognitive Lock-in | Network Lock-in | Cognitive Bridging | Structural Bridging | 5 Year Forward Citations |
|---|---|---|---|---|---|
| Cognitive Lock-in | | | | | 0.166*** |
| | | | | | [0.000] |
| | | | | | (0.001) |
| Cognitive Bridging | | | | | 0.181*** |
| | | | | | [0.000] |
| | | | | | (0.004) |
| Structural Bridging | | | | | 0.201*** |
| | | | | | [0.000] |
| | | | | | (0.009) |
| Network Lock-in | | | | | 0.070*** |
| | | | | | [0.000] |
| | | | | | (0.001) |
| Depth | 1.214*** | 0.034*** | -0.250*** | 0.019*** | -0.089*** |
| | [0.000] | [0.000] | [0.000] | [0.000] | [0.000] |
| | (0.020) | (0.004) | (0.002) | (0.001) | (0.007) |
| Breadth | -0.473*** | -0.041*** | 0.171*** | 0.026*** | 0.047*** |
| | [0.000] | [0.000] | [0.000] | [0.000] | [0.000] |
| | (0.002) | (0.001) | (0.000) | (0.000) | (0.001) |
| Horizontal Location (mean theta) | 0.006*** | -0.001** | 0.003*** | 0.001*** | 0.017*** |
| | [0.000] | [0.007] | [0.000] | [0.000] | [0.000] |
| | (0.002) | (0.000) | (0.000) | (0.000) | (0.001) |
| Ln (Component Familiarity) | 0.134*** | 0.025*** | 0.003*** | 0.000*** | 0.001+ |
| | [0.000] | [0.000] | [0.000] | [0.000] | [0.070] |
| | (0.001) | (0.000) | (0.000) | (0.000) | (0.001) |
| Ln (Combination Familiarity) | -0.057$^{***}$ | 0.001$^{***}$ | -0.029$^{***}$ | -0.006$^{***}$ | 0.024$^{***}$ |
| | [0.000] | [0.000] | [0.000] | [0.000] | [0.000] |
| | (0.001) | (0.000) | (0.000) | (0.000) | (0.000) |
| Number of Tech Classes | 0.028$^{***}$ | -0.009$^{***}$ | -0.005$^{***}$ | -0.004$^{***}$ | 0.013$^{***}$ |
| | [0.000] | [0.000] | [0.000] | [0.000] | [0.000] |
| | (0.001) | (0.000) | (0.000) | (0.000) | (0.000) |
| Number of Prior Art Classes | 0.001$^{***}$ | 0.000$^{***}$ | -0.000$^{***}$ | -0.000$^{***}$ | 0.002$^{***}$ |
| | [0.000] | [0.000] | [0.000] | [0.000] | [0.000] |
| | (0.000) | (0.000) | (0.000) | (0.000) | (0.000) |
| Assigned to Organizations | 0.165$^{***}$ | 0.292$^{***}$ | -0.017$^{***}$ | 0.001$^{***}$ | -0.033$^{***}$ |
| | [0.000] | [0.000] | [0.000] | [0.000] | [0.000] |
| | (0.008) | (0.005) | (0.001) | (0.000) | (0.005) |
| Invented by Team | 0.053$^{***}$ | 0.076$^{***}$ | -0.002$^{***}$ | 0.000 | 0.006$^{**}$ |
| | [0.000] | [0.000] | [0.000] | [0.994] | [0.005] |
| | (0.005) | (0.001) | (0.000) | (0.000) | (0.002) |
| Ln (Inventor Experience) | 0.001 | -0.044$^{***}$ | -0.002$^{***}$ | -0.000$^{***}$ | 0.002$^{**}$ |
| | [0.422] | [0.000] | [0.000] | [0.001] | [0.008] |
| | (0.001) | (0.000) | (0.000) | (0.000) | (0.001) |
| Non Patent References | 0.014$^{***}$ | -0.010$^{***}$ | -0.002$^{***}$ | -0.000$^{***}$ | 0.007$^{***}$ |
| | [0.000] | [0.000] | [0.000] | [0.000] | [0.000] |
| | (0.000) | (0.000) | (0.000) | (0.000) | (0.000) |
| Number of Claims | 0.007$^{***}$ | -0.001$^{***}$ | -0.000$^{***}$ | 0.000 | 0.008$^{***}$ |
| | [0.000] | [0.000] | [0.000] | [0.170] | [0.000] |
| | (0.000) | (0.000) | (0.000) | (0.000) | (0.000) |
| Family Size | -0.102$^{***}$ | 0.028$^{***}$ | 0.000 | 0.001$^{***}$ | -0.110$^{***}$ |
| | [0.000] | [0.000] | [0.224] | [0.000] | [0.000] |
| | (0.004) | (0.001) | (0.000) | (0.000) | (0.001) |
| Year Fixed Effects | YES | YES | YES | YES | YES |
| Constant | 1.634 | 1.163 | 0.608 | 0.152 | 1.091 |
| | [0.000] | [0.000] | [0.000] | [0.000] | [0.000] |
| | (0.130) | (0.008) | (0.025) | (0.008) | (0.028) |
| Vars and Covs | Controlled for Covariates at the Pairwise Level for the Four Mediators | | | | |
| N | 2178535 | | | | |

*Note: P-values are between square brackets. Robust standard errors are in parentheses. All tests are two-tailed.*

We further examine how these effect pathways evolve over time by replicating the same model across four dependent variables: citations received within 0–5, 5–10, 10–15, and 15–20 years after the patent is granted. To ensure comparability, we restrict the sample to patents granted before 2002, so that regressions across all time windows are based on the same dataset.

Figure 6 reports the estimated indirect effects (i.e., depth/breadth to mediator effect × mediator to citation effect) through the four pathways. These results yield several novel insights.

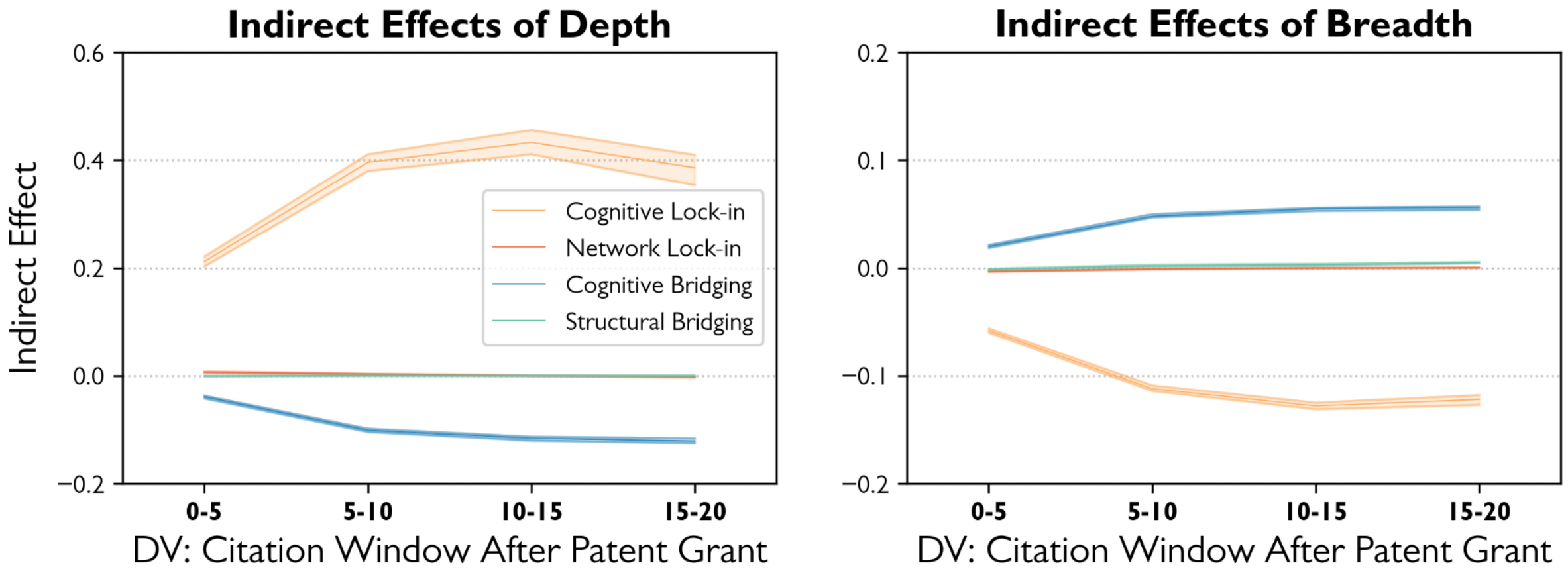

FIGURE 6 Indirect Effects via Four Proposed Mechanisms (CI: 1000 Bootstrap Replications)

First, among the four mechanisms we examine, cognitive mechanisms—specifically cognitive lock-in and cognitive bridging—appear to explain most of the observed effects. While structural bridging also shows a modest association with the impact benefits of breadth in later years, its influence remains limited compared to the cognitive pathways. Second, a substantial portion of the citations that deep patents receive continues to come from local communities (reflected in consistently positive indirect effects through cognitive lock-in), while broad patents display consistent positive associations with cognitive bridging. Notably, the gap between these two indirect effects widens over time for both depth and breadth, suggesting that the citations received by these two types of inventions differ not only in quantity but also in character. Third, depth exhibits a negative direct effect on citations that strengthens over time (from -0.086 to -0.537), whereas breadth shows a positive and growing direct effect (from 0.030 to 0.106). Therefore, it is highly likely that depth and breadth are partially correlated with aspects of patent quality, such as novelty, that become more apparent only several years after a patent's publication. Finally, after controlling for the four mediators in the models, coefficients for both depth and breadth change signs, suggesting that the mediators collectively suppress the effect of search strategies on technological impact. This finding suggests that the early citation patterns of depth and breadth are largely explained by the pathways we identify here. At the same time, the direct effects (with mechanisms controlled) may reflect the portions of these associations that remain after accounting for social and informational context, when evaluations are primarily shaped by inventions' intrinsic features. Notably, these coefficients align in direction with those from the long-term impact models (e.g., 15–20 year model), where lock-in mechanisms appear substantially weaker.

These results suggest that the changing impacts of depth and breadth over time can be partly explained by the multiplex effects of three factors: the decelerating growth of local returns in later years, the growing importance of a broad audience, and the increasing magnitude of direct effects.

## 6. Depth and Breadth Effects on Collective Invention

From corporate and societal standpoints, inventions are not independent products, but weave a collective fabric that conditions future innovation (Shi et al., 2015). To gauge how patent depth and breadth influence the search for technology in subsequent work, we conduct two additional analyses, to further examine how collective contexts have enhanced or hindered the depth and breadth effects.

First, we explore the relationship between depth, breadth, and the knowledge structures of follow-up work. We begin by calculating average depth and breadth of follow-up patents within a 5-year span. We then employ OLS models to estimate the association between depth, breadth, and features of follow-up depth and breadth. As visualized in the left panel of Figure 7 and reported in Appendix Table E7-E8, our results reveal two patterns: (1) depth breeds more depth as breadth breeds more breadth. One unit increase of focal depth leads to 0.414 units increase in followup depth (Model 1), while one unit increase of focal breadth leads to 0.505 units increase in followup breadth (Model 2). Both associations are more pronounced within the same organization ($\beta_{depth_to_depth} = 0.515>0.414$ in Table E7, and $\beta_{breadth_to_breadth} = 0.595>0.505$ in Table E8), likely due to diminished barriers for knowledge exchange. (2) Model in Table E8 (addressing follow-up breadth) surfaces a positive interplay between focal patents' depth and breadth. It shows that depth and breadth enhance and complement one another. This interaction likely implies that a "deep" invention has greater robustness as a reliable component for future recombination with other technologies ($\beta_{depth \times breadth} = 0.404$ in Model 3 and $\beta_{depth \times breadth} = 0.411$ in Model 4). This moderating effect has a larger effect size within organizations, suggesting that without crossing organizational boundaries, invention depth more efficiently prepares the focal invention to become incorporated into future work.

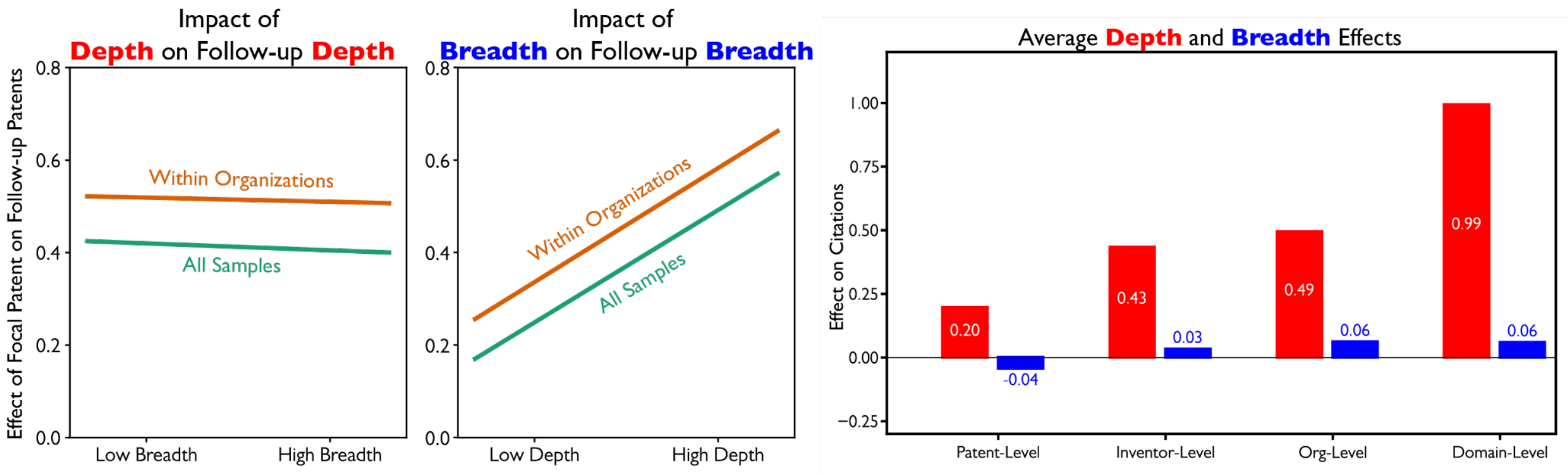

FIGURE 7 Organization and Collective Effects of Depth and Breadth

In order to further examine this collective effect, we conduct an analysis to replicate our results at different analytical levels, consisting of the loci of innovation in technology development from small to large—from inventors to organizations (from the original patent assignees) to technological domains (defined as 5-digit CPC categories). If collective thinking operates via attention—guiding inventors' focus toward relevant components and speeding the process of novel incorporation—we should see an increasing influence of search depth and breadth on citations as we move from the patent to the inventor, organization, and ultimately domain levels, with larger units possess more excess capacity to foster collective intelligence. The right panel of Figure 7 displays this pattern. From patent level models to domain level models, we observe an apparent growing influence of both depth ($\beta_{patent_depth} = 0.200$, $\beta_{inventor_depth} = 0.433$, $\beta_{organization_depth} = 0.494$, $\beta_{domain_depth} = 0.992$, all with $p<10^{-128}$) and breadth ($\beta_{patent_breadth} = -0.035$, $\beta_{inventor_breadth} = 0.033$, $\beta_{organization_breadth} = 0.061$, $\beta_{domain_breadth} = 0.059$, all with $p<10^{-186}$) across levels, supporting our hypothesis regarding the mechanism of collective attention.

## 7. Discussion and Conclusion

### 7.1 Revisiting the Depth vs. Breadth Debate through Computational Analysis

How do depth and breadth in technological search influence the impact of innovations? This question has been debated for decades. Despite its theoretical and practical importance, prior studies have yielded conflicting findings regarding their effects. Kaplan and Vakili (2015a) frame this debate in terms of "foundational" and "tension" views. To advance this line of research, we employ computational methods that model the technological landscape and develop fine-grained measures of recombinant structure and pathways of diffusion. This approach allows us to examine the effects of depth and breadth through the lens of collective intelligence, and to empirically test mechanisms previously difficult to observe.

We first utilize the Poincaré embedding method, based on hyperbolic geometry, to construct a knowledge tree of technological categories within a Poincaré disk. In this representation, the vertical dimension $r$ signifies the continuum from general-purpose technological components to specific applications, the depth of technological combinations, while the horizontal dimension $\theta$ represents the breadth or diversity of cross-domain combinations. Based on this knowledge map, we develop measures for recombination depth and breadth. Contrary to traditional methods that treat these factors as trade-offs, we measure depth as 90th percentile of $\Delta r$ and breadth as 90th percentile of $\Delta\theta$, as defined geometrically in the embedding space. The former measures the extent to which an inventor or company incorporates and revises existing technological foundations within the domain, and the latter the extent to which an inventor or company combines components across domains in unexpected ways. This approach better aligns with the real-world observation that many combinations are both deep and wide, or neither deep nor wide. By using the structure of the entire knowledge system as a reference frame, our model accurately captures technological functions, clearly distinguishes between the dimensions of depth and breadth, enables cross-industry comparison, and considers the upshot in terms of collective intelligence. The hyperbolic representation provides a simple, clear, and compelling framework for understanding the outcomes of different search strategies, while also creating new theoretical opportunities.

Our findings offer new insights into the mechanisms of combinatorial innovation, and carry significant theoretical implications. First, we observe a clear and consistent pattern that depth positively correlates with short-term and negatively with long-term impact. Breadth demonstrates the opposite effect. This suggests that the fundamental and tension views are not contradictory, but capture distinct facets of reality. The foundational view is correct that local search garners more local recognition. The tension view is correct that local search leads to cognitive“lock-in” and constrains creative conflict. Nevertheless, these two outcomes do not occur simultaneously. Local recognition occurs immediately upon patent grant, but the negative constraint emerges and intensifies over time. Given how technological fields vary widely in their pace of evolution—5 years might be “long-term” within the rapidly advancing AI domain but “short-term” in conventional chemistry, previous domain-specific studies may have captured only one aspect of the story, even with identical forward citation metrics, yielding divergent and confounded conclusions. Therefore, our study underscores the role of time in organizational learning, offering a resolution to longstanding debates over depth vs. breadth.

Second, our models on follow-up patent features reveal the unfolding interplay of exploitation and exploration, highlighting their synergistic relationship, especially within companies that can strategically allocate engineering resources to build on prior advances. On the tree of technological knowledge traced by our Poincaré disk, exploitation occurs rapidly when knowledge moves longitudinally. Exploration, by contrast, occurs intermittently as inventors establish connections that cross Poincaré latitudes. Although the immediate advantage of exploitation diminishes in technological influence over time, it primes components for seamless integration into horizontal innovations. This “invisible impact” calls into question the conventional belief that deep recombination is solely associated with exploitation, and suggests the importance for innovative firms to maintain a portfolio of explorations and exploitations that can fuel each other over time.

We acknowledge several limitations of this study. First, although our model provides a mathematical foundation for tracking citation growth, available data does not allow us to fully distinguish among all lock-in and bridging mechanisms. Moreover, the mediators we use are imperfect proxies for the social processes under investigation. Future research using more fine-grained data should explore these mechanisms in greater depth and uncover more precise patterns. Second, our measures of depth and breadth rely on the accurate assignment of technological categories. Although omissions are rare, they may introduce measurement errors when they occur. Third, our embedding relies solely on citation networks rather than textual information. Citation is an imprecise proxy for knowledge inheritance, as it is often shaped by legal and strategic considerations. While our approach significantly reduces computational costs, enhances general applicability, and shows general alignment with text analysis approach (see Appendix A3), it may sacrifice granularity. Finally, our embedding spaces use a 3-year time window, capturing some dynamics of technological evolution but potentially missing rapid changes. For instance, a nascent integrative technology that quickly becomes a platform may shift its position in less than three years. Such rapid developments may not be fully reflected in our current models. Future research could extend our analysis, broaden the range of application cases, and further improve algorithm accuracy.

### 7.2 Implications for R&D Management

Our work also has broad practical implications, particularly for R&D management. First, by revealing the temporal returns of depth versus breadth in search, our work provides guidance for inventors and organizations to make informed R&D decisions based on their priorities. For example, small companies and individual inventors, who often lack the resources for long-term R&D investment and need immediate market returns, may find deep search a safer and more reliable strategy. For these innovators, deep search provides faster market feedback and a clearer path for technological iteration.

Second, our work also suggests another potential strategy, especially for large companies and market incumbents: to alternate between deep and broad search over time. Robust modules developed through deep search can support and enhance future exploration, enabling depth and breadth to mutually reinforce one another. This approach reflects the ambidexterity principle (O'Reilly and Tushman, 2008; Raisch and Birkinshaw, 2008) applied across temporal scales. As this strategy is constrained by R&D budgets, organizational scale, and the maturity of the technological field (McCutchen and Swamidass, 1996; Sørensen and Stuart, 2000; Tsai and Wang, 2005), large firms with sufficient resources and high fault tolerance are better positioned to implement it. A representative example is Pfizer (Figure 8), with a patenting trajectory that shows a clear zoom-in/zoom-out pattern over time.

Overall, our work supports both individual inventors and organizations in making effective R&D decisions and enhancing their competitiveness in the market.

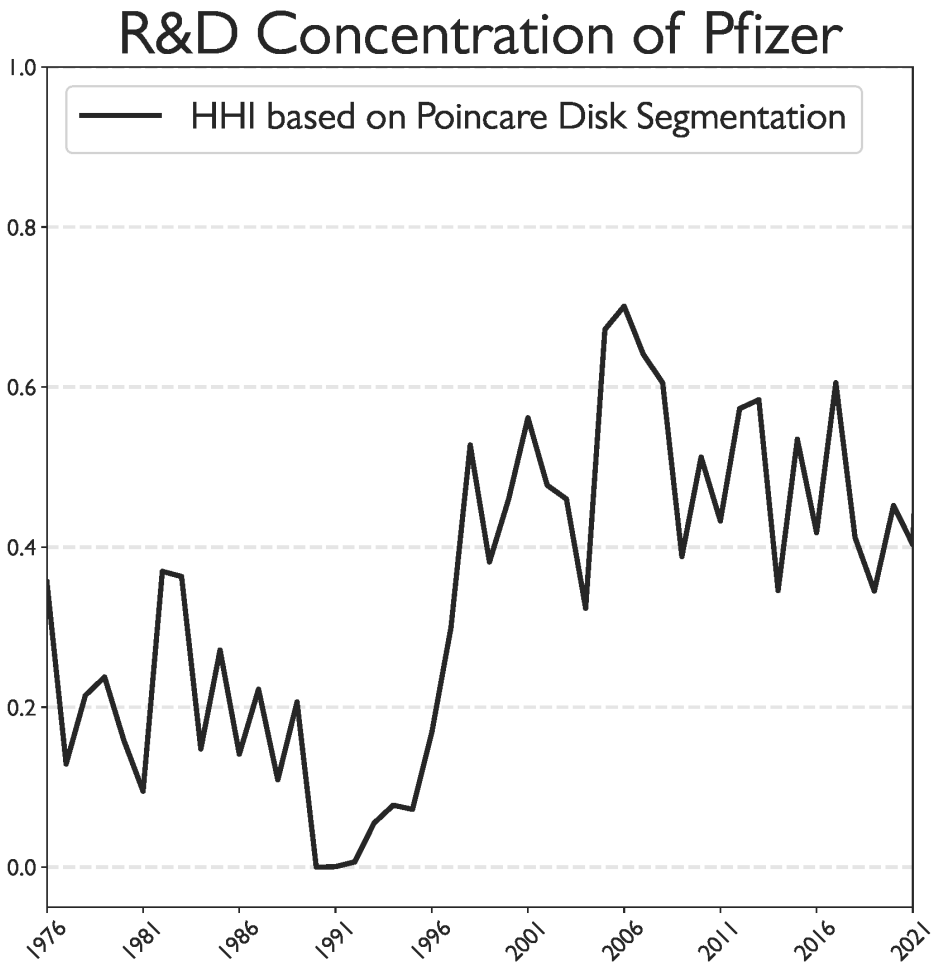

FIGURE 8 Technological Concentration of Pfizer Over Time (HHI by Poincaré Segments)

# Supplementary Materials for

# Depth Versus. Breadth in the Hierarchical Recombination of Technology

## Appendix A. Validation for the Poincaré Embedding Space

Appendix A validates the Poincaré embedding space as a valid representation of the technological tree. To achieve this goal, we (1) visualize the 14 poincaré disks and assess their cross-period stability by correlating r and theta for each technological category across periods; (2) link measures from the Poincaré embedding to knowledge diffusion networks and test their correlation; and (3) demonstrating that our measures depict the inventors' search process more accurately than conventional methods.

### *A1. The 14 Embedding Space and Their Cross-Period Stability*

In this study, we constructed 14 hyperbolic embedding spaces spanning from 1976 to 2017, using three-year intervals (with 1976-1979 as the first period, and 2015-2017 as the last). Each space models the structure of the technological tree for its specific period. Figure A1 visualizes these 14 Poincaré disks. We aligned them using the Procrustes procedure (Hamilton, Leskovec, and Jurafsky, 2016) to enhance readability. The Poincaré disks intuitively display continuity across different periods, underscoring the stability of technological evolution.

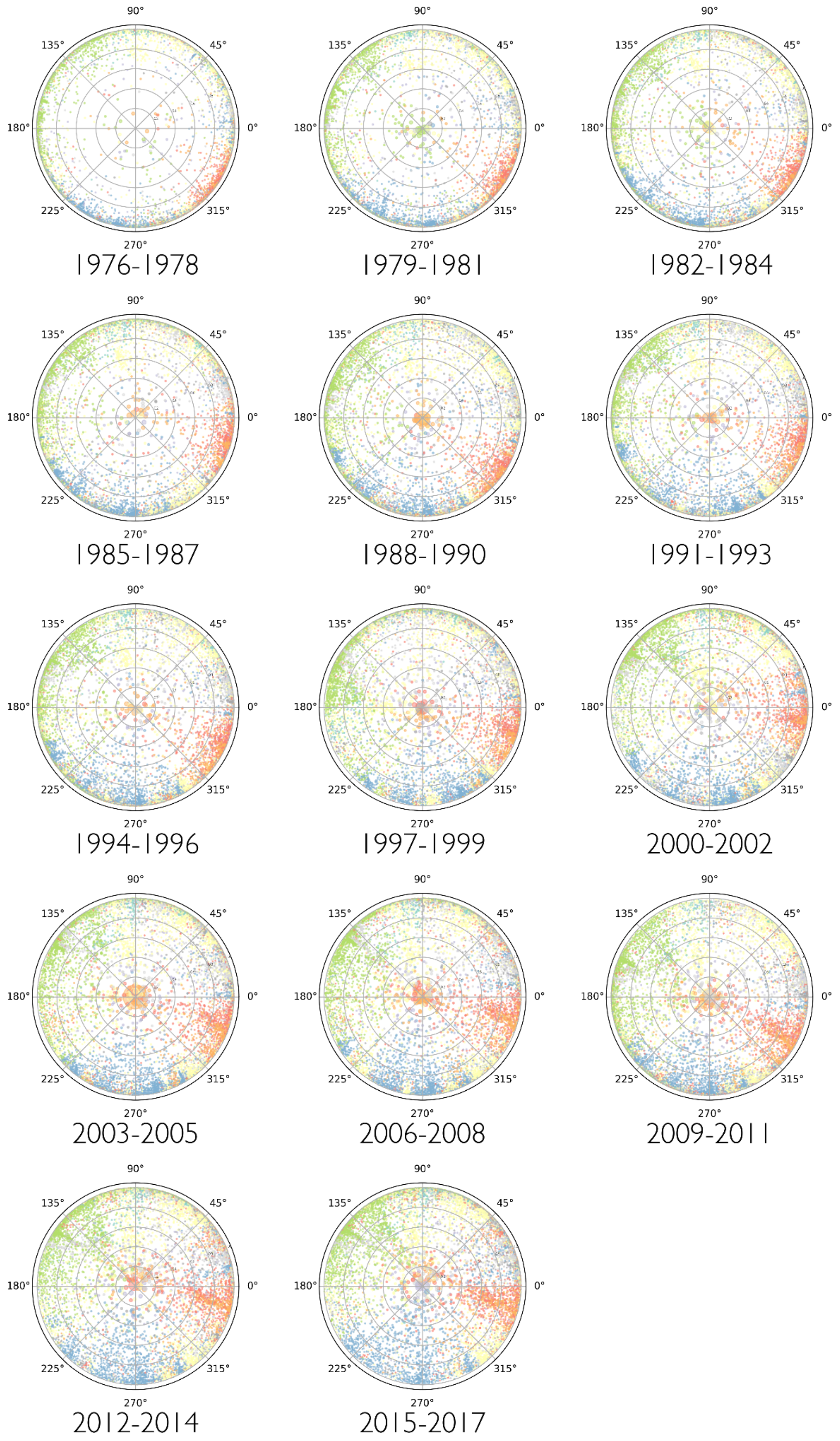

Figure A1. The 14 Embedding Spaces across Periods

We hypothesize that stable technological trajectories over time would result in consistent radial (r) and angular (θ) values across periods, indicating similar levels of generality and horizontal relationships. Furthermore, we expect higher correlation coefficients between adjacent periods. The heatmaps in Figure A2 illustrate these correlations. Our results show stability within the embedding space, with correlation coefficients for r values all exceeding 0.6. When deleting the categories with low frequency, correlation coefficients for θ values mostly exceed 0.5. Significantly, adjacent periods display increased correlation coefficients, reinforcing our initial hypothesis.

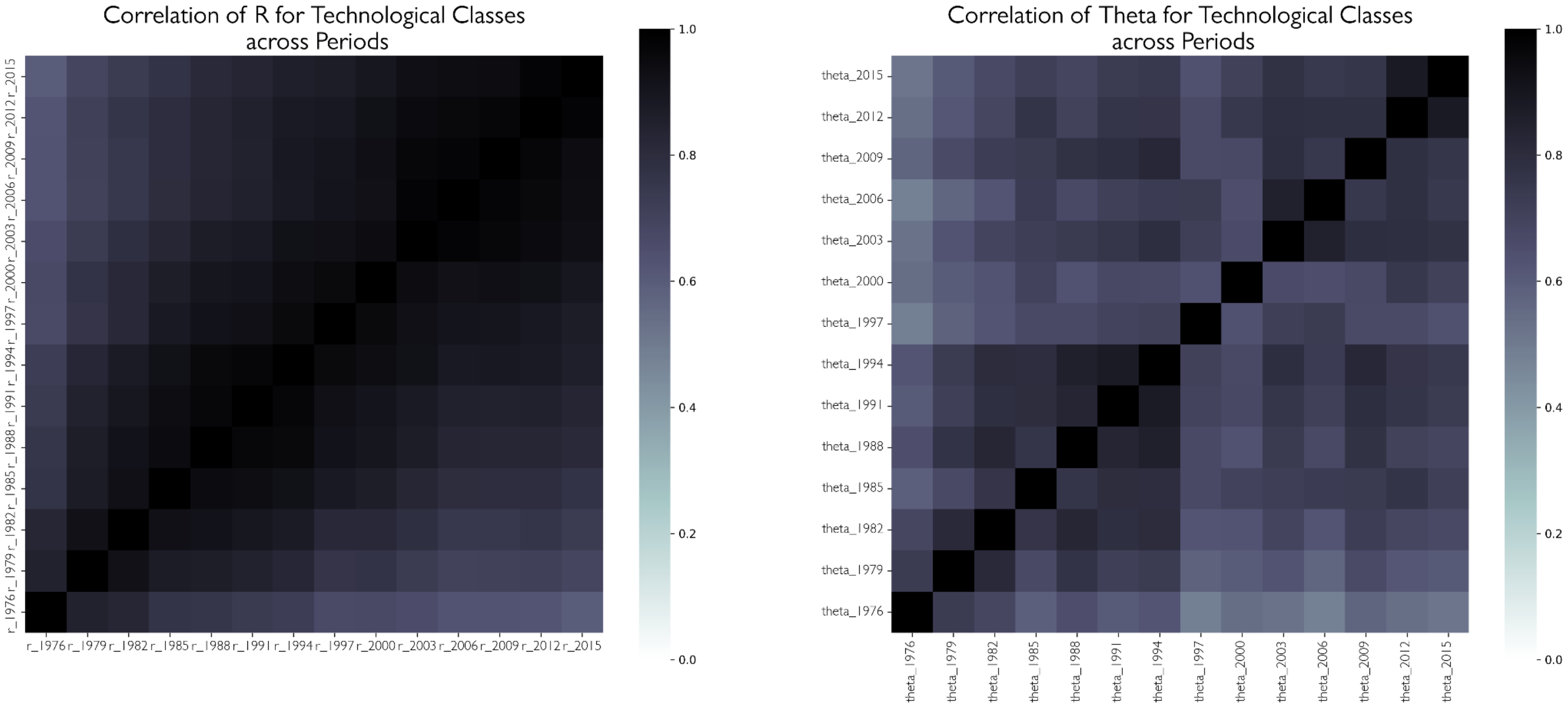

Figure A2. Cross-Period Correlation of Radial (r) and Angular (θ) Values

for Technological Categories

*A2. Correlation between Poincaré-based Measures and Network Measures*

We used the embedding method as a simplified representation of information flow through the citation network. Upon embedding the network within the Poincaré disk, we hypothesized that nodes with higher centrality will be positioned closer to the disk's center, showing a smaller radial distance *r* in polar coordinates. Prior research has shown that such centrally located components play an important role in advancing technology performance improvement rates (Triulzi et al., 2020). Nodes with longer geodesic paths are expected to be located further away from each other, displaying larger angular distances.

In Table A1, we examine the correlation between network centrality measures and radial distance (r). The correlation coefficients between out-degree centrality and radial distance are predominantly greater than 0.6, while correlations between eigenvector centrality and radial distance mostly exceed 0.8. Although the correlation coefficients between closeness centrality and radial distance are somewhat lower, they still attain a baseline of 0.4.

Table A1. Correlation between Centrality Measures and *r* on Poincaré Disk

| | Out-Degree Centrality vs. r | Eigenvector Centrality vs. r | Closeness Centrality vs. r |
|---|---|---|---|
| *1976-1978* | 0.597 [0.000] | 0.590 [0.000] | 0.355 [0.000] |
| *1979-1981* | 0.646 [0.000] | 0.709 [0.000] | 0.402 [0.000] |
| *1982-1984* | 0.677 [0.000] | 0.721 [0.000] | 0.433 [0.000] |
| *1985-1987* | 0.732 [0.000] | 0.807 [0.000] | 0.468 [0.000] |
| *1988-1990* | 0.761 [0.000] | 0.827 [0.000] | 0.475 [0.000] |
| *1991-1993* | 0.773 [0.000] | 0.817 [0.000] | 0.497 [0.000] |
| *1994-1996* | 0.802 [0.000] | 0.841 [0.000] | 0.482 [0.000] |
| *1997-1999* | 0.816 [0.000] | 0.841 [0.000] | 0.517 [0.000] |
| *2000-2002* | 0.834 [0.000] | 0.859 [0.000] | 0.508 [0.000] |
| *2003-2005* | 0.849 [0.000] | 0.875 [0.000] | 0.533 [0.000] |
| *2006-2008* | 0.864 [0.000] | 0.871 [0.000] | 0.542 [0.000] |
| *2009-2011* | 0.871 [0.000] | 0.865 [0.000] | 0.524 [0.000] |
| *2012-2014* | 0.883 [0.000] | 0.857 [0.000] | 0.480 [0.000] |
| *2015-2017* | 0.886 [0.000] | 0.859 [0.000] | 0.473 [0.000] |

*Note: Correlation coefficients based on Pearson correlation. P-values are between square brackets, all tests two-tailed.*

In table A2, we explore the correlation between the geodesic path length and angular distance ($|\Delta\theta|$) for corresponding node pairs. Almost every technological category exhibits a degree of interconnectedness to each other. While some node pairs are connected by only one or two ties, these incidental ties could alter the overall distribution of geodesic distances. To solve this potential problem, we implement various tie weight thresholds. This approach helps maintain the core structure of citation networks. We then correlate the geodesic paths within these networks to angular distance derived from the Poincaré disks.

Table A2. Correlation between Geodesic Distance and $\Delta\theta$ on Poincaré Disk

| | Tie Frequency Thresholds | | | |
|---|---|---|---|---|
| | Top 10% | Top 5% | Top 0.5% | Top 0.1% |
| *1976-1978* | 0.176 [0.000] | 0.204 [0.000] | 0.442 [0.000] | 0.668 [0.000] |
| *1979-1981* | 0.172 [0.000] | 0.151 [0.000] | 0.212 [0.000] | 0.737 [0.000] |
| *1982-1984* | 0.150 [0.000] | 0.206 [0.000] | 0.227 [0.000] | 0.505 [0.000] |
| *1985-1987* | 0.153 [0.000] | 0.173 [0.000] | 0.229 [0.000] | 0.390 [0.000] |
| *1988-1990* | 0.133 [0.000] | 0.136 [0.000] | 0.255 [0.000] | 0.388 [0.000] |
| *1991-1993* | 0.142 [0.000] | 0.161 [0.000] | 0.233 [0.000] | 0.254 [0.000] |
| *1994-1996* | 0.154 [0.000] | 0.159 [0.000] | 0.240 [0.000] | 0.247 [0.000] |
| *1997-1999* | 0.147 [0.000] | 0.151 [0.000] | 0.268 [0.000] | 0.261 [0.000] |
| *2000-2002* | 0.147 [0.000] | 0.151 [0.000] | 0.250 [0.000] | 0.364 [0.000] |
| *2003-2005* | 0.138 [0.000] | 0.145 [0.000] | 0.197 [0.000] | 0.264 [0.000] |
| *2006-2008* | 0.129 [0.000] | 0.150 [0.000] | 0.207 [0.000] | 0.252 [0.000] |
| *2009-2011* | 0.122 [0.000] | 0.136 [0.000] | 0.133 [0.000] | 0.240 [0.000] |
| *2012-2014* | 0.136 [0.000] | 0.182 [0.000] | 0.215 [0.000] | 0.324 [0.000] |
| *2015-2017* | 0.141 [0.000] | 0.166 [0.000] | 0.262 [0.000] | 0.297 [0.000] |

*Note: Correlation coefficients based on Pearson correlation. P-values are between square brackets, all tests two-tailed.*

Table A2 shows a positive correlation between geodesic and angular distances across most time periods. As tie weight thresholds increase, a more precisely defined "core" of the citation network stands out. This core maintains the most important distance information that our embedding space should effectively capture. In the most strictly constructed "core," comprising only the top 0.1% of ties, the correlation between geodesic and angular distances can exceed 0.5.

*A3. Validating Poincaré-based Measures with the Semantic Embeddings from Word2Vec*

Our measures do not directly capture specific semantic distances and may be confounded by the interdisciplinary nature of technological domains. To address this potential confound and validate the Poincaré distance measures, we conducted three additional robustness tests.

We first calculate the position of each patent and technological category (as the average of all related patents) in Euclidean space using semantic embeddings (Word2Vec), and then compare the results with those from Poincaré embeddings (2015-2017 slice). At the level of category pairs—restricted to categories with at least 2,000 patents—we observe a Spearman correlation of approximately 0.57 between distances from Word2Vec and Poincaré embeddings. Given that Word2Vec does not account for temporal dynamics, this represents a strikingly strong correlation. This result suggests that, although our embedding does not incorporate explicit semantic information, it effectively captures technological distance by leveraging knowledge flows derived from citation networks. Based on this evidence, we believe that $\Delta\theta$ serves as a reasonable proxy for semantic distance.

Second, to address the potential confounding effect of interdisciplinarity, we compare the distributions of inter-category and intra-category distances for each CPC category, based on the Word2Vec model. Because our method does not incorporate semantic information, it primarily captures inter-category distances while overlooking intra-category variation. This comparison is presented in Figure A3 below. As shown, inter-category distances are substantially greater than intra-category distances, with a $t$-statistic of 88.946 ($p < 0.001$), indicating a statistically significant difference. Furthermore, the correlation between number of patents within each category and intra-category distance is low (approximately 0.11), suggesting that larger fields do not necessarily exhibit broader cognitive spans. Therefore, although our measure may not capture fine-grained variations in cognitive span introduced by the interdisciplinarity within technology domains, we argue that this limitation does not fundamentally affect the overall distance scale and thus does not alter our main conclusions.

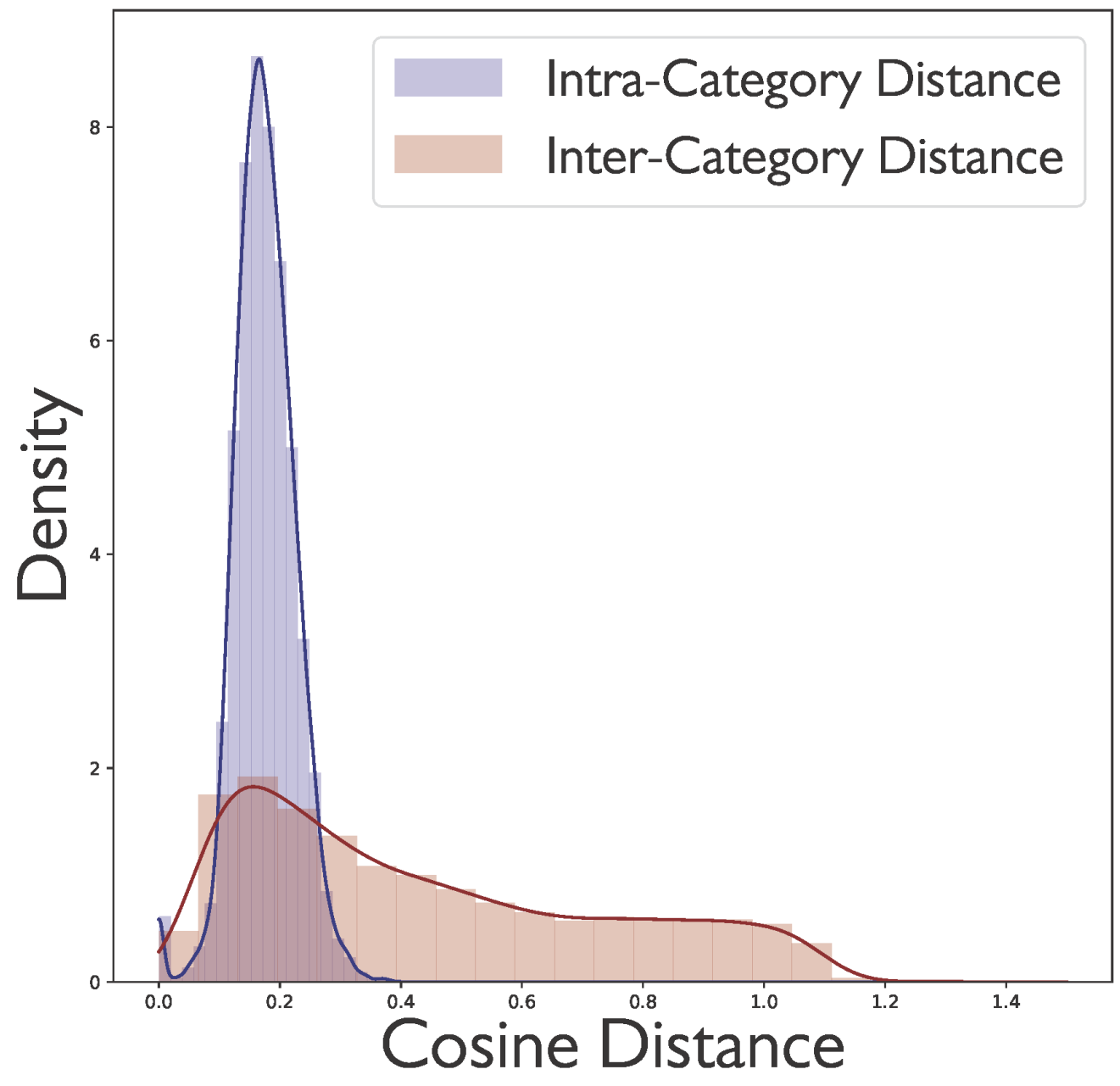

Figure A3. Comparison between Distributions of Intra- and Inter-Category Distances

## Appendix B. Supplementary Models for Main Models and Mechanism Analysis

*B1. Complementary Models based on Different Sample Inclusion Criterion*

In the main text, we selected two sets of samples for model building: all patents granted before 2017 for Model 1, and all patents granted before 2002 for Models 2-5. The earlier patents provide a longer period for constructing citation variables, enabling the estimation of depth versus breadth effects across time. To test the robustness of our models, we also employed two more subsamples: patents granted before 2012, allowing for 0-5 and 5-10 year citations, and patents before 2007, to include 0-5, 5-10, and 10-15 year citations. We then reran the models on these two subsamples. The results are presented in Table B1.

Table B1. Effects of Recombination Depth vs. Breadth on Patent Short- and Long-Term Impact Based on 1976-2007 and 1976-2012 Subsamples

| | **Models with Patents 1976-2007** | | **Models with Patents 1976-2012** | | |
|---|---|---|---|---|---|
| | **Model 1: Predicting 0-5 Year Citation** | **Model 2 Predicting 5-10 Year Citation** | **Model 3 Predicting 0-5 Year Citation** | **Model 4 Predicting 5-10 Year Citation** | **Model 5 Predicting 10-15 Year Citation** |
| ***Main Models*** | | | | | |
| Depth | 0.146 | 0.041 | 0.121 | 0.034 | -0.104 |
| | [0.000] | [0.000] | [0.000] | [0.000] | [0.000] |
| | (0.005) | (0.006) | (0.006) | (0.006) | (0.008) |
| Breadth | -0.055 | -0.036 | -0.054 | -0.041 | -0.018 |
| | [0.000] | [0.000] | [0.000] | [0.000] | [0.000] |
| | (0.001) | (0.001) | (0.001) | (0.001) | (0.001) |
| Horizontal Location (mean theta) | 0.031 | 0.025 | 0.043 | 0.035 | 0.029 |
| | [0.000] | [0.000] | [0.000] | [0.000] | [0.000] |
| | (0.000) | (0.001) | (0.001) | (0.001) | (0.001) |
| Ln (Component Familiarity) | 0.031 | 0.037 | 0.031 | 0.039 | 0.037 |
| | [0.000] | [0.000] | [0.000] | [0.000] | [0.000] |
| | (0.000) | (0.000) | (0.000) | (0.000) | (0.001) |
| Ln (Combination | 0.030 | 0.022 | 0.033 | 0.024 | 0.019 |
| Familiarity) | [0.000] | [0.000] | [0.000] | [0.000] | [0.000] |
| | (0.000) | (0.000) | (0.000) | (0.000) | (0.000) |
| Number of Tech Classes | 0.027 | 0.036 | 0.029 | 0.037 | 0.043 |
| | [0.000] | [0.000] | [0.000] | [0.000] | [0.000] |
| | (0.000) | (0.000) | (0.000) | (0.000) | (0.000) |
| Number of Prior Art | 0.003 | 0.004 | 0.003 | 0.004 | 0.005 |
| Classes | [0.000] | [0.000] | [0.000] | [0.000] | [0.000] |
| | (0.000) | (0.000) | (0.000) | (0.000) | (0.000) |
| Assigned to Organizations | 0.088 | 0.049 | 0.104 | 0.057 | -0.025 |
| | [0.000] | [0.000] | [0.000] | [0.000] | [0.000] |
| | (0.002) | (0.002) | (0.002) | (0.002) | (0.003) |
| Invented by Team | 0.078 | 0.094 | 0.079 | 0.092 | 0.085 |
| | [0.000] | [0.000] | [0.000] | [0.000] | [0.000] |
| | (0.001) | (0.002) | (0.002) | (0.002) | (0.002) |
| Ln (Inventor Experience) | 0.006 | -0.019 | 0.005 | -0.019 | -0.039 |
| | [0.000] | [0.000] | [0.000] | [0.000] | [0.000] |
| | (0.000) | (0.000) | (0.000) | (0.001) | (0.001) |
| Non Patent References | 0.009 | 0.010 | 0.008 | 0.013 | 0.013 |
| | [0.000] | [0.000] | [0.000] | [0.000] | [0.000] |
| | (0.000) | (0.000) | (0.000) | (0.000) | (0.000) |
| Number of Claims | 0.018 | 0.021 | 0.018 | 0.021 | 0.023 |
| | [0.000] | [0.000] | [0.000] | [0.000] | [0.000] |
| | (0.000) | (0.000) | (0.000) | (0.000) | (0.000) |
| Family Size | -0.139 | -0.253 | -0.118 | -0.226 | -0.286 |
| | [0.000] | [0.000] | [0.000] | [0.000] | [0.000] |
| | (0.001) | (0.001) | (0.001) | (0.001) | (0.002) |
| Year Fixed Effects | YES | YES | YES | YES | YES |
| Constant | 0.098 | 0.036 | 0.090 | 0.008 | 0.178** |
| | [0.069] | [0.532] | [0.095] | [0.889] | [0.003] |
| | (0.054) | (0.057) | (0.054) | (0.057) | (0.061) |
| ***Dispersion Models*** | | | | | |
| Depth | -0.011 | -0.074 | 0.085 | 0.002 | -0.101 |
| | [0.145] | [0.000] | [0.000] | [0.840] | [0.000] |
| | (0.008) | (0.007) | (0.010) | (0.008) | (0.008) |
| Breadth | -0.023 | -0.034 | -0.032 | -0.031 | -0.039 |
| | [0.000] | [0.000] | [0.000] | [0.000] | [0.000] |
| | (0.001) | (0.001) | (0.001) | (0.001) | (0.001) |
| Horizontal Location | 0.013 | 0.002 | 0.010 | 0.006 | -0.002 |
| (mean theta) | [0.000] | [0.002] | [0.000] | [0.000] | [0.001] |
| | (0.001) | (0.001) | (0.001) | (0.001) | (0.001) |
| Ln (Component | 0.031 | 0.042 | 0.022 | 0.033 | 0.049 |
| Familiarity) | [0.000] | [0.000] | [0.000] | [0.000] | [0.000] |
| | (0.001) | (0.001) | (0.001) | (0.001) | (0.001) |
| Ln (Combination | -0.016 | -0.004 | -0.021 | -0.009 | 0.003 |
| Familiarity) | [0.000] | [0.000] | [0.000] | [0.000] | [0.000] |
| | (0.000) | (0.000) | (0.000) | (0.000) | (0.000) |
| Number of Tech Classes | 0.002 | 0.002 | -0.000 | 0.000 | -0.001 |
| | [0.000] | [0.000] | [0.228] | [0.480] | [0.002] |
| | (0.000) | (0.000) | (0.000) | (0.000) | (0.000) |
| Number of Prior Art Classes | -0.001 | -0.000 | -0.001 | -0.001 | -0.000 |
| | [0.000] | [0.000] | [0.000] | [0.000] | [0.000] |
| | (0.000) | (0.000) | (0.000) | (0.000) | (0.000) |
| Assigned to Organizations | 0.067 | 0.120 | 0.065 | 0.114 | 0.147 |
| | [0.000] | [0.000] | [0.000] | [0.000] | [0.000] |
| | (0.004) | (0.003) | (0.004) | (0.003) | (0.003) |
| Invented by Team | 0.031 | 0.011 | 0.032 | 0.021 | 0.003 |
| | [0.000] | [0.000] | [0.000] | [0.000] | [0.104] |
| | (0.002) | (0.002) | (0.003) | (0.002) | (0.002) |
| Ln (Inventor Experience) | 0.011 | 0.033 | 0.012 | 0.029 | 0.039 |
| | [0.000] | [0.000] | [0.000] | [0.000] | [0.000] |
| | (0.001) | (0.001) | (0.001) | (0.001) | (0.001) |
| Non Patent References | 0.012 | 0.005 | 0.018 | 0.009 | 0.004 |
| | [0.000] | [0.000] | [0.000] | [0.000] | [0.000] |
| | (0.000) | (0.000) | (0.000) | (0.000) | (0.000) |
| Number of Claims | -0.007 | -0.007 | -0.007 | -0.006 | -0.007 |
| | [0.000] | [0.000] | [0.000] | [0.000] | [0.000] |
| | (0.000) | (0.000) | (0.000) | (0.000) | (0.000) |
| Family Size | -0.025 | 0.024 | -0.030 | -0.000 | 0.049 |
| | [0.000] | [0.000] | [0.000] | [0.991] | [0.000] |
| | (0.002) | (0.002) | (0.002) | (0.002) | (0.002) |
| Year Fixed Effects | YES | YES | YES | YES | YES |
| Constant | -0.039 | -0.021 | -0.045 | -0.012 | 0.075 |
| | [0.673] | [0.819] | [0.626] | [0.893] | [0.417] |
| | (0.092) | (0.092) | (0.092) | (0.092) | (0.093) |
| Observations | 3711809 | 3711809 | 2774652 | 2774652 | 2774652 |

*Note: P-values are between square brackets. Robust standard errors are in parentheses. All tests are two-tailed.*

Table B1 exhibits similar patterns to those observed in Table 3 of the main text. In the short term, depth shows a positive effect while breadth shows a negative effect; in the long term, these effects reverse.

*B2. Depth versus. Breadth Effects within the Same Organization*

We replicated the results of Table 3 using only intra-organization citations. These results are presented in Table B2.

Table B2. Effects of Recombination Depth vs. Breadth on Patent Short- and Long-Term Impact (Within Organization)

| | Model with All Patents: 1976-2017 | Model with Early Patents: 1976-2002 | | | |
|---|---|---|---|---|---|
| | **Model 1: Predicting 0-5 Year Citation** | **Model 2 Predicting 0-5 Year Citation** | **Model 3 Predicting 5-10 Year Citation** | **Model 4 Predicting 10-15 Year Citation** | **Model 5 Predicting 15-20 Year Citation** |
| ***Main Models*** | | | | | |
| Depth | 0.057 | 0.031 | 0.080 | 0.060 | -0.167 |
| | [0.000] | [0.012] | [0.000] | [0.014] | [0.000] |
| | (0.008) | (0.012) | (0.017) | (0.024) | (0.030) |
| Breadth | -0.006 | -0.016 | 0.014 | 0.050 | 0.076 |
| | [0.000] | [0.000] | [0.000] | [0.000] | [0.000] |
| | (0.001) | (0.001) | (0.002) | (0.003) | (0.003) |
| Horizontal Location (mean theta) | 0.012 | 0.017 | 0.028 | 0.024 | 0.034 |
| | [0.000] | [0.000] | [0.000] | [0.000] | [0.000] |
| | (0.001) | (0.001) | (0.001) | (0.002) | (0.002) |
| Ln (Component Familiarity) | 0.021 | 0.012 | 0.031 | 0.046 | 0.042 |
| | [0.000] | [0.000] | [0.000] | [0.000] | [0.000] |
| | (0.001) | (0.001) | (0.001) | (0.002) | (0.002) |
| Ln (Combination Familiarity) | 0.021 | 0.018 | 0.022 | 0.025 | 0.024 |
| | [0.000] | [0.000] | [0.000] | [0.000] | [0.000] |
| | (0.000) | (0.001) | (0.001) | (0.001) | (0.001) |
| Number of Tech Classes | 0.016 | 0.017 | 0.022 | 0.023 | 0.022 |
| | [0.000] | [0.000] | [0.000] | [0.000] | [0.000] |
| | (0.000) | (0.000) | (0.001) | (0.001) | (0.001) |
| Number of Prior Art Classes | 0.003 | 0.002 | 0.002 | 0.002 | 0.002 |
| | [0.000] | [0.000] | [0.000] | [0.000] | [0.000] |
| | (0.000) | (0.000) | (0.000) | (0.000) | (0.000) |
| Assigned to Organizations | 0.000 | 0.000 | 0.000 | 0.000 | 0.000 |
| | [.] | [.] | [.] | [.] | [.] |
| | (.) | (.) | (.) | (.) | (.) |
| Invented by Team | 0.070 | 0.088 | 0.089 | 0.086 | 0.149 |
| | [0.000] | [0.000] | [0.000] | [0.000] | [0.000] |
| | (0.002) | (0.003) | (0.004) | (0.006) | (0.008) |
| Ln (Inventor Experience) | 0.077 | 0.073 | 0.075 | 0.068 | 0.056 |
| | [0.000] | [0.000] | [0.000] | [0.000] | [0.000] |
| | (0.001) | (0.001) | (0.001) | (0.002) | (0.002) |
| Non Patent References | 0.018 | 0.011 | 0.016 | 0.014 | 0.007 |
| | [0.000] | [0.000] | [0.000] | [0.000] | [0.000] |
| | (0.000) | (0.000) | (0.001) | (0.001) | (0.001) |
| Number of Claims | 0.011 | 0.013 | 0.017 | 0.015 | 0.011 |
| | [0.000] | [0.000] | [0.000] | [0.000] | [0.000] |
| | (0.000) | (0.000) | (0.000) | (0.000) | (0.000) |
| Family Size | -0.100 | -0.063 | -0.190 | -0.264 | -0.332 |
| | [0.000] | [0.000] | [0.000] | [0.000] | [0.000] |
| | (0.002) | (0.002) | (0.003) | (0.004) | (0.005) |
| Year Fixed Effects | YES | YES | YES | YES | YES |
| Constant | -0.138 | -0.065 | -0.152 | 0.292 | 0.238 |
| | [0.096] | [0.442] | [0.232] | [0.189] | [0.217] |
| | (0.083) | (0.084) | (0.127) | (0.222) | (0.193) |
| ***Dispersion Models*** | | | | | |
| Depth | 0.029 | 0.093 | 0.068 | 0.071 | -0.070 |
| | [0.042] | [0.002] | [0.018] | [0.036] | [0.084] |
| | (0.014) | (0.030) | (0.029) | (0.034) | (0.040) |
| Breadth | 0.030 | 0.026 | 0.071 | 0.089 | 0.093 |
| | [0.000] | [0.000] | [0.000] | [0.000] | [0.000] |
| | (0.002) | (0.003) | (0.003) | (0.004) | (0.005) |
| Horizontal Location (mean theta) | 0.026 | 0.016 | 0.031 | 0.026 | 0.048 |
| | [0.000] | [0.000] | [0.000] | [0.000] | [0.000] |
| | (0.001) | (0.002) | (0.002) | (0.003) | (0.004) |
| Ln (Component Familiarity) | 0.048 | 0.015 | 0.016 | 0.038 | 0.080 |
| | [0.000] | [0.000] | [0.000] | [0.000] | [0.000] |
| | (0.001) | (0.002) | (0.002) | (0.003) | (0.003) |
| Ln (Combination Familiarity) | 0.017 | 0.009 | 0.008 | 0.015 | 0.022 |
| | [0.000] | [0.000] | [0.000] | [0.000] | [0.000] |
| | (0.001) | (0.001) | (0.001) | (0.002) | (0.002) |
| Number of Tech Classes | 0.005 | 0.016 | 0.017 | 0.016 | 0.018 |
| | [0.000] | [0.000] | [0.000] | [0.000] | [0.000] |
| | (0.000) | (0.001) | (0.001) | (0.001) | (0.001) |
| Number of Prior Art Classes | 0.002 | 0.001 | -0.000 | 0.000 | 0.000 |
| | [0.000] | [0.000] | [0.575] | [0.892] | [0.000] |
| | (0.000) | (0.000) | (0.000) | (0.000) | (0.000) |
| Assigned to Organizations | 0.000 | 0.000 | 0.000 | 0.000 | 0.000 |
| | [.] | [.] | [.] | [.] | [.] |
| | (.) | (.) | (.) | (.) | (.) |
| Invented by Team | 0.060 | 0.043 | 0.036 | 0.033 | 0.184 |
| | [0.000] | [0.000] | [0.000] | [0.000] | [0.000] |
| | (0.004) | (0.008) | (0.008) | (0.009) | (0.012) |
| Ln (Inventor Experience) | 0.042 | 0.060 | 0.055 | 0.042 | 0.054 |
| | [0.000] | [0.000] | [0.000] | [0.000] | [0.000] |
| | (0.001) | (0.002) | (0.002) | (0.003) | (0.003) |
| Non Patent References | -0.001 | 0.010 | 0.000 | -0.006 | -0.001 |
| | [0.000] | [0.000] | [0.530] | [0.000] | [0.223] |
| | (0.000) | (0.001) | (0.001) | (0.001) | (0.001) |
| Number of Claims | 0.008 | 0.011 | 0.008 | 0.006 | 0.007 |
| | [0.000] | [0.000] | [0.000] | [0.000] | [0.000] |
| | (0.000) | (0.000) | (0.000) | (0.000) | (0.001) |
| Family Size | -0.294 | -0.115 | -0.154 | -0.169 | -0.300 |
| | [0.000] | [0.000] | [0.000] | [0.000] | [0.000] |
| | (0.004) | (0.005) | (0.005) | (0.007) | (0.009) |
| Year Fixed Effects | YES | YES | YES | YES | YES |
| Constant | -1.297 | -1.185 | -0.529 | 0.023 | -2.621 |
| | [0.000] | [0.000] | [0.023] | [0.930] | [0.027] |
| | (0.259) | (0.259) | (0.232) | (0.261) | (1.188) |
| Observations | 1440070 | 613180 | 369316 | 204775 | 101397 |

*Note: P-values are between square brackets. Robust standard errors are in parentheses. All tests are two-tailed.*

Table B2 exhibits similar patterns to those observed in Table 3 of the main text. In the short term, depth shows a positive effect while breadth shows a negative effect; in the long term, these effects reverse. It also shows that this reversal occurs more rapidly within organizations. While the positive effect of breadth is only seen 15 years after the grant date outside organizations, this shift can occur as early as 5 years after the grant of focal patent within an organization. This is possibly because information diffusion encounters fewer barriers within organizational boundaries, leading to a faster realization of collective invention.

B3. Robustness Test with Poisson Quasi-Maximum Likelihood Estimator

We replicate Table 3 using poisson quasi-maximum likelihood estimator. These results are presented in Table B3.

Table B3. Effects of Recombination Depth vs. Breadth on Patent Short- and Long-Term Impact (Based on Poisson QMLE)

| | Model with All Patents: 1976-2017 | Model with Early Patents: 1976-2002 | | | |
|---|---|---|---|---|---|
| | **Model 1: Predicting 0-5 Year Citation** | **Model 2 Predicting 0-5 Year Citation** | **Model 3 Predicting 5-10 Year Citation** | **Model 4 Predicting 10-15 Year Citation** | **Model 5 Predicting 15-20 Year Citation** |
| ***Main Models*** | | | | | |
| Depth | 0.278 | 0.122 | 0.117 | -0.053 | -0.395 |
| | [0.000] | [0.000] | [0.000] | [0.000] | [0.000] |
| | (0.008) | (0.007) | (0.009) | (0.011) | (0.015) |
| Breadth | -0.007 | -0.040 | -0.034 | -0.014 | 0.054 |
| | [0.000] | [0.000] | [0.000] | [0.000] | [0.000] |
| | (0.001) | (0.001) | (0.001) | (0.002) | (0.002) |
| Horizontal Location (mean theta) | 0.020 | 0.065 | 0.059 | 0.038 | 0.050 |
| | [0.000] | [0.000] | [0.000] | [0.000] | [0.000] |
| | (0.001) | (0.001) | (0.001) | (0.001) | (0.001) |
| Ln (Component Familiarity) | 0.033 | 0.026 | 0.033 | 0.035 | 0.023 |
| | [0.000] | [0.000] | [0.000] | [0.000] | [0.000] |
| | (0.001) | (0.001) | (0.001) | (0.001) | (0.001) |
| Ln (Combination Familiarity) | 0.032 | 0.035 | 0.026 | 0.025 | 0.026 |
| | [0.000] | [0.000] | [0.000] | [0.000] | [0.000] |
| | (0.004) | (0.004) | (0.004) | (0.001) | (0.001) |
| Number of Tech Classes | 0.027 | 0.031 | 0.038 | 0.042 | 0.045 |
| | [0.003] | [0.003] | [0.003] | [0.004] | [0.001] |
| | (0.000) | (0.000) | (0.000) | (0.001) | (0.001) |
| Number of Prior Art Classes | 0.003 | 0.003 | 0.003 | 0.004 | 0.004 |
| | [0.000] | [0.000] | [0.000] | [0.000] | [0.000] |
| | (0.000) | (0.000) | (0.000) | (0.000) | (0.000) |
| Assigned to Organizations | 0.080 | 0.130 | 0.088 | -0.017 | -0.120 |
| | [0.000] | [0.000] | [0.000] | [0.000] | [0.000] |
| | (0.004) | (0.003) | (0.004) | (0.005) | (0.006) |
| Invented by Team | 0.074 | 0.085 | 0.094 | 0.094 | 0.107 |
| | [0.000] | [0.000] | [0.000] | [0.000] | [0.000] |
| | (0.002) | (0.002) | (0.002) | (0.003) | (0.004) |
| Ln (Inventor Experience) | 0.014 | -0.007 | -0.031 | -0.050 | -0.052 |
| | [0.000] | [0.000] | [0.000] | [0.000] | [0.000] |
| | (0.001) | (0.001) | (0.001) | (0.001) | (0.001) |
| Non Patent References | 0.013 | 0.005 | 0.012 | 0.013 | 0.012 |
| | [0.000] | [0.000] | [0.000] | [0.000] | [0.000] |
| | (0.000) | (0.000) | (0.000) | (0.000) | (0.001) |
| Number of Claims | 0.018 | 0.018 | 0.022 | 0.025 | 0.025 |
| | [0.000] | [0.000] | [0.000] | [0.000] | [0.000] |
| | (0.000) | (0.000) | (0.000) | (0.000) | (0.000) |
| Family Size | -0.220 | -0.124 | -0.248 | -0.336 | -0.384 |
| | [0.000] | [0.000] | [0.000] | [0.000] | [0.000] |
| | (0.002) | (0.002) | (0.002) | (0.003) | (0.004) |
| Year Fixed Effects | YES | YES | YES | YES | YES |
| Constant | -0.034 | 0.075 | -0.070 | 0.104 | 0.416 |
| | [0.775] | [0.274] | [0.511] | [0.476] | [0.042] |
| | (0.118) | (0.068) | (0.106) | (0.146) | (0.204) |
| Observations | 4992915 | 2010031 | 2010031 | 2010031 | 2010031 |

*Note: P-values are between square brackets. Robust standard errors are in parentheses. All tests are two-tailed.*

B4. Robustness Tests Using Alternative Thresholds for Depth and Breadth Measures

In the main model, we use the 90th percentile of r and \theta to construct our depth and breadth measures. This specific threshold may have influenced the model results and, by extension, our main conclusions. To assess the robustness of our findings, we construct alternative depth and breadth measures using the 95th percentile and the maximum value as thresholds, and re-estimate the main models with these new specifications. The results, reported in Table B4, remain consistent with those from the main models.

Table B4. Main Models Using Depth and Breadth Measures with Alternative Thresholds

| | Model with All Patents: 1976-2017 | Model with Early Patents: 1976-2002 | | | |
|---|---|---|---|---|---|
| | **Model 1: Predicting 0-5 Year Citation** | **Model 2 Predicting 0-5 Year Citation** | **Model 3 Predicting 5-10 Year Citation** | **Model 4 Predicting 10-15 Year Citation** | **Model 5 Predicting 15-20 Year Citation** |
| ***Models Using Depth and Breadth Measures with 95% Percentile*** | | | | | |
| Depth | 0.197*** | 0.084*** | -0.001 | -0.160*** | -0.412*** |
| | [0.000] | [0.000] | [0.914] | [0.000] | [0.000] |
| | (0.005) | (0.007) | (0.007) | (0.009) | (0.010) |
| Breadth | -0.045*** | -0.037*** | -0.019*** | 0.000 | 0.040*** |
| | [0.000] | [0.000] | [0.000] | [0.780] | [0.000] |
| | (0.001) | (0.001) | (0.001) | (0.001) | (0.001) |
| Controls (Same as in Table 3 in Main Text) | YES | YES | YES | YES | YES |
| ***Models Using Depth and Breadth Measures with 100% Percentile*** | | | | | |
| Depth | 0.182*** | 0.062*** | -0.006 | -0.150*** | -0.372*** |
| | [0.000] | [0.000] | [0.378] | [0.000] | [0.000] |
| | (0.005) | (0.007) | (0.007) | (0.009) | (0.010) |
| Breadth | -0.049*** | -0.039*** | -0.022*** | -0.004*** | 0.037*** |
| | [0.000] | [0.000] | [0.000] | [0.000] | [0.000] |
| | (0.001) | (0.001) | (0.001) | (0.001) | (0.001) |
| Controls (Same as in Table 3 in Main Text) | YES | YES | YES | YES | YES |
| Observations | 4992915 | 2010031 | 2010031 | 2010031 | 2010031 |

Note: *p*-values are between square brackets. Robust standard errors are in parentheses. All tests are two-tailed.

## B5. Causal Inference for Depth and Breadth Effects

In this section, we conducted two additional analyses to address the endogeneity problem. First, we implemented a US–EU twin-patent design, and second, we employed an instrumental variable approach to strengthen the causal interpretation of our results.

### B5.1 US-EU Twin Patent Tests

In this test, we identified U.S.–European patent pairs to control for patent quality and content. Following prior studies, we retained only valid twin patents filed in both the USPTO and EPO, excluding those with different titles or one-to-many/many-to-one matches. The final sample consists of 84,916 patent pairs.

We use EPO patents, rather than those filed at other national patent offices, as the matching sample for two main reasons. First is the issue of language. For example, many patents filed with the German Patent and Trademark Office (DPMA) are written exclusively in German, with only limited content available in English. This reduces cross-sample comparability and creates challenges for both matching and text analysis. Second, is the issue of sample size. The matched USPTO–EPO dataset provides a substantially larger sample than a matched USPTO–DPMA dataset would offer.

For patents granted by the USPTO, citations may be added both by applicants and examiners, while patents granted by EPO only receive citations added by examiners. To make the pairwise samples comparable, we only count citations added by examiners for both USPTO and EPO samples. Distinct from our first analysis, here we calculated depth and breadth for the US patents based on the USPC classification system and for EPO patents based on the IPC system in order to intentionally keep different operationalizations between the twin samples.

European categories are embedded in separate Poincaré disks within 3-year windows. We restricted our sample to patents granted between 2001-2012, as 2001 is the first year when data on examiner-added citations is available for the USPTO, and 2013 is the year when USPTO changed the standardized classification system from USPC to CPC, which shifts the comparison baseline.

An invention's citation counts at the USPTO can be specified as:

$$log(Citations_{i,US}) = \beta_1 \times Depth_{i,US} + \beta_2 \times Breadth_{i,US} + \beta_3 \times Controls_{i,US} + Quality_i + \epsilon_{i,US}$$

while the impact of its patent twin at the EPO can be expressed as:

$$log(Citations_{i,EU}) = \gamma_1 \times Depth_{i,EU} + \gamma_2 \times Breadth_{i,EU} + \gamma_3 \times Controls_{i,EU} + Quality_i + \epsilon_{i,EU}$$

We substitute the second equation into the first equation. The resulting equation has the following form:

$$log(Citations_{i,US}) = \alpha_1 \times Depth_{i,US} + \alpha_2 \times Breadth_{i,US} + \alpha_3 \times Controls_{i,US}$$
$$+\alpha_4 \times log(Citations_{i,EU}) - \alpha_5 \times Depth_{i,EU} - \alpha_6 \times Breadth_{i,EU} - \alpha_7 \times Controls_{i,EU}$$
$$+\epsilon_{i,US} - \epsilon_{i,EU}$$

We estimate this model on the pairwise samples to get the depth and breadth effects when controlling for patent quality and other individual-level heterogeneities as a robustness test. By substituting the European citation equation into the original model, we strictly control for patent quality, and so estimate the perception of depth and breadth effects independent of the inventive process.

We report the results of this model in Table B5. If our argument is robust, we adopt the standards set by Kovács and colleagues (2021) and expect (1) the depth and breadth effects to persist with the same signs, as in the previous models from Table 3; (2) US and European citation counts to be positively correlated in each of the target periods; and (3) within the US citation model, the European independent variables (e.g., European depth and breadth for the same patent) will have opposite signs as the US variables, suggesting that they manifest the same effects, but flipped by our model specification (see Section 4.2). The results, as reported in Table B5, meet these expectations, and show the robustness of our model from main text Table 3 when strictly controlling for the quality of the underlying invention.

Table B5. Patent-Twin Tests:
0–5Y Forward Citations at USPTO Controlling for EPO Measures

| | Model 1 | Model 2 | Model 3 | Model 4 |
|---|---|---|---|---|
| | Model: Poisson QMLE | | Model: Negative Binomial | |
| USPTO Depth | 0.750*** | 0.624*** | 0.548*** | 0.571*** |
| | [0.000] | [0.000] | [0.000] | [0.000] |
| | (0.038) | (0.040) | (0.031) | (0.036) |
| USPTO Breadth | -0.053*** | -0.050*** | -0.020*** | -0.043*** |
| | [0.000] | [0.000] | [0.000] | [0.000] |
| | (0.007) | (0.007) | (0.005) | (0.006) |
| European Five-Year Forward Citation | | 0.079*** | | 0.103*** |
| | | [0.000] | | [0.000] |
| | | (0.003) | | (0.003) |
| European Depth | | -0.075** | | -0.063* |
| | | [0.006] | | [0.010] |
| | | (0.027) | | (0.025) |
| European Breadth | | 0.012 | | 0.008 |
| | | [0.065] | | [0.166] |
| | | (0.006) | | (0.006) |
| US/EU Patent Controls | YES | YES | YES | YES |
| US/EU Year Fixed Effects | YES | YES | YES | YES |
| Dispersion Model For Negative Binomial | | | YES | YES |
| N | 84,956 | 84,956 | 84,956 | 84,956 |
| Log-Likelihood | -137492.823 | -131271.191 | -118030.06 | -116809.3 |

*Note: P-values are between square brackets. Robust standard errors are in parentheses. All tests are two-tailed.*

## B5.2 Instrumental Variable Approach

Next, we employ an instrumental variable test to further rule out confounding effects and address the potential endogeneity issue. For this purpose, we select an instrumental variable, the reclassification of codings in prior art categories before focal patent filing.

Reclassification occurs when patent administration adjusts the CPC system. Specifically, reclassification of prior art categories takes place when a focal patent cites other patents and the CPC categories of those cited patents are updated. This variable satisfies the two key criteria for a valid IV: first, it is relevant to the independent variables (depth and breadth), as it shapes how focal inventors gather information and navigate the technological space. Second, it does not directly affect the outcome (citations of the focal patent), since it primarily reflects an external, system-level shock. By the time the focal patent is filed, both its backward citations and the reclassification of prior art categories have already occurred, making it unlikely that its forward citations (that focal patent receives) are affected by these adjustments.

For each focal patent, we measure what proportion of its prior art categories are under adjustment by USPTO. Since the data on CPC taxonomy changes (collected from https://www.uspto.gov/web/patents/classification/cpc/html/cpc-notices-of-changes.html) only begins in 2013, we estimate the IV models using patents granted between 2013 and 2017. Because these samples do not provide a full 10-year window to observe forward citations (or longer periods), we limit the outcome variable to 0–5 year forward citations. Nevertheless, this analysis helps strengthen our causal inference and mitigates the influence of potential confounders.

Based on the IV we described above, we estimated the model using two-stage least squares (2SLS) with robust standard errors. Given we only have one instrumental variable, we use the IV to test depth and breadth separately. For the depth model, the Kleibergen-Paap rk Wald F statistic is 5703.794 ($p < 0.001$), and the Cragg-Donald Wald F statistic is 6059.688 ($p < 0.001$). For the breadth model, the Kleibergen-Paap rk Wald F statistic is 296.505 ($p < 0.001$), and the Cragg-Donald Wald F statistic is 298.091 ($p < 0.001$). All the F-statistics are far above the highest Stock-Yogo threshold in both cases, suggesting that our IV satisfies the relevance condition.

Results of both the original models and 2SLS estimations are reported in Table B6. The findings show that the IV estimates are largely consistent with the original model results, indicating that the effects of search depth and breadth on forward citations remain robust after accounting for potential endogeneity, and confirming that our main conclusions regarding the influence of search strategies on patent impact are not driven by omitted variable bias or reverse causality.

Table B6. 2SLS Analysis of Depth and Breadth Effects
(Data: Patents between 2013-2017, DV: 0-5 Year Forward Citations)

| | **Original Model** | **Original Model** | **2SLS: Depth** | | **2SLS: Breadth** | |
|---|---|---|---|---|---|---|
| | Negative Binomial | OLS | **Stage 1: DV= Depth** | **Stage 2: DV=Citations** | **Stage 1: DV=Breadth** | **Stage 2: DV=Citations** |
| **Instrumental: Reclassification of Prior Art Categories** | | | -0.018$^{***}$ | | 0.027$^{***}$ | |
| | | | [0.000] | | [0.000] | |
| | | | (0.000) | | (0.002) | |
| Depth / **Predicted Depth** | 0.546$^{***}$ | 5.120$^{***}$ | | **17.282**$^{***}$ | 2.408$^{***}$ | 24.395$^{***}$ |
| | [0.000] | [0.000] | | **[0.000]** | [0.000] | [0.000] |
| | (0.016) | (0.149) | | **(1.662)** | (0.006) | (2.808) |
| Breadth **/ Predicted Breadth** | -0.021$^{***}$ | -0.078$^{***}$ | 0.051$^{***}$ | -0.704$^{***}$ | | **-8.104**$^{***}$ |
| | [0.000] | [0.000] | [0.000] | [0.000] | | **[0.000]** |
| | (0.002) | (0.022) | (0.000) | (0.085) | | **(1.152)** |
| Horizontal Location (mean theta) | 0.022$^{***}$ | 0.056$^{***}$ | -0.001$^{***}$ | 0.064$^{***}$ | -0.022$^{***}$ | -0.124$^{***}$ |
| | [0.000] | [0.000] | [0.000] | [0.000] | [0.000] | [0.000] |
| | (0.002) | (0.013) | (0.000) | (0.006) | (0.000) | (0.028) |
| Ln (Component Familiarity) | 0.029$^{***}$ | 0.066$^{***}$ | -0.003$^{***}$ | 0.106$^{***}$ | -0.100$^{***}$ | -0.736$^{***}$ |
| | [0.000] | [0.000] | [0.000] | [0.000] | [0.000] | [0.000] |
| | (0.001) | (0.008) | (0.000) | (0.007) | (0.000) | (0.115) |
| Ln (Combination Familiarity) | 0.020$^{***}$ | 0.117$^{***}$ | -0.003$^{***}$ | 0.150$^{***}$ | -0.017$^{***}$ | -0.019 |
| | [0.000] | [0.000] | [0.000] | [0.000] | [0.000] | [0.336] |
| | (0.001) | (0.005) | (0.000) | (0.008) | (0.000) | (0.020) |
| Number of Tech Classes | 0.029$^{***}$ | 0.123$^{***}$ | -0.001$^{***}$ | 0.135$^{***}$ | -0.001$^{***}$ | 0.111$^{***}$ |
| | [0.000] | [0.000] | [0.000] | [0.000] | [0.000] | [0.000] |
| | (0.000) | (0.004) | (0.000) | (0.005) | (0.000) | (0.005) |
| Number of Prior Art Classes | 0.004$^{***}$ | 0.016$^{***}$ | 0.000$^{***}$ | 0.011$^{***}$ | 0.001$^{***}$ | 0.027$^{***}$ |
| | [0.000] | [0.000] | [0.000] | [0.000] | [0.000] | [0.000] |
| | (0.000) | (0.000) | (0.000) | (0.001) | (0.000) | (0.002) |
| Assigned to Organizations | -0.079*** | -0.155* | -0.007* | -0.071 | -0.111* | -1.048*** |
| | [0.000] | [0.016] | [0.000] | [0.061] | [0.000] | [0.000] |
| | (0.007) | (0.065) | (0.000) | (0.038) | (0.003) | (0.136) |
| Invented by Team | 0.060*** | 0.117*** | -0.002*** | 0.140*** | -0.007*** | 0.057 |
| | [0.000] | [0.000] | [0.000] | [0.000] | [0.000] | [0.068] |
| | (0.004) | (0.033) | (0.000) | (0.028) | (0.001) | (0.031) |
| Ln (Inventor Experience) | 0.019*** | 0.121*** | -0.000*** | 0.125*** | -0.003*** | 0.090*** |
| | [0.000] | [0.000] | [0.000] | [0.000] | [0.000] | [0.000] |
| | (0.001) | (0.008) | (0.000) | (0.009) | (0.000) | (0.009) |
| Non Patent References | 0.011*** | 0.094*** | 0.000*** | 0.091*** | -0.004*** | 0.063*** |
| | [0.000] | [0.000] | [0.000] | [0.000] | [0.000] | [0.000] |
| | (0.000) | (0.002) | (0.000) | (0.003) | (0.000) | (0.005) |
| Number of Claims | 0.022*** | 0.051*** | -0.000** | 0.052*** | -0.001** | 0.046*** |
| | [0.000] | [0.000] | [0.000] | [0.000] | [0.000] | [0.000] |
| | (0.000) | (0.002) | (0.005) | (0.002) | (0.000) | (0.002) |
| Family Size | -0.423*** | -1.070*** | -0.005*** | -1.012*** | -0.003* | -1.091*** |
| | [0.000] | [0.000] | [0.000] | [0.000] | [0.012] | [0.000] |
| | (0.003) | (0.027) | (0.000) | (0.018) | (0.001) | (0.019) |
| Year Fixed Effects | YES | YES | YES | YES | YES | YES |
| Constant | -0.282*** | -1.719*** | 0.325*** | -5.639*** | 3.119*** | 23.371*** |
| | [0.000] | [0.000] | [0.000] | [0.000] | [0.000] | [0.000] |
| | (0.019) | (0.161) | (0.001) | (0.543) | (0.007) | (3.594) |
| Observations | 1281106 | 1281106 | 1281106 | 1281106 | 1281106 | 1281106 |

Note: *p*-values are between square brackets. Robust standard errors are in parentheses. All tests are two-tailed.

**Appendix C. Comparing Poincaré Embedding with CPC classification system in Capturing Distances between Technological Categories**

The distinction between knowledge and its functions is well established in the literature on technological innovation (Benson and Magee, 2015; Magee et al., 2016). In this paper, we argue that the conventional classification system does not adequately capture the practical relevance among technological categories, as it is grounded in knowledge similarity rather than functional relationships. To provide empirical evidence for this argument, we implement a statistical test.

For each of the 14 periods of our embedding, we first identify all the technological categories that have appeared in the patents. Assuming the number of eligible categories is N, then we get N(N-1)/2 pairs of categories. For each pair, we calculate the co-appearance by pointwise mutual information:

$$PMI(i,j)=log2(\frac{P(i,j)}{P(i)*P(j)})$$

A higher Pointwise Mutual Information (PMI) score indicates a greater propensity for a pair to be utilized conjointly in patents, thereby suggesting a smaller distance between them. In this study, PMI is used as a benchmark against which our measures are compared. A measure that accurately encapsulates this distance is more effective in defining the search scope.

We evaluate two candidates in this regard: our proposed hyperbolic embedding and the conventional CPC classification system.

- For hyperbolic embedding, we calculate the distance between two points as:

$$d(u,v) = arcosh(1 + 2\frac{||u-v||^2}{(1-||u||^2)(1-||v||^2)})$$

As defined by (Nickle & Kiela, 2017). We calculate similarity between two points as:

$similarity(u,v) = \frac{1}{1+d(u,v)}$

Following the source code of the Gensim package on Poincaré embedding.

- ❖ For traditional hierarchical CPC classification, we use the approach by (Trajtenberg et al., 1997). In the Cooperative Patent Classification ("CPC") system, each patent is assigned one or more four-level classifications. For instance, the classification A01B33 belongs to "A" at the top level, "A01" at the second, "A01B" at the third level, and "A01B33" at the lowest level. For each pair of technological categories, we set the tech distance $d_{i,j}$ to be 0.25 if they are in the same three-digit class (e.g., "A01B33" and "A01B35"), 0.5 if they are in the same two-digit class (e.g., "A01B33" and "A01F12"), and 1 if they are in different one-digit classes or have no prior art (e.g., "A01B33" and "B01B1"). Technological similarity is simply set as 1-$d_{i,j}$.

Having established the empirical benchmark and the two measures of pairwise technological relevance, we proceed to calculate the correlation among these three measures. To do this, we employ Spearman's rank correlation coefficient, given that the distance based on the CPC hierarchy is ordinal in nature. The findings from this analysis are presented in Table C1.

The findings demonstrate that, across all periods and all inclusion thresholds, the similarity measure based on our Poincaré embedding algorithm outperforms the traditional CPC classification in capturing the practical technological relevance. Notably, an increase in the threshold leads to a higher performance disparity. This highlights the strength of our method in accurately representing the "skeleton" of knowledge flow, by yielding a better fit when there is more information in the data input.

Table C1. Spearman Correlation between Benchmark Co-Appearance and Two Similarity Measures

| **Frequency Threshold for Tech Categories** | **0** | | **100** | | **200** | | **500** | | **1000** | |
|---|---|---|---|---|---|---|---|---|---|---|
| Measure | CPC Tree | Poincaré | CPC Tree | Poincaré | CPC Tree | Poincaré | CPC Tree | Poincaré | CPC Tree | Poincaré |
| *1976-1978* | 0.144 | 0.224 | 0.183 | 0.293 | 0.233 | 0.395 | 0.276 | 0.529 | 0.399 | 0.643 |
| *1979-1981* | 0.143 | 0.236 | 0.180 | 0.308 | 0.233 | 0.413 | 0.279 | 0.556 | 0.482 | 0.711 |
| *1982-1984* | 0.138 | 0.232 | 0.175 | 0.307 | 0.217 | 0.416 | 0.274 | 0.532 | 0.385 | 0.678 |
| *1985-1987* | 0.133 | 0.225 | 0.165 | 0.291 | 0.215 | 0.386 | 0.273 | 0.482 | 0.261 | 0.513 |
| *1988-1990* | 0.136 | 0.227 | 0.164 | 0.283 | 0.207 | 0.371 | 0.249 | 0.476 | 0.233 | 0.488 |
| *1991-1993* | 0.137 | 0.232 | 0.170 | 0.292 | 0.209 | 0.375 | 0.261 | 0.494 | 0.266 | 0.511 |
| *1994-1996* | 0.143 | 0.238 | 0.169 | 0.295 | 0.212 | 0.379 | 0.264 | 0.494 | 0.278 | 0.552 |
| *1997-1999* | 0.141 | 0.234 | 0.168 | 0.285 | 0.203 | 0.357 | 0.254 | 0.482 | 0.305 | 0.569 |
| *2000-2002* | 0.141 | 0.235 | 0.167 | 0.282 | 0.194 | 0.345 | 0.243 | 0.456 | 0.290 | 0.543 |
| *2003-2005* | 0.144 | 0.244 | 0.172 | 0.293 | 0.201 | 0.360 | 0.248 | 0.456 | 0.307 | 0.534 |
| *2006-2008* | 0.149 | 0.251 | 0.175 | 0.301 | 0.207 | 0.366 | 0.267 | 0.471 | 0.320 | 0.546 |
| *2009-2011* | 0.150 | 0.258 | 0.174 | 0.304 | 0.202 | 0.367 | 0.252 | 0.459 | 0.290 | 0.538 |
| *2012-2014* | 0.148 | 0.261 | 0.169 | 0.301 | 0.196 | 0.359 | 0.249 | 0.453 | 0.293 | 0.520 |
| *2015-2017* | 0.150 | 0.280 | 0.169 | 0.318 | 0.195 | 0.371 | 0.225 | 0.452 | 0.275 | 0.529 |

## Appendix D. Technical Details of Measure Construction

*D1. Details of Traditional Measures for Recombination Breadth*

In section 3.3.4, we compare our Poincaré disk-based measures to other most widely applied measures in prior research, including technological diversity, technological distances, original combination indicator, and combination familiarity. A detailed explanation of each measure is provided below.

First, we employ the technological categories of prior art to compute the Technological diversity based on Herfindahl index and the technological distances.

- *Technological Diversity*: Initially proposed by (Hall et al., 2001), this measure considers a focal patent i that cites n technological classes, each with a frequency $S_{i,j}$ , where j = 1, 2, 3...n. The Herfindahl diversity for focal patent i is calculated as: $1- \sum_{l}^{n} \quad S_{i,j}$ . To account for potential downward bias in the patents with fewer prior art, we introduce a multiplier as an adjustment. Consequently, the refined technological diversity measure is defined as:

$$\text{Technological diversity} = \frac{number\ of\ prior\ art}{number\ of\ prior\ art-1}(1- \sum_{l}^{n} \quad S_{ij})$$

- *Technological Distance*: Technological Distance: In line with the approach by (Trajtenberg et al., 1997), we rely on the hierarchical organization of knowledge domains within the USPTO classification system, rather than constructing our own knowledge hierarchy representation. In the Cooperative Patent Classification ("CPC") system, each patent is assigned one or more four-level classifications. For instance, the classification

A01B33 belongs to "A" at the top level, "A01" at the second, "A01B" at the third level, and "A01B33" at the lowest level. Our focus lies on the distance between prior art categories. For each pair of technological classes in prior art, we set the tech distance $d_{i,j}$ to be 0 if prior art classes i and j belong to the same four-digit technological class (e.g., both "A01B33"), 0.25 if they are in the same three-digit class (e.g., "A01B33" and "A01B35"), 0.5 if they are in the same two-digit class (e.g., "A01B33" and "A01F12"), and 1 if they are in different one-digit classes or have no prior art (e.g., "A01B33" and "B01B1"). After obtaining all the distances, we compute their average value. It is important to note that we have adjusted the original measure proposed by Trajtenberg et al. by (1) focusing on the distances between technological categories instead of individual patents, and (2) utilizing the CPC system, which features four levels of technological classes, as opposed to three levels. Consequently, the technological distance is defined as: Technological diversity = $\sum_{i,j}^{n^2} d_{i,j}$ , where n is the total number of prior art categories.

In the second step, we examine two additional classical measures that draw upon the technological categories of the focal patent, rather than prior art categories. These measures evaluate the familiarity or novelty of combinations. Due to the large number of prior art categories, prior art technological combinations generally lean towards being less familiar and more original. This tendency may lead to a significant bias towards over-novelty along the continuum. To address this issue, we follow previous research and utilize the technological classes of the focal patents instead.

- The first measure, the original combination indicator, is relatively straightforward and is determined by whether the combination appears for the first time in our dataset. If the combination is entirely novel, the originality dummy is assigned a value of 1. Conversely, if the combination has been previously encountered, the dummy is assigned a value of 0.

- The second measure, combination familiarity, is adopted from (Fleming, 2001). Combination familiarity for patent i =

$$\sum 1\{\text{patent k uses identical combination of subclasses as i}\} \times e^{-\left(\frac{application\ date\ of\ i\ -\ application\ date\ of\ k}{time\ constant\ of\ knowledge\ loss}\right)}$$

The assumption underlying the second measure is that a combination used more frequently in the past would be more familiar. This measure also takes into account that the familiarity of a combination may decline over time as collective memories fade. In accordance with Fleming (2001), we set the time constant for knowledge loss at five years.

It is worth noting that, although our component familiarity variable generates a distribution highly similar to that observed in Fleming (2001) and Kaplan and Vakili (2015), our combination familiarity and cumulative combination usage exhibit a considerably higher mean and variance. This discrepancy can be attributed to our larger sample size compared to both of these studies. As the number of components increases linearly with the sample size, the number of combinations grows quadratically, thereby increasing the likelihood of each combination attaining a higher familiarity value in the history.

*D2. Comparison with Canonical Measures of Recombination Depth and Breadth*

We next correlate our measure with classic measures for depth and breadth from previous literature. The most widely applied measures include Herfindahl diversity (Hall et al., 2001; Kaplan and Vakili, 2015a), technological distances (Trajtenberg et al., 1997), combination familiarity (Fleming, 2001) and a dummy for whether the combination is new to the world (Verhoeven et al., 2016c). Details of these traditional measures are reported in the Appendix C and Appendix D, and correlation results are reported in Table D1.

Table D1. Correlation of Poincaré Depth and Breadth with Previous Measures.

| | Herfindahl Diversity | Technological Distances | Original Combinations (dummy) | Combination Familiarity |
|---|---|---|---|---|
| Depth (max $r$) | 0.376 | 0.272 | 0.104 | -0.242 |
| Breadth (max $\theta$) | 0.415 | 0.410 | 0.096 | -0.160 |

*Note: N=4,992,915, all p-values<0.001.*

Results show that both our measures for depth and breadth exhibit varying positive correlations with conventional operationalizations. A significant proportion of recombinant breadth identified in prior research stems from the horizontal (cross-domain) dimension of Poincaré embedding. This suggests that when inventors combine multiple technological categories, they often bridge several branches of the knowledge hierarchy. Paradoxically, combinatorial novelty displays a larger correlation with depth than breadth ($r_{depth}$=0.104 > $r_{breadth}$=0.096). This demonstrates that most new combinations arise from exploitative search within a single knowledge domain, confirming patterns discovered by Kaplan and Vakili (2015).

*D3. Two Examples*

To explain our measures intuitively, we illustrate two examples for recombination with the Poincaré disk in Figure D1 by showing the location of the two patents and the prior art categories they combine. In the left panel, patent US8837824, was filed by Microsoft Corporation for a method of choosing the best encoding codec for images, which reduces network bandwidth consumption in communication. Among the prior art technology categories underlying this patent, H04N/L/W (pictorial communication, transmission of digital information, communication networks) and G06F/Q (digital data processing, information technology for managerial and administrative purposes) are both categories frequently related to digital information. G06F/Q lies upstream of H04N/L/W, with methods to increase the efficiency and quality of communication (H04N/L/W) improving overall data processing (G06F/Q). This patent exemplifies deep but narrow recombination, as all categories fall along the same branch of the technology tree and draw upon technology components well-used within the specific domains of digital image analysis.

In contrast, the right panel of Figure D1 shows patent US9513224 (assigned to Labrador Diagnostics LLC), which provides an image analysis and measurement method for biological samples using specialized sample holders and systems for examining samples in different settings. This patent (1) aims to improve image data processing (G06T); and (2) targets laboratory settings, focusing on the chemical and physical properties of materials (G01N and B01L). This exemplifies wide but shallow recombination with elements spanning distinct branches of the technology tree, unexpectedly combining general designs across candidate technologies, rather than drawing upon nuanced technical implementations that lie deep in the thickets of a specific application. In this way, Figure D1 illustrates the rationale and inspiration for our measures.

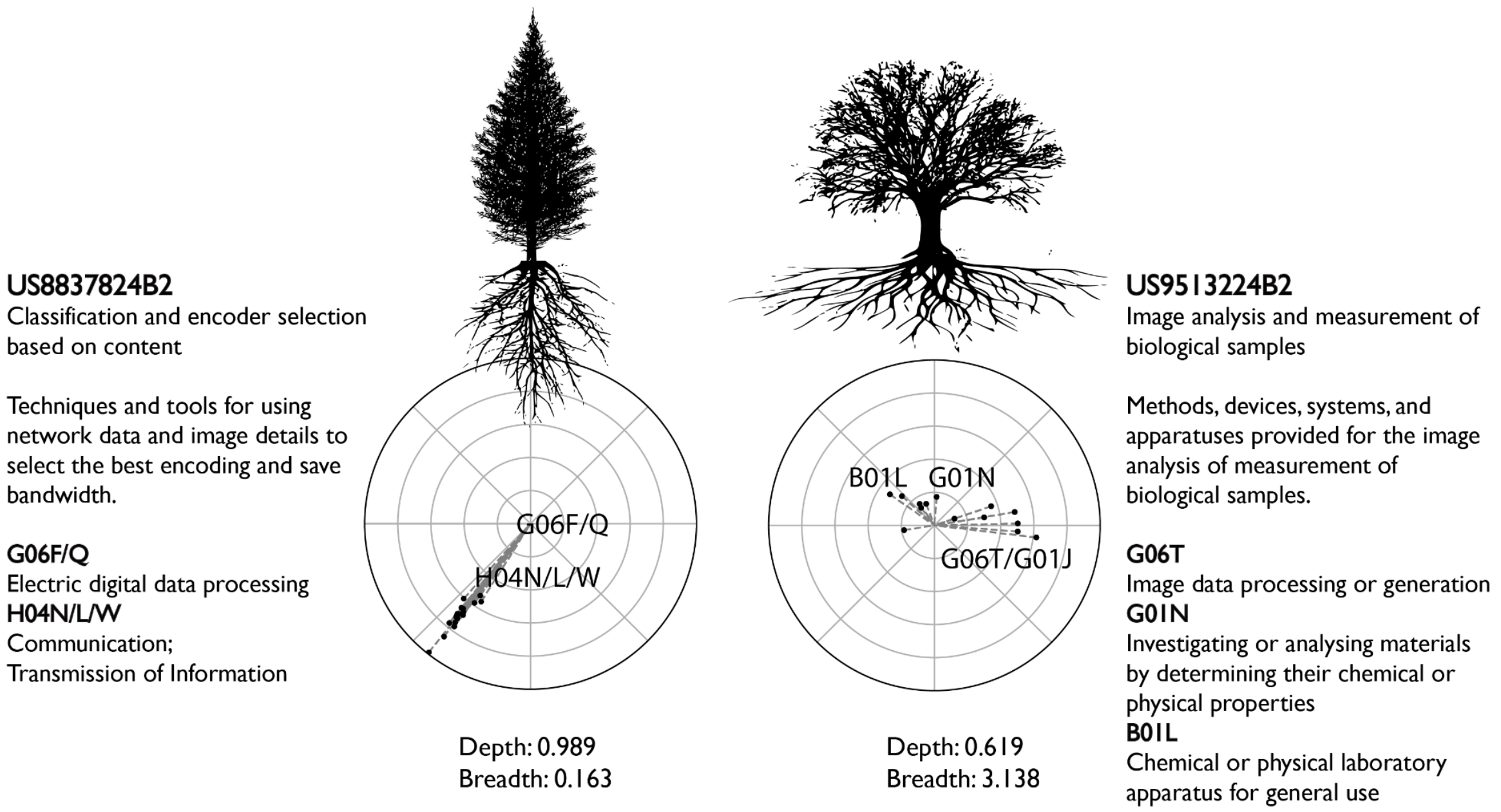

Figure D1. Examples of Deep and Wide Recombinations

*D4. Details of Δθ in the Mechanism Analysis*

In the analysis for mechanism, we use $\Delta\theta$ to measure the diffusion of knowledge on the Poincaré disk, after a focal patent has been published. We use an example (Figure D2) to explain this more intuitively.

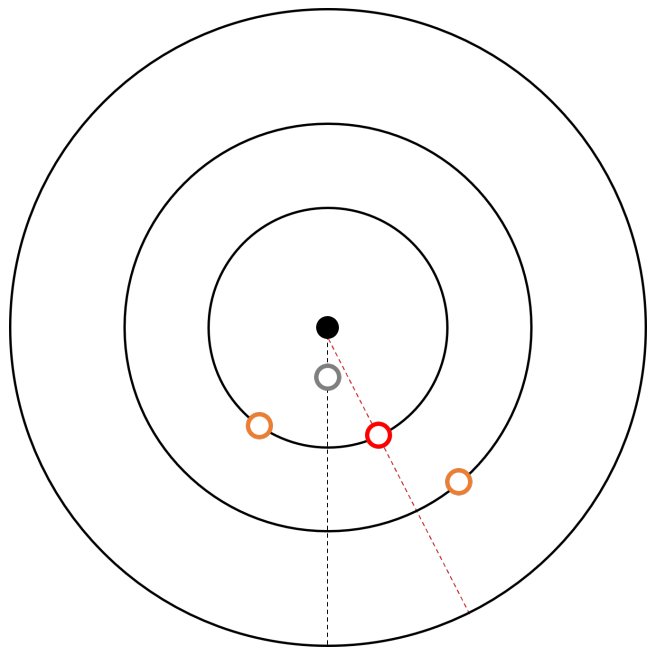

Figure D2. Conceptual Figure for Knowledge Diffusion on the Poincaré Disk

Assume we have constructed a Poincaré embedding space. Upon the publication of a patent, its position on the disk is denoted by a hollow, gray-colored point. Within three years following its publication, a subsequent work citing the focal patent emerges, indicated by a red point. Between three and six years after the initial publication, two additional follow-up works citing the gray patent appear; and we color them with orange.

First, we compare the gray and red points. Their distance can be measured by $\Delta r$ on the vertical dimension and $|\Delta\theta|$ on the horizontal dimension. The horizontal difference measure, $|\Delta\theta|$, reflects the influence exerted by the technological components on other domains. We use standard deviation of $|\Delta\theta|$ as a proxy for dispersion in the mechanism analysis; and at the patent level, we calculate the average values of $|\Delta\theta|$ for all follow-up works, which serve as the dependent variables for Table E8 of the main text. Similarly, we use average values of $\Delta r$ for all follow-up works as dependent variables for Table E7. It is important to note that $\Delta r$ can be either positive (indicating downstream movement) or negative (indicating upstream movement), whereas $\Delta\theta$ does not have a direction, so we use its absolute value. Consequently, the distribution of follow-up works along the $\Delta r$ dimension follows a normal distribution, while the distribution of follow-up works along the $|\Delta\theta|$ dimension is highly skewed.

## Appendix E: Details of Models in Mechanism Analysis

We conducted a series of analyses to reveal the temporal changes in effect paths based on generalized structural equation modeling. The results are reported in Table E1-E5. Two additional analyses, (1) the effects of depth and breadth on patent impact at the inventor, organization, and domain levels, and (2) the impact of depth and breadth on features of follow-up patents, are reported in Tables E6 to E8. Finally, we estimate a generalized negative binomial regression (mean-dispersion model). The mean model and dispersion models have exactly the same set of predictors. We report the mean model in the main text, and report the dispersion model in Table E9.

Table E1. Mediation Analysis for Impact Accumulation
(0-5 Years, with Patent Sample before 2002) based on GSEM

| | Cognitive Lock-in | Network Lock-in | Cognitive Bridging | Structural Bridging | 0-5 Year Forward Citations |
|---|---|---|---|---|---|
| Cognitive Lock-in | | | | | 0.133 |
| | | | | | [0.000] |
| | | | | | (0.001) |
| Cognitive Bridging | | | | | 0.129 |
| | | | | | [0.000] |
| | | | | | (0.004) |
| Structural Bridging | | | | | -0.056 |
| | | | | | [0.000] |
| | | | | | (0.011) |
| Network Lock-in | | | | | 0.078 |
| | | | | | [0.000] |
| | | | | | (0.001) |
| Depth | 1.592 | 0.086 | -0.305 | 0.013 | -0.086 |
| | [0.000] | [0.000] | [0.000] | [0.000] | [0.000] |
| | (0.024) | (0.007) | (0.003) | (0.001) | (0.008) |
| Breadth | -0.438 | -0.041 | 0.155 | 0.023 | 0.030 |
| | [0.000] | [0.000] | [0.000] | [0.000] | [0.000] |
| | (0.003) | (0.001) | (0.000) | (0.000) | (0.001) |
| Horizontal Location (mean theta) | 0.030 | 0.018 | -0.005 | 0.001 | 0.038 |
| | [0.000] | [0.000] | [0.000] | [0.000] | [0.000] |
| | (0.002) | (0.001) | (0.000) | (0.000) | (0.001) |
| Ln (Component Familiarity) | 0.111 | 0.035 | 0.013 | 0.002 | -0.003 |
| | [0.000] | [0.000] | [0.000] | [0.000] | [0.000] |
| | (0.002) | (0.001) | (0.000) | (0.000) | (0.001) |
| Ln (Combination Familiarity) | -0.059 | 0.006 | -0.035 | -0.008 | 0.023 |
| | [0.000] | [0.000] | [0.000] | [0.000] | [0.000] |
| | (0.001) | (0.000) | (0.000) | (0.000) | (0.000) |
| Number of Tech Classes | 0.032 | -0.004 | -0.007 | -0.005 | 0.016 |
| | [0.000] | [0.000] | [0.000] | [0.000] | [0.000] |
| | (0.001) | (0.000) | (0.000) | (0.000) | (0.000) |
| Number of Prior Art Classes | 0.002 | 0.000 | -0.000 | -0.000 | 0.001 |
| | [0.000] | [0.000] | [0.000] | [0.000] | [0.000] |
| | (0.000) | (0.000) | (0.000) | (0.000) | (0.000) |
| Assigned to Organizations | 0.178 | 0.410 | -0.016 | 0.002 | -0.034 |
| | [0.000] | [0.000] | [0.000] | [0.000] | [0.000] |
| | (0.009) | (0.007) | (0.001) | (0.000) | (0.006) |
| Invented by Team | 0.053 | 0.094 | -0.007 | -0.000 | 0.002 |
| | [0.000] | [0.000] | [0.000] | [0.057] | [0.289] |
| | (0.006) | (0.002) | (0.001) | (0.000) | (0.002) |
| Ln (Inventor Experience) | -0.001 | -0.028 | -0.002 | -0.000 | -0.015 |
| | [0.685] | [0.000] | [0.000] | [0.566] | [0.000] |
| | (0.002) | (0.001) | (0.000) | (0.000) | (0.001) |
| Non Patent References | 0.018 | -0.005 | -0.003 | 0.000 | 0.006 |
| | [0.000] | [0.000] | [0.000] | [0.000] | [0.000] |
| | (0.001) | (0.000) | (0.000) | (0.000) | (0.000) |
| Number of Claims | 0.007 | -0.003 | -0.000 | 0.000 | 0.010 |
| | [0.000] | [0.000] | [0.000] | [0.180] | [0.000] |
| | (0.000) | (0.000) | (0.000) | (0.000) | (0.000) |
| Family Size | -0.065 | 0.033 | -0.003 | 0.000 | -0.079 |
| | [0.000] | [0.000] | [0.000] | [0.301] | [0.000] |
| | (0.004) | (0.001) | (0.001) | (0.000) | (0.002) |
| Year Fixed Effects | YES | YES | YES | YES | YES |
| Constant | 1.405 | 1.049 | 0.618 | 0.153 | 0.859 |
| | [0.000] | [0.000] | [0.000] | [0.000] | [0.000] |
| | (0.131) | (0.013) | (0.025) | (0.008) | (0.014) |
| Vars and Covs | Controlled for Covariates at the Pairwise Level for the Four Mediators | | | | |

*Note: P-values are between square brackets. Robust standard errors are in parentheses. All tests are two-tailed.*

Table E2. Mediation Analysis for Impact Accumulation

(5-10 Years, with Patent Sample before 2002) based on GSEM

| | Cognitive Lock-in | Network Lock-in | Cognitive Bridging | Structural Bridging | 5-10 Year Forward Citations |
|---|---|---|---|---|---|
| Cognitive Lock-in | | | | | 0.275 |
| | | | | | [0.000] |
| | | | | | (0.001) |
| Cognitive Bridging | | | | | 0.297 |
| | | | | | [0.000] |
| | | | | | (0.004) |
| Structural Bridging | | | | | 0.066 |
| | | | | | [0.000] |
| | | | | | (0.011) |
| Network Lock-in | | | | | 0.041 |
| | | | | | [0.000] |
| | | | | | (0.002) |
| Depth | 1.441 | 0.075 | -0.340 | 0.006 | -0.232 |
| | [0.000] | [0.000] | [0.000] | [0.000] | [0.000] |
| | (0.021) | (0.005) | (0.003) | (0.001) | (0.008) |
| Breadth | -0.406 | -0.026 | 0.162 | 0.024 | 0.068 |
| | [0.000] | [0.000] | [0.000] | [0.000] | [0.000] |
| | (0.003) | (0.001) | (0.000) | (0.000) | (0.001) |
| Horizontal Location (mean theta) | 0.039 | 0.004 | -0.006 | 0.000 | 0.031 |
| | [0.000] | [0.000] | [0.000] | [0.000] | [0.000] |
| | (0.002) | (0.001) | (0.000) | (0.000) | (0.001) |
| Ln (Component Familiarity) | 0.112 | 0.014 | 0.009 | 0.000 | -0.010 |
| | [0.000] | [0.000] | [0.000] | [0.000] | [0.000] |
| | (0.002) | (0.001) | (0.000) | (0.000) | (0.001) |
| Ln (Combination Familiarity) | -0.047 | -0.002 | -0.034 | -0.008 | 0.026 |
| | [0.000] | [0.000] | [0.000] | [0.000] | [0.000] |
| | (0.001) | (0.000) | (0.000) | (0.000) | (0.000) |
| Number of Tech Classes | 0.026 | -0.003 | -0.007 | -0.005 | 0.019 |
| | [0.000] | [0.000] | [0.000] | [0.000] | [0.000] |
| | (0.001) | (0.000) | (0.000) | (0.000) | (0.000) |
| Number of Prior Art Classes | 0.002 | 0.000 | -0.000 | -0.000 | 0.002 |
| | [0.000] | [0.000] | [0.000] | [0.000] | [0.000] |
| | (0.000) | (0.000) | (0.000) | (0.000) | (0.000) |
| Assigned to Organizations | 0.144 | 0.255 | -0.013 | 0.003 | -0.078 |
| | [0.000] | [0.000] | [0.000] | [0.000] | [0.000] |
| | (0.008) | (0.006) | (0.001) | (0.000) | (0.006) |
| Invented by Team | 0.057 | 0.075 | -0.006 | 0.000 | 0.003 |
| | [0.000] | [0.000] | [0.000] | [0.627] | [0.202] |
| | (0.005) | (0.002) | (0.001) | (0.000) | (0.002) |
| Ln (Inventor Experience) | -0.009 | -0.040 | -0.002 | 0.001 | -0.021 |
| | [0.000] | [0.000] | [0.000] | [0.000] | [0.000] |
| | (0.002) | (0.001) | (0.000) | (0.000) | (0.001) |
| Non Patent References | 0.022 | -0.004 | -0.003 | 0.000 | 0.007 |
| | [0.000] | [0.000] | [0.000] | [0.000] | [0.000] |
| | (0.001) | (0.000) | (0.000) | (0.000) | (0.000) |
| Number of Claims | 0.008 | -0.002 | -0.000 | 0.000 | 0.011 |
| | [0.000] | [0.000] | [0.000] | [0.041] | [0.000] |
| | (0.000) | (0.000) | (0.000) | (0.000) | (0.000) |
| Family Size | -0.091 | -0.011 | -0.004 | -0.001 | -0.132 |
| | [0.000] | [0.000] | [0.000] | [0.005] | [0.000] |
| | (0.004) | (0.001) | (0.001) | (0.000) | (0.002) |
| Year Fixed Effects | YES | YES | YES | YES | YES |
| Constant | 1.135 | 1.256 | 0.692 | 0.178 | 0.519 |
| | [0.000] | [0.000] | [0.000] | [0.000] | [0.000] |
| | (0.116) | (0.127) | (0.023) | (0.008) | (0.102) |
| Vars and Covs | Controlled for Covariates at the Pairwise Level for the Four Mediators | | | | |

*Note: P-values are between square brackets. Robust standard errors are in parentheses. All tests are two-tailed.*

Table E3. Mediation Analysis for Impact Accumulation

(10-15 Years, with Patent Sample before 2002) based on GSEM

| | Cognitive Lock-in | Network Lock-in | Cognitive Bridging | Structural Bridging | 10-15 Year Forward Citations |
|---|---|---|---|---|---|
| Cognitive Lock-in | | | | | 0.335 |
| | | | | | [0.000] |
| | | | | | (0.002) |
| Cognitive Bridging | | | | | 0.329 |
| | | | | | [0.000] |
| | | | | | (0.005) |
| Structural Bridging | | | | | 0.118 |
| | | | | | [0.000] |
| | | | | | (0.012) |
| Network Lock-in | | | | | 0.010 |
| | | | | | [0.000] |
| | | | | | (0.003) |
| Depth | 1.294 | 0.069 | -0.354 | 0.004 | -0.371 |
| | [0.000] | [0.000] | [0.000] | [0.001] | [0.000] |
| | (0.026) | (0.005) | (0.003) | (0.001) | (0.010) |
| Breadth | -0.383 | -0.013 | 0.167 | 0.025 | 0.085 |
| | [0.000] | [0.000] | [0.000] | [0.000] | [0.000] |
| | (0.003) | (0.001) | (0.000) | (0.000) | (0.002) |
| Horizontal Location (mean theta) | 0.040 | 0.001 | -0.007 | 0.000 | 0.020 |
| | [0.000] | [0.004] | [0.000] | [0.388] | [0.000] |
| | (0.002) | (0.001) | (0.000) | (0.000) | (0.001) |
| Ln (Component Familiarity) | 0.114 | 0.002 | 0.007 | -0.001 | -0.013 |
| | [0.000] | [0.000] | [0.000] | [0.000] | [0.000] |
| | (0.002) | (0.000) | (0.000) | (0.000) | (0.001) |
| Ln (Combination Familiarity) | -0.046 | -0.005 | -0.033 | -0.007 | 0.026 |
| | [0.000] | [0.000] | [0.000] | [0.000] | [0.000] |
| | (0.001) | (0.000) | (0.000) | (0.000) | (0.001) |
| Number of Tech Classes | 0.024 | -0.002 | -0.007 | -0.004 | 0.019 |
| | [0.000] | [0.000] | [0.000] | [0.000] | [0.000] |
| | (0.001) | (0.000) | (0.000) | (0.000) | (0.000) |
| Number of Prior Art Classes | 0.002 | 0.000 | -0.000 | -0.000 | 0.002 |
| | [0.000] | [0.910] | [0.000] | [0.000] | [0.000] |
| | (0.000) | (0.000) | (0.000) | (0.000) | (0.000) |
| Assigned to Organizations | 0.097 | 0.132 | -0.009 | 0.004 | -0.127 |
| | [0.000] | [0.000] | [0.000] | [0.000] | [0.000] |
| | (0.009) | (0.005) | (0.001) | (0.000) | (0.006) |
| Invented by Team | 0.057 | 0.057 | -0.007 | -0.000 | -0.004 |
| | [0.000] | [0.000] | [0.000] | [0.622] | [0.164] |
| | (0.006) | (0.002) | (0.001) | (0.000) | (0.003) |
| Ln (Inventor Experience) | -0.009 | -0.036 | -0.001 | 0.001 | -0.025 |
| | [0.000] | [0.000] | [0.009] | [0.000] | [0.000] |
| | (0.002) | (0.000) | (0.000) | (0.000) | (0.001) |
| Non Patent References | 0.024 | -0.003 | -0.003 | 0.000 | 0.005 |
| | [0.000] | [0.000] | [0.000] | [0.000] | [0.000] |
| | (0.001) | (0.000) | (0.000) | (0.000) | (0.000) |
| Number of Claims | 0.007 | -0.002 | -0.000 | 0.000 | 0.011 |
| | [0.000] | [0.000] | [0.000] | [0.008] | [0.000] |
| | (0.000) | (0.000) | (0.000) | (0.000) | (0.000) |
| Family Size | -0.112 | -0.018 | -0.003 | -0.001 | -0.151 |
| | [0.000] | [0.000] | [0.000] | [0.001] | [0.000] |
| | (0.005) | (0.001) | (0.001) | (0.000) | (0.002) |
| Year Fixed Effects | YES | YES | YES | YES | YES |
| Constant | 1.763 | 2.013 | 0.742 | 0.196 | 0.579 |
| | [0.000] | [0.000] | [0.000] | [0.000] | [0.000] |
| | (0.195) | (0.123) | (0.026) | (0.008) | (0.124) |
| Vars and Covs | Controlled for Covariates at the Pairwise Level for the Four Mediators | | | | |

*Note: P-values are between square brackets. Robust standard errors are in parentheses. All tests are two-tailed.*

Table E4. Mediation Analysis for Impact Accumulation

(15-20 Years, with Patent Sample before 2002) based on GSEM

| | Cognitive Lock-in | Network Lock-in | Cognitive Bridging | Structural Bridging | 15-20 Year Forward Citations |
|---|---|---|---|---|---|
| Cognitive Lock-in | | | | | 0.376 |
| | | | | | [0.000] |
| | | | | | (0.002) |
| Cognitive Bridging | | | | | 0.330 |
| | | | | | [0.000] |
| | | | | | (0.005) |
| Structural Bridging | | | | | 0.217 |
| | | | | | [0.000] |
| | | | | | (0.014) |
| Network Lock-in | | | | | -0.026 |
| | | | | | [0.000] |
| | | | | | (0.004) |
| Depth | 1.027 | 0.059 | -0.366 | 0.001 | -0.537 |
| | [0.000] | [0.000] | [0.000] | [0.423] | [0.000] |
| | (0.031) | (0.005) | (0.003) | (0.001) | (0.012) |
| Breadth | -0.326 | -0.005 | 0.169 | 0.024 | 0.106 |
| | [0.000] | [0.000] | [0.000] | [0.000] | [0.000] |
| | (0.004) | (0.001) | (0.000) | (0.000) | (0.002) |
| Horizontal Location (mean theta) | 0.056 | -0.001 | -0.012 | -0.001 | 0.020 |
| | [0.000] | [0.110] | [0.000] | [0.000] | [0.000] |
| | (0.002) | (0.000) | (0.000) | (0.000) | (0.001) |
| Ln (Component Familiarity) | 0.106 | -0.004 | 0.005 | -0.001 | -0.016 |
| | [0.000] | [0.000] | [0.000] | [0.000] | [0.000] |
| | (0.002) | (0.000) | (0.000) | (0.000) | (0.001) |
| Ln (Combination Familiarity) | -0.044 | -0.005 | -0.033 | -0.006 | 0.027 |
| | [0.000] | [0.000] | [0.000] | [0.000] | [0.000] |
| | (0.001) | (0.000) | (0.000) | (0.000) | (0.001) |
| Number of Tech Classes | 0.021 | -0.001 | -0.008 | -0.004 | 0.019 |
| | [0.000] | [0.000] | [0.000] | [0.000] | [0.000] |
| | (0.001) | (0.000) | (0.000) | (0.000) | (0.000) |
| Number of Prior Art Classes | 0.002 | -0.000 | -0.000 | -0.000 | 0.002 |
| | [0.000] | [0.000] | [0.000] | [0.000] | [0.000] |
| | (0.000) | (0.000) | (0.000) | (0.000) | (0.000) |
| Assigned to Organizations | 0.081 | 0.061 | -0.008 | 0.003 | -0.146 |
| | [0.000] | [0.000] | [0.000] | [0.000] | [0.000] |
| | (0.010) | (0.004) | (0.001) | (0.000) | (0.007) |
| Invented by Team | 0.049 | 0.039 | -0.008 | -0.001 | -0.000 |
| | [0.000] | [0.000] | [0.000] | [0.029] | [0.991] |
| | (0.008) | (0.002) | (0.001) | (0.000) | (0.004) |
| Ln (Inventor Experience) | -0.003 | -0.032 | -0.000 | 0.001 | -0.024 |
| | [0.249] | [0.000] | [0.080] | [0.000] | [0.000] |
| | (0.002) | (0.000) | (0.000) | (0.000) | (0.001) |
| Non Patent References | 0.021 | -0.002 | -0.003 | 0.000 | 0.004 |
| | [0.000] | [0.000] | [0.000] | [0.000] | [0.000] |
| | (0.001) | (0.000) | (0.000) | (0.000) | (0.000) |
| Number of Claims | 0.006 | -0.001 | -0.000 | 0.000 | 0.009 |
| | [0.000] | [0.000] | [0.000] | [0.067] | [0.000] |
| | (0.000) | (0.000) | (0.000) | (0.000) | (0.000) |
| Family Size | -0.097 | -0.007 | -0.001 | 0.000 | -0.152 |
| | [0.000] | [0.000] | [0.275] | [0.674] | [0.000] |
| | (0.006) | (0.001) | (0.001) | (0.000) | (0.003) |
| Year Fixed Effects | YES | YES | YES | YES | YES |
| Constant | 2.424 | 2.224 | 0.792 | 0.206 | 0.562 |
| | [0.000] | [0.000] | [0.000] | [0.000] | [0.000] |
| | (0.367) | (0.094) | (0.028) | (0.010) | (0.112) |
| Vars and Covs | Controlled for Covariates at the Pairwise Level for the Four Mediators | | | | |

*Note: P-values are between square brackets. Robust standard errors are in parentheses. All tests are two-tailed.*

Table E5. Direct and Indirect Effects Across Four Pathways (Bootstrapped, 1000 Replications)

| | Direct Effect | Cognitive Lock-in | Network Lock-in | Cognitive Bridging | Structural Bridging | Full |
|---|---|---|---|---|---|---|
| **Depth** | | | | | | |
| All Patents (1976-2017)<br>DV: 0-5 Year Citations | -0.089***<br>[-0.103, -0.076] | 0.201***<br>[0.193, 0.209] | 0.003***<br>[0.002, 0.003] | -0.045***<br>[-0.047, -0.043] | 0.004***<br>[0.003, 0.005] | 0.073***<br>[0.058, 0.088] |
| Early Patents (1976-2002)<br>DV: 0-5 Year Citations | -0.086***<br>[-0.103, -0.070] | 0.212***<br>[0.203, 0.221] | 0.007***<br>[0.006, 0.008] | -0.039***<br>[-0.042, -0.037] | -0.001**<br>[-0.001,0.000] | 0.093***<br>[0.076, 0.110] |
| Early Patents (1976-2002)<br>DV: 5-10 Year Citations | -0.232***<br>[-0.248, -0.215] | 0.396***<br>[0.380, 0.411] | 0.003***<br>[0.003, 0.004] | -0.101***<br>[-0.104, -0.098] | 0.000+<br>[-0.000, 0.001] | 0.067***<br>[0.045, 0.088] |
| Early Patents (1976-2002)<br>DV: 10-15 Year Citations | -0.371***<br>[-0.391, -0.350] | 0.433***<br>[0.411, 0.456] | 0.001***<br>[-0.000, 0.001] | -0.116***<br>[-0.120, -0.113] | 0.000+<br>[-0.000, 0.001] | -0.053***<br>[-0.081, -0.024] |
| Early Patents (1976-2002)<br>DV: 15-20 Year Citations | -0.537***<br>[-0.562, -0.513] | 0.386***<br>[0.354, 0.410] | -0.002***<br>[-0.002, -0.001] | -0.121***<br>[-0.125, -0.116] | 0.000<br>[-0.001, 0.001] | -0.274***<br>[-0.314, -0.233] |
| **Breadth** | | | | | | |
| All Patents (1976-2017)<br>DV: 0-5 Year Citations | 0.047***<br>[0.045, 0.049] | -0.078***<br>[-0.080, -0.077] | -0.003***<br>[-0.003, -0.003] | 0.031***<br>[0.030, 0.032] | 0.005***<br>[0.005, 0.006] | 0.002<br>[0.000, 0.004] |
| Early Patents (1976-2002)<br>DV: 0-5 Year Citations | 0.030***<br>[0.028, 0.033] | -0.058***<br>[-0.060, -0.057] | -0.003***<br>[-0.003, -0.003] | 0.020***<br>[0.019, 0.021] | -0.001***<br>[-0.002, -0.001] | -0.013***<br>[-0.015, -0.010] |
| Early Patents (1976-2002)<br>DV: 5-10 Year Citations | 0.068***<br>[0.065, 0.071] | -0.112***<br>[-0.114, -0.109] | -0.001***<br>[-0.001, -0.001] | 0.048***<br>[0.047, 0.050] | 0.002***<br>[0.001, 0.002] | 0.005***<br>[0.002, 0.008] |
| Early Patents (1976-2002)<br>DV: 10-15 Year Citations | 0.085***<br>[0.082, 0.089] | -0.128***<br>[-0.131, -0.125] | -0.000***<br>[-0.000, -0.000] | 0.055***<br>[0.053, 0.056] | 0.003***<br>[0.002, 0.004] | 0.015***<br>[0.011, 0.019] |
| Early Patents (1976-2002)<br>DV: 15-20 Year Citations | 0.106***<br>[0.102, 0.109] | -0.122***<br>[-0.127, -0.118] | 0.000***<br>[0.000, 0.000] | 0.056***<br>[0.054, 0.057] | 0.005***<br>[0.005, 0.006] | 0.044***<br>[0.039, 0.050] |

*Note: P-values are between square brackets. Robust standard errors are in parentheses. All tests are two-tailed.*

Table E6. Effects of Depth and Breadth on Inventor, Firm and Technological Domain Level

| | **Patent Level Effects (Negative Binomial-baseline)** | **Inventor Level with Inventor-Year (xtNBREG)** | **Organization Level with Organization-Year (xtNBREG)** | **Domain Level with Domain-Year (xtNBREG)** |
|---|---|---|---|---|
| | **Model 1** | **Model 2** | **Model 3** | **Model 4** |
| Depth (Maximum) | 0.200 | 0.433 | 0.494 | 0.992 |
| | [0.000] | [0.000] | [0.000] | [0.000] |
| | (0.005) | (0.004) | (0.014) | (0.041) |
| Breadth (Maximum) | -0.035 | 0.033 | 0.061 | 0.059 |
| | [0.000] | [0.000] | [0.000] | [0.000] |
| | (0.001) | (0.001) | (0.002) | (0.002) |
| Horizontal Location (mean theta) | 0.023 | 0.005 | 0.001 | -0.008 |
| (average) | [0.000] | [0.000] | [0.493] | [0.000] |
| | (0.000) | (0.000) | (0.001) | (0.001) |
| Ln (Component Familiarity) | 0.031 | -0.031 | -0.042 | -0.017 |
| (average) | [0.000] | [0.000] | [0.000] | [0.000] |
| | (0.000) | (0.000) | (0.001) | (0.001) |
| Ln (Combination Familiarity) | 0.028 | 0.011 | 0.009 | -0.025 |
| (average) | [0.000] | [0.000] | [0.000] | [0.000] |
| | (0.000) | (0.000) | (0.001) | (0.003) |
| Number of Tech Classes | 0.027 | 0.021 | 0.021 | 0.020 |
| (average) | [0.000] | [0.000] | [0.000] | [0.000] |
| | (0.000) | (0.000) | (0.000) | (0.001) |
| Number of Prior Art Classes | 0.004 | 0.001 | 0.001 | 0.001 |
| (average) | [0.000] | [0.000] | [0.000] | [0.000] |
| | (0.000) | (0.000) | (0.000) | (0.000) |
| Assigned to Organizations (average) | 0.061 | 0.023 | 0.000 | 0.170 |
| | [0.000] | [0.000] | [.] | [0.000] |
| | (0.002) | (0.003) | (.) | (0.008) |
| Invented by Team | 0.072 | 0.083 | 0.059 | 0.086 |
| (average) | [0.000] | [0.000] | [0.000] | [0.000] |
| | (0.001) | (0.002) | (0.004) | (0.006) |
| Ln (Inventor Experience) | 0.013 | -0.047 | -0.046 | -0.023 |
| (average) | [0.000] | [0.000] | [0.000] | [0.000] |
| | (0.000) | (0.000) | (0.001) | (0.002) |
| Non Patent References | 0.010 | -0.000 | -0.002 | 0.000 |
| (average) | [0.000] | [0.001] | [0.000] | [0.969] |
| | (0.000) | (0.000) | (0.000) | (0.000) |
| Number of Claims | 0.019 | 0.014 | 0.011 | 0.021 |
| (average) | [0.000] | [0.000] | [0.000] | [0.000] |
| | (0.000) | (0.000) | (0.000) | (0.000) |
| Family Size | -0.205 | 0.115 | 0.066 | -0.122 |
| (average) | [0.000] | [0.000] | [0.000] | [0.000] |
| | (0.001) | (0.001) | (0.005) | (0.005) |
| Log(Yearly Patent Count) | | 0.230 | 0.153 | 0.090 |
| | | [0.000] | [0.000] | [0.000] |
| | | (0.001) | (0.002) | (0.002) |
| Inventor/Organization/Domain Fixed Effect | | YES | YES | YES |
| Year Fixed Effect | | YES | YES | YES |
| Career Year Fixed Effect | | YES | YES | YES |
| Constant | 0.076 | -0.250 | 0.255 | -0.021 |
| | [0.217] | [0.000] | [0.000] | [0.626] |
| | (0.062) | (0.006) | (0.018) | (0.044) |
| N | 4992915 | 6068972 | 681500 | 334648 |

Note: P-values are between square brackets. Robust standard errors are in parentheses. All tests are two-tailed.

Table E7. Effects of Focal Patent Structure on Depth of Follow-Up Technologies

| | Model 1<br><br>Predicting Follow-up Depth | Model 2<br><br>Predicting Follow-up Depth within Same Organization | Model 3<br><br>Predicting Follow-up Depth | Model 4<br><br>Predicting Follow-up Depth within Same Organization |
|---|---|---|---|---|
| Depth | 0.414 | 0.515 | 0.425 | 0.522 |
| | [0.000] | [0.000] | [0.000] | [0.000] |
| | (0.000) | (0.001) | (0.001) | (0.001) |
| Breadth | -0.018 | -0.017 | -0.012 | -0.013 |
| | [0.000] | [0.000] | [0.000] | [0.000] |
| | (0.000) | (0.000) | (0.000) | (0.001) |
| Interaction: Depth*Breadth | | | -0.008 | -0.005 |
| | | | [0.000] | [0.000] |
| | | | (0.000) | (0.001) |
| Horizontal Location (mean theta) | -0.002 | -0.002 | -0.002 | -0.002 |
| | [0.000] | [0.000] | [0.000] | [0.000] |
| | (0.000) | (0.000) | (0.000) | (0.000) |
| Ln (Component Familiarity) | -0.012 | -0.012 | -0.012 | -0.012 |
| | [0.000] | [0.000] | [0.000] | [0.000] |
| | (0.000) | (0.000) | (0.000) | (0.000) |
| Ln (Combination Familiarity) | -0.004 | -0.003 | -0.004 | -0.003 |
| | [0.000] | [0.000] | [0.000] | [0.000] |
| | (0.000) | (0.000) | (0.000) | (0.000) |
| Number of Tech Classes | 0.000 | 0.000 | 0.000 | 0.000 |
| | [0.000] | [0.095] | [0.000] | [0.146] |
| | (0.000) | (0.000) | (0.000) | (0.000) |
| Number of Prior Art Classes | -0.000 | -0.000 | -0.000 | -0.000 |
| | [0.000] | [0.000] | [0.000] | [0.000] |
| | (0.000) | (0.000) | (0.000) | (0.000) |
| Assigned to Organizations | -0.004 | 0.000 | -0.004 | 0.000 |
| | [0.000] | [.] | [0.000] | [.] |
| | (0.000) | (.) | (0.000) | (.) |
| Invented by Team | -0.002 | -0.002 | -0.002 | -0.002 |
| | [0.000] | [0.000] | [0.000] | [0.000] |
| | (0.000) | (0.000) | (0.000) | (0.000) |
| Ln (Inventor Experience) | -0.000 | 0.001 | -0.000 | 0.001 |
| | [0.000] | [0.000] | [0.000] | [0.000] |
| | (0.000) | (0.000) | (0.000) | (0.000) |
| Non Patent References | 0.001 | 0.001 | 0.001 | 0.001 |
| | [0.000] | [0.000] | [0.000] | [0.000] |
| | (0.000) | (0.000) | (0.000) | (0.000) |
| Number of Claims | 0.000 | 0.000 | 0.000 | 0.000 |
| | [0.000] | [0.000] | [0.000] | [0.000] |
| | (0.000) | (0.000) | (0.000) | (0.000) |
| Family Size | -0.005 | -0.008 | -0.005 | -0.008 |
| | [0.000] | [0.000] | [0.000] | [0.000] |
| | (0.000) | (0.000) | (0.000) | (0.000) |
| Year Fixed Effects | YES | YES | YES | YES |
| Constant | 0.641 | 0.540 | 0.631 | 0.534 |
| | [0.000] | [0.000] | [0.000] | [0.000] |
| | (0.005) | (0.009) | (0.005) | (0.009) |
| Adjusted R-squared | 0.489 | 0.524 | 0.489 | 0.524 |
| N | 2925660 | 929483 | 2925660 | 929483 |

*Note: P-values are between square brackets. Robust standard errors are in parentheses. All tests are two-tailed.*

Table E8. Effects of Focal Patent Structure on Breadth of Follow-Up Technologies

| | Model 1 | Model 2 | Model 3 | Model 4 |
|---|---|---|---|---|
| | Predicting Follow-up Breadth | Predicting Follow-up Breadth within Same Organization | Predicting Follow-up Breadth | Predicting Follow-up Breadth within Same Organization |
| Depth | -1.271 | -1.020 | -1.858 | -1.615 |
| | [0.000] | [0.000] | [0.000] | [0.000] |
| | (0.003) | (0.006) | (0.005) | (0.009) |
| Breadth | 0.505 | 0.595 | 0.168 | 0.254 |
| | [0.000] | [0.000] | [0.000] | [0.000] |
| | (0.000) | (0.001) | (0.002) | (0.004) |
| Interaction: Depth*Breadth | | | 0.404 | 0.411 |
| | | | [0.000] | [0.000] |
| | | | (0.003) | (0.005) |
| Horizontal Location (mean theta) | 0.006 | 0.015 | 0.007 | 0.015 |
| | [0.000] | [0.000] | [0.000] | [0.000] |
| | (0.000) | (0.000) | (0.000) | (0.000) |
| Ln (Component Familiarity) | -0.024 | -0.017 | -0.024 | -0.017 |
| | [0.000] | [0.000] | [0.000] | [0.000] |
| | (0.000) | (0.000) | (0.000) | (0.000) |
| Ln (Combination Familiarity) | -0.023 | -0.017 | -0.023 | -0.017 |
| | [0.000] | [0.000] | [0.000] | [0.000] |
| | (0.000) | (0.000) | (0.000) | (0.000) |
| Number of Tech Classes | 0.002 | 0.000 | 0.003 | 0.001 |
| | [0.000] | [0.163] | [0.000] | [0.000] |
| | (0.000) | (0.000) | (0.000) | (0.000) |
| Number of Prior Art Classes | 0.000 | 0.001 | 0.000 | 0.001 |
| | [0.000] | [0.000] | [0.000] | [0.000] |
| | (0.000) | (0.000) | (0.000) | (0.000) |
| Assigned to Organizations | -0.108 | 0.000 | -0.103 | 0.000 |
| | [0.000] | [.] | [0.000] | [.] |
| | (0.001) | (.) | (0.001) | (.) |
| Invented by Team | -0.006 | -0.009 | -0.005 | -0.007 |
| | [0.000] | [0.000] | [0.000] | [0.000] |
| | (0.001) | (0.002) | (0.001) | (0.002) |
| Ln (Inventor Experience) | 0.002 | 0.009 | 0.003 | 0.010 |
| | [0.000] | [0.000] | [0.000] | [0.000] |
| | (0.000) | (0.000) | (0.000) | (0.000) |
| Non Patent References | -0.005 | -0.004 | -0.005 | -0.004 |
| | [0.000] | [0.000] | [0.000] | [0.000] |
| | (0.000) | (0.000) | (0.000) | (0.000) |
| Number of Claims | -0.001 | 0.001 | -0.001 | 0.001 |
| | [0.000] | [0.000] | [0.000] | [0.000] |
| | (0.000) | (0.000) | (0.000) | (0.000) |
| Family Size | -0.012 | -0.037 | -0.011 | -0.037 |
| | [0.000] | [0.000] | [0.000] | [0.000] |
| | (0.001) | (0.001) | (0.001) | (0.001) |
| Year Fixed Effects | YES | YES | YES | YES |
| Constant | 2.030 | 1.524 | 2.547 | 2.052 |
| | [0.000] | [0.000] | [0.000] | [0.000] |
| | (0.038) | (0.066) | (0.038) | (0.066) |
| Adjusted R-squared | 0.425 | 0.487 | 0.429 | 0.491 |
| N | 2925660 | 929483 | 2925660 | 929483 |

*Note: P-values are between square brackets. Robust standard errors are in parentheses. All tests are two-tailed.*

Table E9. Dispersion Model for Generalized Negative Binomial Regressions
(Following Table 3 in the Main Text)

| ***Mean Model*** | ***With Same Independent Variables, Reported in Table 3*** | | | | |
|---|---|---|---|---|---|
| ***Dispersion Model*** | Following Model 1 | Following Model 2 | Following Model 3 | Following Model 4 | Following Model 5 |
| Depth | -0.117 | 0.111 | 0.086 | -0.008 | -0.037 |
| | [0.000] | [0.000] | [0.000] | [0.364] | [0.000] |
| | (0.006) | (0.011) | (0.010) | (0.009) | (0.009) |
| Breadth | -0.031 | -0.052 | -0.049 | -0.053 | -0.055 |
| | [0.000] | [0.000] | [0.000] | [0.000] | [0.000] |
| | (0.001) | (0.002) | (0.001) | (0.001) | (0.001) |
| Horizontal Location | 0.021 | 0.007 | 0.014 | -0.000 | 0.014 |
| (mean theta) | [0.000] | [0.000] | [0.000] | [0.946] | [0.000] |
| | (0.001) | (0.001) | (0.001) | (0.001) | (0.001) |
| Ln (Component | 0.036 | 0.016 | 0.019 | 0.037 | 0.055 |
| Familiarity) | [0.000] | [0.000] | [0.000] | [0.000] | [0.000] |
| | (0.001) | (0.001) | (0.001) | (0.001) | (0.001) |
| Ln (Combination | -0.011 | -0.022 | -0.014 | -0.003 | 0.009 |
| Familiarity) | [0.000] | [0.000] | [0.000] | [0.000] | [0.000] |
| | (0.000) | (0.001) | (0.001) | (0.000) | (0.000) |
| Number of Tech Classes | -0.000 | -0.001 | -0.001 | -0.003 | -0.005 |
| | [0.844] | [0.252] | [0.016] | [0.000] | [0.000] |
| | (0.000) | (0.000) | (0.000) | (0.000) | (0.000) |
| Number of Prior Art | -0.000 | -0.001 | -0.001 | -0.000 | -0.000 |
| Classes | [0.000] | [0.000] | [0.000] | [0.000] | [0.000] |
| | (0.000) | (0.000) | (0.000) | (0.000) | (0.000) |
| Assigned to Organizations | 0.085 | 0.054 | 0.109 | 0.131 | 0.174 |
| | [0.000] | [0.000] | [0.000] | [0.000] | [0.000] |
| | (0.003) | (0.005) | (0.004) | (0.004) | (0.004) |
| Invented by Team | 0.022 | 0.033 | 0.030 | 0.018 | 0.021 |
| | [0.000] | [0.000] | [0.000] | [0.000] | [0.000] |
| | (0.002) | (0.003) | (0.003) | (0.003) | (0.003) |
| Ln (Inventor Experience) | 0.016 | 0.012 | 0.030 | 0.038 | 0.050 |
| | [0.000] | [0.000] | [0.000] | [0.000] | [0.000] |
| | (0.001) | (0.001) | (0.001) | (0.001) | (0.001) |
| Non Patent References | 0.008 | 0.019 | 0.011 | 0.006 | 0.006 |
| | [0.000] | [0.000] | [0.000] | [0.000] | [0.000] |
| | (0.000) | (0.000) | (0.000) | (0.000) | (0.000) |
| Number of Claims | -0.008 | -0.005 | -0.005 | -0.005 | -0.007 |
| | [0.000] | [0.000] | [0.000] | [0.000] | [0.000] |
| | (0.000) | (0.000) | (0.000) | (0.000) | (0.000) |
| Family Size | -0.021 | -0.023 | -0.003 | 0.025 | 0.071 |
| | [0.000] | [0.000] | [0.096] | [0.000] | [0.000] |
| | (0.002) | (0.002) | (0.002) | (0.002) | (0.002) |
| Year Fixed Effects | YES | YES | YES | YES | YES |
| Constant | 0.017 | -0.003 | 0.006 | 0.081 | -0.152 |
| | [0.850] | [0.976] | [0.945] | [0.386] | [0.136] |
| | (0.092) | (0.092) | (0.093) | (0.093) | (0.102) |

*Note: P-values are between square brackets. Robust standard errors are in parentheses. All tests are two-tailed.*

The majority of existing literature on technological evolution and recombination employs the United States Patent Classification ("USPC") system (Fleming and Sorenson, 2004; Petralia, 2020) or the International Patent Classification ("IPC") system (Corrocher et al., 2007; Fontana et al., 2008) as primary classification systems. In 2013, USPTO introduced the Cooperative Patent Classification ("CPC") system and gradually replaced the USPC. Based on the IPC, the system used by the European patent system, the CPC system boasts greater granularity than USPC and is favored by the USPTO for future use in its efforts to defend U.S. intellectual property in alignment with Europe. Empirical evidence reveals high consistency between measures built on USPC and CPC classifications (Lobo and Strumsky, 2019).